\documentclass{cernrep}
\usepackage{tikz}
\usepackage{amsmath}
\usepackage{amssymb}
\usepackage{epic,eepic,pspicture,mathptmx,times,graphpap}
\usepackage{tikz}

\usetikzlibrary{decorations}
\usetikzlibrary{decorations.pathmorphing}
\usetikzlibrary{decorations.pathreplacing}
\usetikzlibrary{arrows,backgrounds,automata}
\usepgflibrary{plotmarks}
\usetikzlibrary{trees}
\usetikzlibrary{patterns}
\usepackage[small,bf]{caption}

\begin{document}
\newcommand{\eqn}[3]{\parbox{9cm}{$ #1 $}\hspace{1cm}\parbox{5cm}{\raggedright\bf #2}\hfill\parbox[b]{8mm}{\begin{equation} \label{#3} \end{equation}}}
\newcommand{\meqn}[3]{\noindent\parbox{9cm}{\begin{align*} #1 \end{align*}}\hspace{1cm}\parbox{5cm}{\raggedright\bf #2}\hfill\parbox[b]{8mm}{\begin{equation} \label{#3} \end{equation}}}
\newcommand{\meqnbis}[3]{\noindent\parbox{9cm}{\begin{align*} #1 \end{align*}}\hspace{1cm}\parbox{4cm}{\raggedright\bf #2}\hfill\parbox[b]{8mm}{\begin{equation} \label{#3} \end{equation}}}
\newcommand{\meqnter}[3]{\noindent\parbox{9cm}{\begin{align*} #1 \end{align*}}\hspace{1cm}\parbox{4cm}{\raggedright\bf #2}\hfill\parbox[b]{8mm}{\begin{equation} \label{#3} \end{equation}}}
\newcommand{\dd}{\mathrm{d}}
\newcommand{\twovector}[2]{\begin{pmatrix} \displaystyle #1 \vspace{0.2cm} \\ \displaystyle #2 \end{pmatrix}}
\newcommand{\threevector}[3]{\begin{pmatrix} \displaystyle #1 \vspace{0.2cm}\\ \displaystyle #2 \vspace{0.2cm}\\ \displaystyle #3 \end{pmatrix}} %

\title{RF Basics I and II}
\author{Frank Gerigk}
\institute{CERN, Geneva, Switzerland}
\maketitle

\begin{abstract}
Maxwell's equations are introduced in their general form, together
with a basic set of mathematical operations needed to work with
them. After simplifying and adapting the equations for application
to radio frequency problems, we derive the most important formulae
and characteristic quantities for cavities and waveguides. Several
practical examples are given to demonstrate the use of the derived
equations and to explain the importance of the most common figures
of merit.
\end{abstract}

\section{Introduction to Maxwell's equations}

\subsection{Maxwell's equations}
Maxwell's equations were published in their earliest form in
1861--1862 in a paper entitled ``On physical lines of force'' by
the Scottish physicist and mathematician James Clerk Maxwell. They
represent a uniquely complete set of equations that covers all
areas of electrostatic and magnetostatic problems, as well as
electrodynamic problems, of which radio frequency (RF) engineering
is only a subset. Surprisingly, they include the effects of
relativity even though they were conceived much earlier than
Einstein's theories.

In differential form, Maxwell's equations can be written as\\
\begin{align}
    &\nabla \times \mathbf{H} = \mathbf{J} + \frac{\partial\mathbf{D}}{\partial t} \,, \label{eq:rotH}\\
    &\nabla \times \mathbf{E} = -\frac{\partial\mathbf{B}}{\partial t} \,, \label{eq:rotE} &&\mbox{\bf Maxwell's equations}\\
    &\nabla \cdot \mathbf{D} = q_v \,, \label{eq:divD} \\
    &\nabla \cdot \mathbf{B} = 0  \,, \label{eq:divB}
\intertext{where the field components and constants are defined as
follows:}
    &\mathbf{E} \,, &&\mbox{{\bf electric field} (V/m)} \nonumber \\
    &\mathbf{D} = \varepsilon_0\varepsilon_\mathrm{r} \mathbf{E} \,, &&\mbox{{\bf dielectric displacement} (A~s/m$^2$)} \\
    &\mathbf{B} \,, &&\mbox{{\bf magnetic induction, magnetic flux density} (T)} \nonumber \\
    &\mathbf{H} = \frac{1}{\mu_0\mu_\mathrm{r}}\mathbf{B} \,, &&\mbox{{\bf magnetic field strength or field intensity} (A/m)} \\
    &\mathbf{J} = \kappa\mathbf{E} \,, &&\mbox{{\bf electric current density} (A/m$^2$)} \\
    &\frac{\dd}{\dd t}\mathbf{D} \,, &&\mbox{{\bf displacement current} (A/m$^2$)} \nonumber \\
    &\varepsilon_0 = 8.854 \cdot 10^{-12} \,, &&\mbox{{\bf electric field constant} (F/m)} \\
    &\varepsilon_\mathrm{r} \,, &&\mbox{\bf relative dielectric constant} \nonumber \\
    &\mu_0 = 4\pi \cdot 10^{-7} \,, &&\mbox{{\bf magnetic field constant} (H/m)} \\
    &\mu_\mathrm{r} \,, &&\mbox{\bf relative permeability constant} \nonumber\\
    &\kappa \,. &&\mbox{{\bf electrical conductivity} (S/m)} \nonumber
\end{align}


In the following sections, we shall see that most of the important
RF formulae can be derived in a few lines from Maxwell's
equations.

\subsection{Basic vector analysis and its application to Maxwell's equations}
In order to make efficient use of Maxwell's equations, some basic
vector analysis is needed, which is introduced in this section.
More detailed introductions can be found in a number of textbooks,
such as for instance the excellent \emph{Feynman Lectures on
Physics} \cite{bib:feynman}.

\subsubsection*{Gradient of a potential}
The gradient of a potential $\phi$ is the derivative of the
potential function $\phi(x,y,z)$ in all directions of a particular
coordinate system (e.g., $x$, $y$, $z$). The result is a vector
that tells us how much the potential changes in different
directions. Applied to the geographical profile of a mountain
landscape, the gradient describes the slope of the landscape in
all directions. The mathematical sign that is used for the
gradient of a potential is the `nabla operator'; applied to a
Cartesian coordinate system, one can
 write\\
\meqn{
    \nabla \Phi = \begin{pmatrix} \frac{\partial}{\partial x} \\[1ex] \frac{\partial}{\partial y} \\[1ex] \frac{\partial}{\partial z} \end{pmatrix}
    \Phi = \begin{pmatrix} \frac{\partial\Phi}{\partial x} \\[1ex] \frac{\partial\Phi}{\partial y} \\[1ex] \frac{\partial\Phi}{\partial z}
    \end{pmatrix} \,.
}{gradient of a potential}{gradient}

\noindent The expressions for the gradient in cylindrical and
spherical coordinate systems are given in Appendices
\ref{sec:cylindrical} and \ref{sec:spherical}.

\subsubsection*{Divergence of a vector field}

The divergence of a vector field $\mathbf{a}$ tells us if the
vector field has a source. If the resulting scalar expression is
zero, we have a `source-free' vector field, as in the case of the
magnetic field. From basic physics, we know that there are no
magnetic monopoles, which is why magnetic field lines are always
closed. In Maxwell's equations \eqref{eq:divB}, this property is
included by means of the fact that the divergence of the magnetic
induction $\mathbf{B}$ equals zero.

In Cartesian coordinates, the divergence of a vector field is defined as\\
\meqn{\displaystyle
    \nabla \cdot \mathbf{a} = \begin{pmatrix} \frac{\partial}{\partial x} \\[1ex] \frac{\partial}{\partial y} \\[1ex] \frac{\partial}{\partial z}
     \end{pmatrix} \cdot \begin{pmatrix} a_x \\[1ex] a_y \\[1ex] a_z \end{pmatrix} = \frac{\partial a_x}{\partial x} +
     \frac{\partial a_y}{\partial y} + \frac{\partial a_z}{\partial
     z} \,.
}{divergence of a vector}{eq:divergence}

\noindent The expressions for cylindrical and spherical coordinate
systems are given in Appendices \ref{sec:cylindrical} and
\ref{sec:spherical}.

\subsubsection*{Curl of a vector field}
When we form the curl of a vector, we are interested in knowing if there are any curls or eddies in the field. Let us
imagine that we are looking at the flow of water in a cooling pipe. To check for curls, we can use a stick around which
a ball can rotate freely. We position a Cartesian coordinate system at an arbitrary origin and align the stick first with the
 $x$ axis, and then with the $y$ and $z$ axes. If the ball starts rotating in any of these positions, then we know
that the curl of the vector field describing the water flow is
non-zero in the direction of the respective axis. The curl of a
vector $\mathbf{a}$ is therefore also a vector, because its
information is direction-specific. Its mathematical form in
Cartesian coordinates is defined as\\
\meqn{
    \nabla \times \mathbf{a} &= \begin{pmatrix} \frac{\partial}{\partial x} \\[1ex] \frac{\partial}{\partial y} \\[1ex] \frac{\partial}{\partial z}
     \end{pmatrix} \times \begin{pmatrix} a_x \\[1ex] a_y \\[1ex] a_z \end{pmatrix} \\
     &= \mathrm{det} \begin{pmatrix} \mathbf{u}_x & \mathbf{u}_y & \mathbf{u}_z \\[1ex]
     \frac{\partial}{\partial x} & \frac{\partial}{\partial y} & \frac{\partial}{\partial z} \\[1ex] a_x & a_y & a_z \end{pmatrix}
     =  \begin{pmatrix} \frac{\partial a_z}{\partial y} -
     \frac{\partial a_y}{\partial z} \\[1ex] \frac{\partial a_x}{\partial z} -
     \frac{\partial a_z}{\partial x} \\[1ex] \frac{\partial a_y}{\partial x} - \frac{\partial a_x}{\partial y}
     \end{pmatrix} \,.
}{curl of a vector}{eq:curl}

\noindent The \textit{unit} vectors $\mathbf{u}_n$ have no
physical meaning and simply point in the $x$, $y$, and $z$
directions. They have a constant length of 1. The expressions for
cylindrical and spherical coordinate systems can be found in
Appendices \ref{sec:cylindrical} and \ref{sec:spherical}.

\subsubsection*{Second derivatives}
In some instances, we have to make use of second derivatives. One
of the expressions that is used regularly in electrodynamics is
the Laplace operator $\Delta = \nabla^2$, which---since the
operator itself is scalar---can be applied to both scalar fields
and vector fields:

\vspace*{-0.6cm}
\meqn{
    \displaystyle \Delta \phi = \nabla \cdot \left( \nabla \phi \right) = \nabla^2 \phi
     = \frac{\partial^2 \phi }{\partial x^2} + \frac{\partial^2 \phi}{\partial y^2} + \frac{\partial^2 \phi}{\partial z^2}
 \,.}{Laplace operator}{eq:laplace} The expressions for cylindrical
and spherical coordinate systems can be found in Appendices
\ref{sec:cylindrical} and \ref{sec:spherical}.

We also introduce two interesting identities:
\begin{align}
    \nabla \times (\nabla \phi) &= 0 \,, \label{eq:potential0}\\
    \nabla\cdot (\nabla \times \mathbf{a}) &= 0 \,. \label{eq:vector0}
\end{align}
Equation~\eqref{eq:potential0} tells us that if the curl of a
vector equals zero, then this vector can be written as the
gradient of a potential. This feature can save us a lot of writing
when we are dealing with complicated three-dimensional expressions
for electric and magnetic fields, and we shall use this principle
later on to define non-physical potential functions that can
describe (via derivatives) complete three-dimensional vector
functions.

In the same way, Eq.~\eqref{eq:vector0} can (and will) be used to
describe divergence-free fields with simple `vector potentials'.

\subsection{Useful theorems by Gauss and Stokes}
The theorems of Gauss and Stokes are some of the most commonly
used transformations in this chapter, and therefore we shall take
a moment to explain the concepts of them.

\subsubsection*{Gauss's theorem}

Gauss's theorem not only saves us a lot of mathematics but also
has a very useful physical interpretation when applied to
Maxwell's equations. Mathematically speaking, we transform a
volume integral over the divergence of a vector into a surface
integral over the vector itself:

\vspace*{-0.6cm}
\meqn{
    \int\limits_V \underbrace{\nabla \cdot \mathbf{a}}_{\mbox{`sources'}} \dd V = \oint\limits_S^{\textcolor{white}{C}} \mathbf{a} \cdot \dd \mathbf{S}
 \,.}{Gauss's theorem}{eq:Gauss}

\noindent The surface on the right-hand side of the theorem is the
one that surrounds the volume on the left-hand side. If we
remember that the divergence of a vector field is equal to its
sources, Gauss's theorem tells us that:

\begin{itemize}
    \item The vector flux through a closed surface equals the sources of flux within the enclosed volume.
    \item If there are no sources, the amounts of flux entering and leaving the volume must be equal.
\end{itemize}

These statements can be applied directly to Maxwell's equations.
Using Eq.~\eqref{eq:divD} and applying Gauss's theorem, we obtain
 \begin{equation}
 \int\limits_V \nabla \cdot \mathbf{E} \, \dd V = \oint\limits_S \mathbf{E} \cdot \dd \mathbf{S} = \frac{Q}{\varepsilon}
 \end{equation}
(Fig. 1), which means that one can calculate the amount of charge
in a volume simply by integrating the electric flux lines over any
closed surface that surrounds the charge, or vice versa.

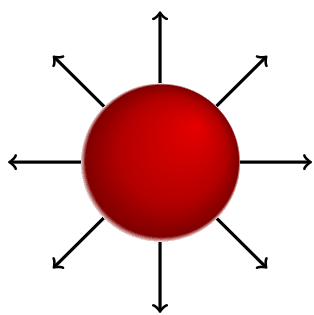
\begin{figure}[h!]
    \centering
    \begin{tikzpicture}
    \draw[->,very thick] (0,0) -- (2,0);
    \draw[->,very thick] (0,0) -- (-2,0);
    \draw[->,very thick] (0,0) -- (0,2);
    \draw[->,very thick] (0,0) -- (0,-2);
    \draw[->,very thick] (0,0) -- (1.41,1.41);
    \draw[->,very thick] (0,0) -- (1.41,-1.41);
    \draw[->,very thick] (0,0) -- (-1.41,1.41);
    \draw[->,very thick] (0,0) -- (-1.41,-1.41);
    \pgfdeclareradialshading{sphere}{\pgfpoint{0.5cm}{0.5cm}}%
    {rgb(0cm)=(0.9,0,0);
    rgb(0.7cm)=(0.7,0,0);
    rgb(1cm)=(0.5,0,0);
    rgb(1.05cm)=(1,1,1)}
    \pgfuseshading{sphere}
    \end{tikzpicture}
 \caption{Example of electric flux lines emanating from electric charge in the centre of a sphere}
\end{figure}

The same trick can be applied to the source-free magnetic field.
Here, we use Eq.~\eqref{eq:divB} and obtain
 \begin{equation}
 \int\limits_V \nabla \cdot \mathbf{B} \, \dd V = \oint\limits_S \mathbf{B} \cdot \dd \mathbf{S} =
0 \,.
 \label{eq:GaussB}
 \end{equation}
Equation~\eqref{eq:GaussB} gives us the proof of what was already
stated earlier: magnetic field lines have no sources
($\nabla\cdot\mathbf{B}=0$), and therefore the magnetic flux lines
are always closed and have neither sources nor sinks. If magnetic
flux lines enter a volume, then they also have to leave that
volume (Fig. 2).

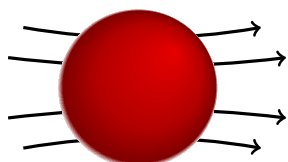
\begin{figure}[h!]
    \centering
    \begin{tikzpicture}
    \draw[->,very thick] (-1.5,0.8) arc (260:280:9);
    \draw[->,very thick] (-1.7,0.4) arc (263:277:15);
    \draw[->,very thick] (-1.5,-0.8) arc (100:80:9);
    \draw[->,very thick] (-1.7,-0.4) arc (97:83:15);
    \pgfdeclareradialshading{sphere}{\pgfpoint{0.5cm}{0.5cm}}
    {rgb(0cm)=(0.9,0,0);
    rgb(0.7cm)=(0.7,0,0);
    rgb(1cm)=(0.5,0,0);
    rgb(1.05cm)=(1,1,1)}
    \pgfuseshading{sphere}
    \end{tikzpicture}

    \vspace*{1ex}
    \caption{Example of magnetic flux lines penetrating a sphere}
\end{figure}

\subsubsection*{Stokes's theorem}
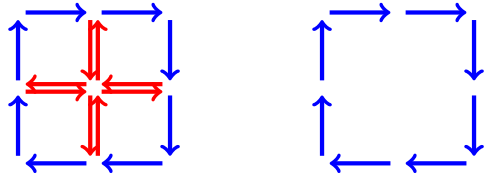
\begin{figure}[h!]
    \centering
    \begin{tikzpicture}[->,scale=1, ultra thick, blue]
    \draw (0,0.1) -- (0,0.9);
    \draw (0,1.1) -- (0,1.9);
    \draw (0.1,2) -- (0.9,2);
    \draw (1.1,2) -- (1.9,2);
    \draw (2,1.9) -- (2,1.1);
    \draw (2,0.9) -- (2,0.1);
    \draw (1.9,0) -- (1.1,0);
    \draw (0.9,0) -- (0.1,0);

    \draw[red] (0.1,0.95) -- (0.9,.95);
    \draw[red] (0.9,1.05) -- (0.1,1.05);
    \draw[red] (0.95,0.9) -- (0.95,0.1);
    \draw[red] (1.05,0.1) -- (1.05,0.9);
    \draw[red] (1.1,0.95) -- (1.9, 0.95);
    \draw[red] (1.9,1.05) -- (1.1,1.05);
    \draw[red] (1.05,1.1) -- (1.05,1.9);
    \draw[red] (0.95,1.9) -- (0.95,1.1);

    \draw (4,0.1) -- (4,0.9);
    \draw (4,1.1) -- (4,1.9);
    \draw (4.1,2) -- (4.9,2);
    \draw (5.1,2) -- (5.9,2);
    \draw (6,1.9) -- (6,1.1);
    \draw (6,0.9) -- (6,0.1);
    \draw (5.9,0) -- (5.1,0);
    \draw (4.9,0) -- (4.1,0);

    \end{tikzpicture}
    \caption{Illustration of Stokes's theorem}
    \label{fig:Stokes}
\end{figure}

Whereas Gauss's theorem is useful for equations involving the
divergence of a vector, Stokes's theorem offers a similar
simplification for equations that contain the curl of a vector.
With Stokes's theorem, we can transform surface integrals over the
curl of a vector into closed line integrals over the vector
itself:

\vspace*{-0.6cm}
\meqn{
    \int\limits_A \left(\nabla \times \mathbf{a} \right) \cdot \dd \mathbf{A} = \oint\limits_C^{\textcolor{white}{C}} \mathbf{a} \cdot \dd \mathbf{l}
\,. }{Stokes's theorem}{eq:Stokes}
 One can interpret Stokes's
theorem with the help of Fig.~\ref{fig:Stokes} as follows:

\begin{itemize}
    \item the area integral over the curl of a vector field can be calculated from a line integral along its closed borders, or
    \item the field lines of a vector field with non-zero curl must be closed contours.
\end{itemize}

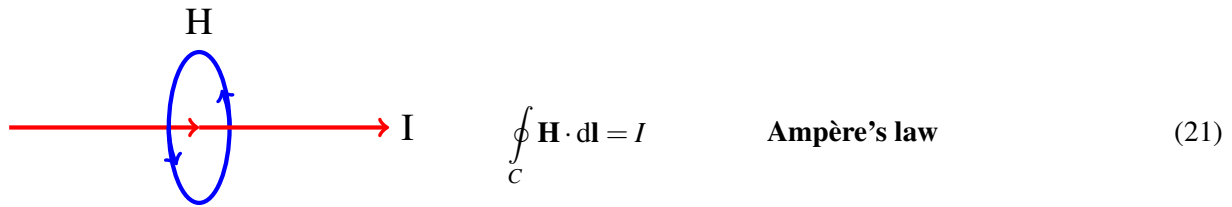
\begin{figure}[h!]
\parbox{6cm}{
    \begin{tikzpicture}[scale=1]
    \draw[->,ultra thick,red] (0,0) -- (2.5,0);
    \draw[->,ultra thick, blue] (2.5,0) ellipse (0.4 and 1);
    \draw (2.5,1.1) node [above, black] {\Large H};
    \draw[->,ultra thick,blue] (2.1,0) arc (180:210:0.8 and 1);
    \draw[->,ultra thick,blue] (2.9,0) arc (0:30:0.8 and 1);
    \draw[->,ultra thick,red] (2.5,0) -- (5,0) node [right, black] {\Large I};
    \end{tikzpicture}
    \caption{\label{fig:Amp1}Illustration of Amp\`{e}re's law}}\parbox{3cm}{
    \begin{align*} \oint\limits_C^{\textcolor{white}{C}} \mathbf{H} \cdot \dd \mathbf{l} =
    I \end{align*}}\hspace{1cm}\parbox{5cm}{\raggedright\bf Amp\`{e}re's
    law}\hfill\parbox[b]{8mm}{\begin{equation} \label{eq:Ampere} \end{equation}}
\end{figure}

The meaning of these statements becomes immediately clear when we
apply Stokes's theorem to Maxwell's equation~\eqref{eq:rotH}:
\begin{equation}
    \int\limits_A \left(\nabla\times\mathbf{H}\right)\cdot \dd \mathbf{A} = \oint\limits_C \mathbf{H} \cdot \dd \mathbf{l}
    = \int\limits_A \left( \mathbf{J}+\frac{\dd \mathbf{D}}{\dd t} \right) \cdot \dd
    \mathbf{A} \,.
\end{equation}
In the electrostatic case, the time derivative disappears and the
area integral over the current density may, for instance, be the
current flowing in an electric wire as shown in
Fig.~\ref{fig:Amp1}. This means that with a one-line manipulation
of Maxwell's equations, we have derived Amp\`{e}re's law, which
tells us that every current induces a circular magnetic field
around itself, whose strength can be be calculated from a simple
closed line integral along a circular path with the current at its
centre.

With similar ease, we can derive Faraday's induction law, which is
the basis of every electric motor and generator. All we have to do
is apply Stokes's theorem to Maxwell's equation~\eqref{eq:rotE}:
\begin{equation}
    \int\limits_A\left(\nabla\times\mathbf{E}\right) \cdot \dd \mathbf{A} =
    \underbrace{\oint\limits_c \mathbf{E} \cdot \dd \mathbf{l}}_{V_i} =
    - \underbrace{\frac{\dd}{\dd t}\int\limits_A \mathbf{B}\cdot \dd \mathbf{A}}_{\displaystyle\frac{\dd \psi_m}{\dd
    t}} \,,
\end{equation}

\noindent and again, after one line, we obtain one of the
fundamental laws of electrical engineering.
\begin{figure}[h!]
\parbox{6cm}{
    \centering
    \begin{tikzpicture}[scale=1, ultra thick, blue]
    \draw (0,0) circle (1);
    \foreach \x in {-1.5,-1,-0.5,0,0.5,1,1.5}
        \foreach \y in {-1.5,-1,-0.5,0,0.5,1,1.5}
            \node [red] at (\x,\y) {x};
    \draw [->] (0,1) arc (90:45:1);
    \node [fill,white] at (1.0,1.2) {\Large\textcolor{black}{$V_i$}};
    \end{tikzpicture}
    \caption{\label{fig:Faraday}Illustration of Faraday's induction law}}\parbox{3cm}{
    \begin{align*}  V_i = -\frac{\dd \psi_m}{\dd t} \end{align*}}\hspace{1cm}\parbox{5cm}{\raggedright\bf Faraday's induction
     law}\hfill\parbox[b]{8mm}{\begin{equation} \label{eq:Faraday} \end{equation}}
\end{figure}

Faraday's law tells us that an electric voltage is induced in a
loop if the magnetic flux $\psi$ penetrating the loop changes over
time, as shown in Fig.~\ref{fig:Faraday}. Alternatively, one can
change the flux by moving the loop in or out of a static magnetic
field.

I hope that these examples have convinced you that Maxwell's
equations are indeed very powerful, and that with a bit of vector
analysis we really can derive everything we need for RF
engineering (although maybe not always in one line \ldots).

\subsection{Displacement current}
Although most people have an idea of what electric and magnetic
fields are, the displacement current $\dd \mathbf{D}/\dd t$ is
often not so well understood. Since it is vital for the
propagation of electromagnetic waves, we shall spend a few lines
studying this quantity. We start by deriving and interpreting the
continuity equation, and then look at a simple practical example.

We apply the divergence to Maxwell's equation~\eqref{eq:rotH}:
\begin{equation}
    \underbrace{\nabla \cdot \left( \nabla \times \mathbf{H} \right)}_{\displaystyle\equiv 0} =
    \nabla \cdot \mathbf{J} + \underbrace{\nabla \cdot \frac{\dd \mathbf{D}}{\dd t}}_{\displaystyle\frac{\dd}{\dd t}
    \rho_v} \,.
    \label{eq:pre-continuity}
\end{equation}

\noindent Using Maxwell's equation~\eqref{eq:divD}, we have made
an association between the `sources of the displacement current'
$\nabla \cdot ({\dd \mathbf{D}}/{\dd t})$ and the `rate of change
of electric charge' $({\dd}/{\dd t}) \rho_v$. Using the
identity~\eqref{eq:vector0}, we obtain the continuity equation

\vspace*{-0.5cm}
\meqn{
    \displaystyle \nabla\cdot\mathbf{J} = -\frac{\dd}{\dd t} \rho_v
\,,}{continuity equation}{eq:continuity} to which we apply a
volume integral and Gauss's theorem~\eqref{eq:Gauss}:

\vspace*{-0.3cm} \meqn{ \int\limits_V\nabla \cdot \mathbf{J} \,
\dd V = \oint\limits_S \mathbf{J} \cdot \dd \mathbf{S} = \sum I_n
= -\frac{\dd}{\dd t} \int\limits_V \rho_v \, \dd V \,.}{continuity
equation}{eq:continuityV} In this form, the interpretation is very
straightforward, and we can state that:
\begin{itemize}
    \item if the amount of electric charge in a volume is changing over time, a current needs to flow; or, more poignantly, electric
 charges cannot be destroyed.
\end{itemize}

\begin{figure}[h!]
\centering
    \begin{tikzpicture}[scale=1, ultra thick]
    \draw[->,red] (0,0) -- (0.5,0) node[red,above] {I};
    \draw[red] (0.5,0) -- (2,0);
    \draw[->,red] (3,0) -- (5,0);
    \draw[line width = 4pt] (2,-1) -- (2,1);
    \draw[line width = 4pt] (3,-1) -- (3,1);
    \draw[->] (2.2,0) -- (2.8,0);
    \draw[->] (2.2,0.5) -- (2.8,0.5);
    \draw[->] (2.2,-0.5) -- (2.8,-0.5);
    \draw[->] (2.2,1) arc (102:78:1.5);
    \draw[->] (2.2,-1) arc (258:282:1.5);
    \draw[dashed] (1.8,0) ellipse (0.7 and 1.7);
    \draw (1.6,-0.6) node {$V$};
    \draw (1.8,-1.7) -- (1.4,-2.4) node [below,left] {$S$};
    \draw (2.7,1.9) node {$\displaystyle\frac{\dd \mathbf{D}}{\dd t}$};
    \end{tikzpicture}
\caption{Example of a displacement current: charging of a
capacitor} \label{fig:capacitor}
\end{figure}
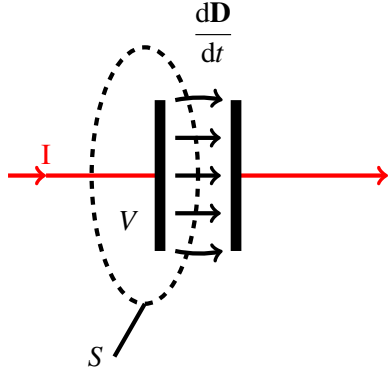

Now, it is good to know that electric charges cannot be destroyed,
but that does not yet help us to understand the displacement
current. For this purpose, we go back to
Eq.~\eqref{eq:pre-continuity} and this time we do not replace the
expression for the displacement current. Instead, we apply a
volume integral and Gauss's theorem and obtain
\begin{equation}
    \oint\limits_S \mathbf{J}\cdot \dd \mathbf{S} = \sum I_n = -\frac{\dd}{\dd t}\oint\limits_S \mathbf{D} \, \dd S =
    -\frac{\dd}{\dd t}\int\limits_V \rho_v \, \dd V \,,
    \label{eq:displacement-example}
\end{equation}
which we can apply to the simple geometry of a capacitor shown in Fig.~\ref{fig:capacitor}, which is charged by a static current $I$.

If we assume a small volume with a surface $S$ around one of the
capacitor plates, then we can directly interpret
Eq.~\eqref{eq:displacement-example}: the current $I$, which enters
the volume $V$ on the left, equals the flux integral of the
displacement current $-({\dd}/{\dd t}) \mathbf{D}$, which leaves
the volume $V$ on the right. This means that the displacement
current can be understood as a `current without charge transport',
which in this case can only exist because of the rate of change of
the electric charge ($-({\dd}/{\dd t})\int\limits_V \rho_v \, \dd
V$) on the left capacitor plate.

\subsection{Boundary conditions}
\label{sec:boundary} Before we try to calculate electromagnetic
fields in accelerating cavities, we need to understand how these
fields behave close to material boundaries, for example the
electrically conducting walls of a cavity. Using Stokes's and
Gauss's theorems, we can quickly derive these boundary conditions.

\subsubsection*{Field components parallel to a material boundary}
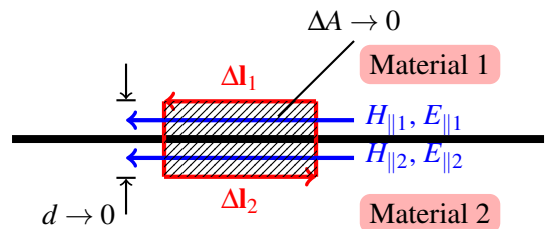
\begin{figure}[h!]
    \centering
    \begin{tikzpicture}[scale=1, ultra thick]
    \draw[line width = 3pt] (-0.5,0) -- (6.5,0);
    \draw[->,red] (3.5,0.5) -- (1.5,0.5) node [above,midway] {$\Delta \mathbf{l}_1$};
    \draw [red] (1.5,0.5) -- (1.5,-0.5);
    \draw[->,red] (1.5,-0.5) -- (3.5,-0.5) node [below,midway] {$\Delta \mathbf{l}_2$};
    \draw[red] (3.5,-0.5) -- (3.5,0.5);

    \draw[<-, blue] (1,0.25) -- (4,0.25) node [right] {$H_{\parallel 1}$,  $E_{\parallel 1}$};
    \draw[<-, blue] (1,-0.25) -- (4,-0.25) node [right] {$H_{\parallel 2}$,  $E_{\parallel 2}$};

    \draw[->|,thick] (1,1.0) -- (1,0.5);
    \draw[->|,thick] (1,-1.0) node[left] {$d \rightarrow 0$} -- (1,-0.5);
    \fill[pattern=north east lines] (3.5,0.5) -- (1.5,0.5) -- (1.5,-0.5) -- (3.5,-0.5) -- (3.5,0.5);
    \draw[thick] (3,0.3) -- (4,1.3) node[above] {$\Delta A \rightarrow 0$};
    \draw (5,0.7) node[fill=red!30,right,above,rounded corners] {Material 1};
    \draw (5,-0.7) node[fill=red!30,right,below,rounded corners] {Material 2};
    \end{tikzpicture}
    \caption{Boundary conditions parallel to a material boundary}
    \label{fig:boundary-parallel}
\end{figure}

We start with the field components ($E_{\parallel}$,
$H_{\parallel}$) parallel to a surface between two materials, as
depicted in Fig.~\ref{fig:boundary-parallel}. We define a small
surface $\Delta A$, which is perpendicular to the boundary and
encloses a small cross-section of the boundary area. Then we
integrate Maxwell's equations~\eqref{eq:rotH} and \eqref{eq:rotE}
over this area and apply Stokes's theorem:
\begin{align}
    \int\limits_A \nabla \times \mathbf{H} \cdot \dd \mathbf{A} &=  \oint\limits_C \mathbf{H}\cdot \dd \mathbf{l} =
    \underbrace{\int\limits_A \mathbf{J}\cdot \dd \mathbf{A}}_{=i'\Delta l} +
    \underbrace{\frac{\dd}{\dd t} \int\limits_A \mathbf{D} \cdot \dd \mathbf{A} \,,}_{\rightarrow 0 \mbox{ for } \mathbf{A} \rightarrow 0}\\
    \int\limits_A \nabla \times \mathbf{E} \cdot \dd \mathbf{E} &= \oint\limits_C \mathbf{E} \cdot \dd\mathbf{l} =
    -\underbrace{\frac{\dd}{\dd t}\int\limits_A \mathbf{B} \cdot \dd \mathbf{A}\,.}_{\rightarrow 0 \mbox{ for } \mathbf{A}\rightarrow 0}
\end{align}

Using Stokes's theorem, the area integrals over $A$ are
transformed into line integrals around the contour $C$ of the
area. If the width $d$ of the area (see
Fig.~\ref{fig:boundary-parallel}) is now reduced to zero, the
calculation of the contour integral simplifies to a multiplication
of the field components $E_{\parallel}$ and $H_{\parallel}$ by the
path elements $\Delta l$. The area integrals over $\mathbf{D}$ and
$\mathbf{B}$ vanish and the area integral over the current density
$\mathbf{J}$ is replaced by a surface current, which may flow in
the boundary plane between the two materials, times the path
element $\Delta l$. This results in the following boundary
conditions:

\vspace*{-0.5cm}
\meqn{
    &H_{\parallel 1} - H_{\parallel 2} = i' \,, \\
    &E_{\parallel 1} = E_{\parallel 2}
\,.}{conditions for magnetic and electric fields parallel to a
material boundary}{eq:generalboundaries} In the case of a
waveguide or an accelerator cavity, we generally assume one of the
materials (e.g., material 2) to be an ideal electrical conductor,
and in that case the electric and magnetic field components in
this material vanish, so that we obtain

\vspace*{-0.5cm}
\meqn{
    H_{\parallel 1} &= i' \,, \\
     E_{\parallel 1} &=0
\,.}{conditions for magnetic and electric fields parallel to ideal
electric surfaces}{eq:electricboundary}

\subsubsection*{Field components perpendicular to a material boundary}

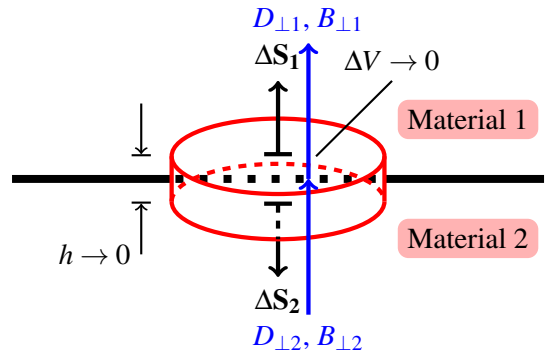
\begin{figure}[h!]
\centering
    \begin{tikzpicture}[scale=1, ultra thick]
    \draw[line width = 3pt] (-1,0) -- (1.1,0);
    \draw[line width = 3pt] (3.9,0) -- (6,0);
    \draw[loosely dashed, line width = 3pt] (1.15,0) -- (3.85,0);
    \draw[red] (2.5,0.3) ellipse (1.4 and 0.5);
    \draw[dashed,red] (1.1,-0.3) arc (180:0:1.4 and 0.5);
    \draw[->] (2.5,-0.8) -- (2.5, -1.3) node [below] {$\Delta\mathbf{S_2}$};
    \draw[red] (1.1,-0.3) arc (180:360:1.4 and 0.5);
    \draw[red] (1.1,0.3) -- (1.1,-0.3);
    \draw[red] (3.9,0.3) -- (3.9,-0.3);
    \draw[|->] (2.5,0.3) -- (2.5,1.3)node [above] {$\Delta\mathbf{S_1}$};
    \draw[|-,dashed] (2.5,-0.3) -- (2.5,-0.8);

    \draw[blue,->] (2.9, 0) -- (2.9, 1.8) node [above] {$D_{\perp 1}$, $B_{\perp 1}$};
    \draw[blue,<-] (2.9, 0) -- (2.9, -1.8) node [below] {$D_{\perp 2}$, $B_{\perp 2}$};

    \draw[->|,thick] (0.7,1.0) -- (0.7,0.3);
    \draw[->|,thick] (0.7,-1.0) node[left] {$h \rightarrow 0$} -- (0.7,-0.3);
    \draw[thick] (3,0.3) -- (4,1.3) node[above] {$\Delta V \rightarrow 0$};
    \draw (5,0.5) node[fill=red!30,right,above,rounded corners] {Material 1};
    \draw (5,-0.5) node[fill=red!30,right,below,rounded corners] {Material 2};
    \end{tikzpicture}
    \caption{Boundary conditions perpendicular to a material boundary}
    \label{fig:normalboundary}
\end{figure}

In a very similar way, we can derive the boundary conditions for
fields ($D_{\perp}$, $B_{\perp}$) that are perpendicular to a
boundary surface between two materials. This time, however, we do
not define an area but a small cylinder with a volume $\Delta V$
around the boundary, as shown in Fig.~\ref{fig:normalboundary}. We
form a volume integral from Maxwell's equations \eqref{eq:divD}
and \eqref{eq:divB} over the volume of the cylinder and apply
Gauss's theorem to transform the volume integrals into surface
integrals:
\begin{align}
    \int\limits_V\nabla\cdot\mathbf{D} \, \dd V &= \oint\limits_S\mathbf{D} \cdot \dd \mathbf{S} = \int\limits_V q_v \, \dd V \,, \\
    \int\limits_V\nabla\cdot\mathbf{B} \, \dd B &= \oint\limits_S\mathbf{B}\cdot \dd \mathbf{S} =
    0 \,.
\end{align}

In the following step, we reduce the height of the cylinder to
zero, so that we end up with two surfaces, one on each side of the
boundary. And now it becomes clear why we have to start with a
volume integral. Since the surface element $d\mathbf{S}$ is
perpendicular to the surface of the cylinder, the `dot product' in
the integrals basically reduces the vector fields $\mathbf{D}$ and
$\mathbf{B}$ to the components perpendicular to the surface of the
cylinder. This means that above equations can now be written as\\
\meqn{
    & D_{\perp 1} - D_{\perp 2} = q_s \,, \\
    & B_{\perp 1} = B_{\perp 2}
\,,}{conditions for dielectric displacement and magnetic induction
perpendicular to a material boundary}{eq:perpboundary}
 where $q_s$
is a surface charge (measured in units of C/m$^2$) that may exist
on the boundary surface. In the case where material 2 is an ideal
conductor, we obtain

\vspace*{-0.5cm}
\meqn{
    D_{\perp 1} &= q_s \,, \\
    B_{\perp 1} &= 0
\,.}{conditions for dielectric displacement and magnetic induction
perpendicular to ideal electric surface}{eq:perboundaryel}

We note that when the fields are parallel to a boundary surface,
the electric and magnetic fields are used in the boundary
conditions, whereas when they are perpendicular to the boundary
surface, we have a condition for the dielectric displacement and
the magnetic induction.  This means that, for instance, the
tangential electric field $E_{\parallel}$ may be smooth across a
boundary but there will be a jump in the dielectric displacement
$D_{\parallel}$ if there are different relative dielectric
constants $\varepsilon_\mathrm{r}$ in the two materials.
Similarly, the component of the magnetic induction $B_{\perp}$
perpendicular to a surface may be smooth, whereas the magnetic
field $H_{\perp}$ will jump if the two materials have different
relative magnetic field constants $\mu_\mathrm{r}$.

\section{Electromagnetic waves}

In this section, we shall derive the general form of the wave
equation and then restrict ourselves to phenomena that are
harmonic in time. Since RF systems mostly deal with sinusoidal
waves, we shall be able to explain and understand most of the
relevant phenomena with this approach. This includes the `skin
effect', the propagation of energy, RF losses, and acceleration
via travelling waves.

\subsection{The wave equation}
We start with the simplification of looking only at homogeneous,
isotropic media, meaning we assume that the electromagnetic fields
`see' the same material constants ($\mu$, $\varepsilon$, $\kappa$)
in all directions. With this assumption, Maxwell's equations can
be conveniently expressed in terms of only $E$ and $H$:
\begin{align}
    \nabla \times \mathbf{H} &= \kappa\mathbf{E} + \varepsilon\frac{\partial\mathbf{E}}{\partial t} \,, \label{eq:rotHb}\\
    \nabla \times \mathbf{E} &= -\mu\frac{\partial\mathbf{H}}{\partial t} \,, \label{eq:rotEb} &\hspace*{1.5cm}&\mbox{\bf Maxwell's equations}\\
    \nabla \cdot \mathbf{E} &= \frac{q_v}{\varepsilon} \,,  \label{eq:divE} \\
    \nabla \cdot \mathbf{H} &= 0 \,.  \label{eq:divH}
\end{align}
The curl of Eq.~\eqref{eq:rotEb} together with
Eq.~\eqref{eq:rotHb}, and the curl of Eq.~\eqref{eq:rotHb}
together with Eqs.~\eqref{eq:rotEb}
and \eqref{eq:divE} result in the general wave equations for a homogeneous medium\\
\meqn{
    \nabla^2\mathbf{E} -\nabla\left(\nabla\cdot\mathbf{E}\right) &=
    \mu\kappa\frac{\dd}{\dd t}\mathbf{E} + \mu\epsilon\frac{\dd^2}{\dd t^2}\mathbf{E} \,,\\
    \nabla^2\mathbf{H} &= \mu\kappa\frac{\dd}{\dd t}\mathbf{H} + \mu\epsilon\frac{\dd^2}{\dd t^2}\mathbf{H}
\,.}{wave equations in a homogeneous medium}{eq:waveshomogeneous}
In the case of waveguides and cavities, we can simplify these
equations even further by considering only the fields inside the
waveguide or cavity, which exist in
a non-conducting medium ($\kappa =0$) and a charge-free volume ($\nabla\cdot E = 0$):\\
\meqn{
    \nabla^2\mathbf{E} &= \mu\epsilon\frac{\dd^2}{\dd t^2}\mathbf{E}\,,\\
    \nabla^2\mathbf{H} &= \mu\epsilon\frac{\dd^2}{\dd t^2}\mathbf{H}
\,.}{wave equations in a non-conducting, charge-free homogeneous
medium}{eq:wave-eq}

\subsection{Complex notation for time-harmonic fields}
The already compact wave equations in Eq.~\eqref{eq:wave-eq} can
be simplified even further by taking into account the fact that in
RF engineering one usually deals with time-harmonic signals, which
are sometimes modulated in phase or amplitude. We can therefore
introduce the complex notation for electric and magnetic fields.
We start by assuming a time-harmonic electric field with amplitude
$E_0$ and phase $\varphi$,
\begin{equation}
    E(t) = E_0 \cos\left(\omega t+\varphi\right) \,, 
 \end{equation}
\noindent which we can interpret as the real part of a complex
expression,
\begin{equation}
    E(t) = \Re \left\{E_0 e^{i\varphi} e^{i\omega t} \right\} = \Re \left\{E_0\cos\left(\omega t + \varphi\right) +
 iE_0\sin\left(\omega t + \varphi\right)\right\} \,.
  \end{equation}
\noindent In this form, we can easily separate the harmonic time
dependence $\omega t$ from the phase information $\varphi$. The
phase information can be merged into the amplitude by defining a
`complex amplitude' or `phasor'
\begin{equation}
    \tilde{E} = E_0 e^{i\varphi} \,. 
      \end{equation}
\noindent We keep in mind that the real physical fields are
obtained as the real part of the complex amplitude times
$e^{i\omega t}$:
\begin{equation}
    E_0 \cos\left(\omega t + \varphi\right) = \Re \left\{ \tilde{E} e^{i\omega t} \right\} \,. 
 \end{equation}
\indent To simplify our writing, we skip the part with
the harmonic time dependence and omit the tilde, which means that
from now on all field quantities are written as complex
amplitudes. In order to convince you that this really is a
simplification, let us consider what happens to time derivatives
when complex notation is used:
\begin{equation}
    \frac{\dd}{\dd t} \tilde{E}e^{i\omega t} = i\omega \tilde{E} e^{i\omega
    t} \,.
              \end{equation}

This means that all time derivatives in Maxwell's equations and
also in the wave equations can simply be replaced by a
multiplication by $i\omega$, and we are able to do this because
the time dependence is always harmonic. Only when we have to deal
with transient events, such as the switching on of an RF amplifier
or the sudden arrival of a beam in a cavity, do we have to go back
the non-harmonic general equations.

As our first application of the complex notation, we rewrite
Maxwell's equations as follows:
\begin{align}
    \nabla\times\mathbf{H} &= i\omega\underline{\varepsilon}\mathbf{E} \,,  \\
    \nabla\times\mathbf{E} &= -i\omega\mu\mathbf{H} \,,   &\hspace*{2.2cm}&\mbox{\bf Maxwell's equations in}\\
    \nabla\cdot\mathbf{E} &= \frac{\rho_V}{\varepsilon} \,,  &&\mbox{\bf complex notation}\\
    \nabla\cdot\mathbf{H} &= 0 \,,  \\
\intertext{where the complex dielectric constant
$\underline{\varepsilon}$ is defined as}
    \underline{\varepsilon} &= \varepsilon' - i\varepsilon'' = \varepsilon\left( 1-i\frac{\kappa}{\omega\varepsilon}\right) \,.
    &&\mbox{\bf complex dielectric constant} \label{eq:complexdielectric}
\intertext{We note that $\underline{\varepsilon}$ is complex only
in a conducting medium. We can now proceed to write the
 general wave equations in complex form:}
    \nabla^2\mathbf{E}-\nabla\left(\nabla\cdot\mathbf{E}\right) &= -\underline{k}^2\mathbf{E} \,,
      &&\mbox{\bf general complex} \label{eq:complexwaveeqE} \\
    \nabla^2\mathbf{H} &= -\underline{k}^2\mathbf{H} \,.  && \mbox{\bf wave equations} \label{eq:complexwaveeqH}
\intertext{Here also, we note that the complex wavenumber
$\underline{k}$ becomes real in the case of a non-conducting
medium:}
    \underline{k}^2 &= \omega^2\mu\underline{\varepsilon} = \omega^2\mu\varepsilon\left( 1-i\frac{\kappa}{\omega\varepsilon}\right) \,.
    && \mbox{\bf complex wavenumber} \label{eq:complexwavenumber}
\intertext{Finally, we simplify the wave equations again for the
case of a non-conducting, charge-free medium and obtain}
    \nabla^2\mathbf{E} &= -k^2\mathbf{E} \,,  &&\mbox{\bf complex wave equations in a non-}\\
    \nabla^2\mathbf{H} &= -k^2\mathbf{H} \,,  && \mbox{\bf conducting, charge-free medium}
\intertext{with}
    k^2 &= \omega^2\mu\varepsilon = \frac{\omega^2}{c^2} \,.  && \mbox{\bf free-space wavenumber} \label{eq:freespacek}
\end{align}

On the way, we have also introduced a simple definition for the
speed of light, $c=1/\sqrt{\mu\varepsilon}$, in
Eq.~\eqref{eq:freespacek}.

\subsection{Plane waves}
\label{sec:PlaneWaves} As an introduction to the theory of
electromagnetic waves, we look at a very simple case, that of
so-called plane waves. We assume again that we are in a
homogeneous, isotropic, linear medium and that there are no
charges or currents, which means that
Eqs.~\eqref{eq:complexwaveeqE} and \eqref{eq:complexwaveeqH}
apply. Furthermore---for a plane wave---we assume that the field
components depend only on one coordinate (e.g., $z$). The solution
of the harmonic wave equations \eqref{eq:complexwaveeqE} and
\eqref{eq:complexwaveeqH} can then be written as a superposition
of two waves

\vspace*{-0.1cm}
\meqn{
    E_x(z) &= \underline{C}_1 e^{-\underline{\gamma}z} + \underline{C}_2 e^{+\underline{\gamma}z} \,, \\
    H_y(z) &= \frac{1}{\underline{Z}}\left(\underline{C}_1 e^{-\underline{\gamma}z} +  \underline{C}_2
     e^{+\underline{\gamma}z} \right)
\,,}{}{eq:solharmwave}\\
one of which propagates in the positive and one in the negative
$z$ direction. The complex propagation constant
 $\underline{\gamma}$ has a real component $\alpha$, which describes the damping in a lossy material, and
 a complex component $i\beta$, which describes the propagation of the wave. The relation between the
 propagation constant $\gamma$ and the wavenumber $k$ is

\vspace*{-0.7cm}
\meqn{
    \underline{\gamma} = \alpha + i\beta = i\underline{k} = i\omega\sqrt{\mu\underline{\varepsilon}}\mbox{.}
}{propagation constant}{eq:propconst}

We already know that time-harmonic electric and magnetic fields
are linked via Maxwell's equations, which means that their
amplitudes have a certain fixed ratio to each other. This ratio
has been introduced in Eq.~\eqref{eq:solharmwave} as the wave
impedance $\underline{Z}$, the ratio between the electric and
magnetic field amplitudes

\vspace*{-0.5cm}
\meqn{
    \underline{Z} = \frac{E_y}{H_z} = \sqrt{\frac{\mu}{\underline{\varepsilon}}}\mbox{,}
}{complex wave impedance}{} which becomes real in the absence of
lossy material. The wave impedance of free space is given by

\vspace*{-0.5cm}
\meqn{
    Z_0 = \sqrt{\frac{\mu_0}{\varepsilon_0}} \approx 377 \, \Omega
\,.}{free-space wave impedance}{eq:ZfreeSpace}

\subsection{Skin depth}
\label{sec:skindepth} When electromagnetic waves encounter a
conducting (lossy) material, we have to evaluate the boundary
conditions (see Section~\ref{sec:boundary}), and we find that the
wave amplitudes are attenuated suddenly by an attenuation constant
$\alpha$. In the RF case we can assume that
 \begin{align}
    \frac{\kappa}{\omega\varepsilon} &\gg 1 \,,
 \intertext{which means that the complex wavenumber \eqref{eq:complexwavenumber} and, obviously, also the complex dielectric constant
  \eqref{eq:complexdielectric} are dominated by their imaginary parts, so that we can write}
\underline{\varepsilon} \approx -i \varepsilon'' &= -i \frac{\kappa}{\omega} \hspace*{0.5cm}\mbox{or}\hspace*{0.5cm}
    \underline{k}^2 = -i\omega\mu\kappa \,,
\intertext{which is actually equivalent to neglecting the
displacement current. Using Eq.~\eqref{eq:propconst}, we can then
write the propagation constant as}
    \gamma &= \alpha + i\beta = i\underline{k} = i\omega\sqrt{\frac{-i\mu\kappa}{\omega}} =
    (1+i)\sqrt{\frac{\kappa\mu\omega}{2}}\,,
\end{align}
which defines the attenuation constant $\alpha$. The `skin depth'
is then defined as the distance after which the wave amplitudes
have been attenuated by a factor $1/e \approx 36.8\%$:

\vspace*{-0.5cm}
\meqn{
    \delta_\mathrm{s} = \frac{1}{\alpha} = \sqrt{\frac{2}{\omega\mu\kappa}}
\,.}{skin depth}{eq:SkinDepth}

\begin{figure}[h!]
    \centering
    \includegraphics[angle=270,width=0.65\textwidth]{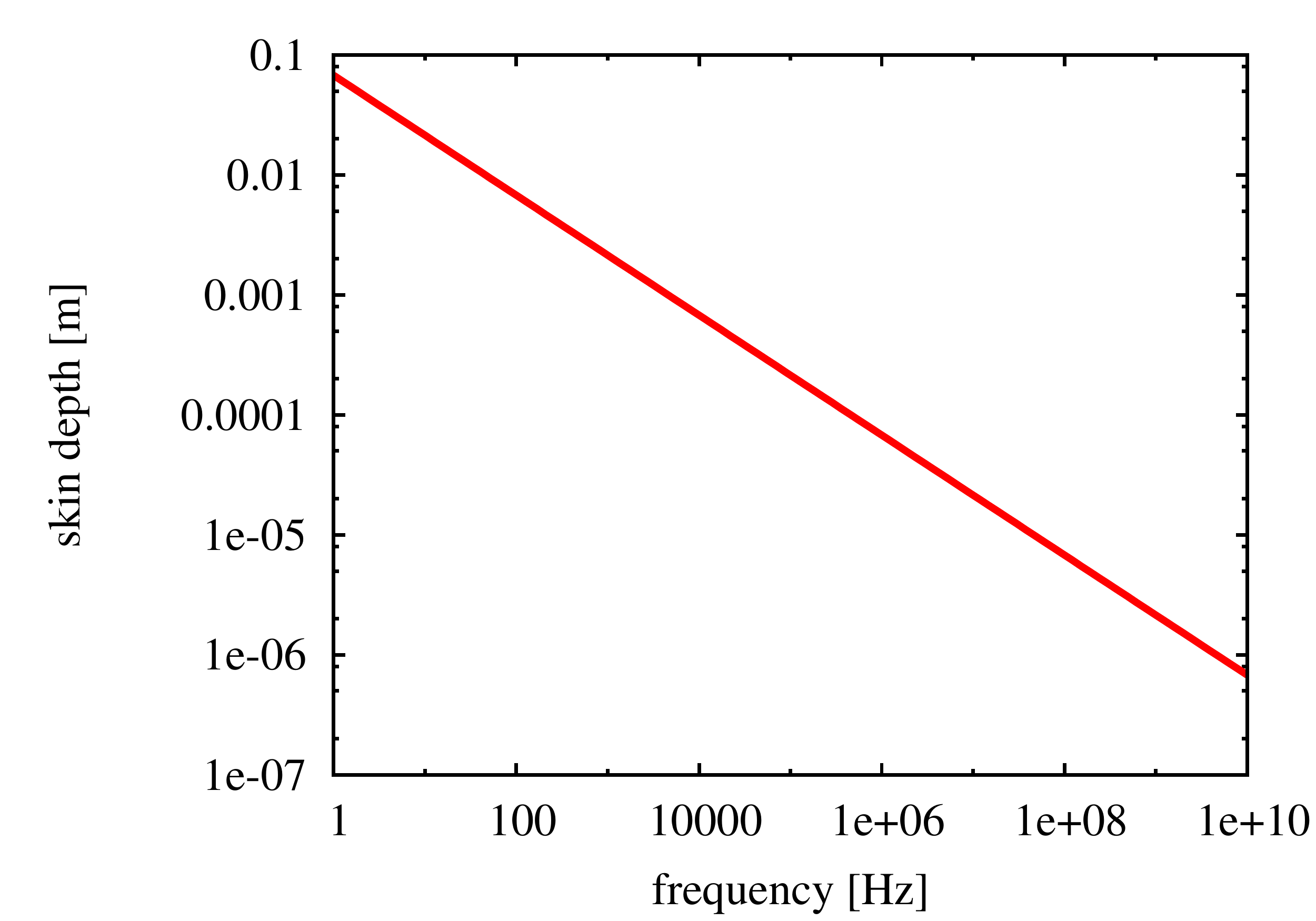}
    \caption{Skin depth versus RF frequency}
    \label{fig:skinVsFreq}
\end{figure}

Knowing the value of the skin depth is crucial for the design of
RF equipment. Let us assume that we want to build an accelerating
cavity that resonates at 500\,MHz. Since high-quality copper is
quite expensive, we consider the possibility of constructing the
cavity out of steel and then copper-plating the interior in order
to obtain a good quality factor and reduce the losses in the
surface. From Eq.~\eqref{eq:SkinDepth}, we calculate that the skin
depth in copper is approximately 3~\textmu m. Depending on how
well the copper plating is done by the plating company, we can now
define the thickness of the copper layer that is needed on the
inside of the cavity. Typically, around 10--20 times the skin
depth is chosen as the plating thickness.
Figure~\ref{fig:skinVsFreq} shows the dependence of the skin depth
on the RF frequency.

\begin{figure}[h!]
\begin{center}
\begin{tikzpicture}[scale=1, ultra thick]
\draw[line width = 3pt] (0,-2) -- (0,2); \fill[pattern=horizontal
lines light gray] (0,-2) rectangle (3,2);
 \draw (-3.5,-1) cos
(-3,0) sin (-2.5,1) cos (-2,0) sin (-1.5,-1) cos (-1,0) sin
(-0.5,1) cos (0,0);
 \draw[dashed] (0,0) sin (0.125,-0.61) cos
(0.25,0) sin (0.375,0.37) cos (0.5,0) sin (0.625,-0.22) cos
(0.75,0) sin (0.875,0.14) cos (1,0) sin (1.125,-0.08) cos (1.25,0)
sin (1.375,0.05) cos (1.5,0) sin (1.625,-0.03) cos (1.75,0) sin
(1.875,0.018) cos (2,0);
 \draw[dashed] (0.37,-2) -- (0.37,2);
 \draw[->] (-0.5,2.2) -- (0,2.2);
  \draw (-0.5,2.2) node[left]
{$\delta_\mathrm{s}$};
 \draw[->] (0.87,2.2) -- (0.37,2.2);
 \draw[->, line width = 3 pt,red] (0.185,0.4) -- (0.185,1.8);
 \draw
(-0.1,1.4) node[left] {\Large\textcolor{red}{$I_S$}};

\draw (1.6,-1.5) node[fill=red!30,rounded corners] {$\mu_0$, $\varepsilon_0$, $\kappa\neq 0$};
\draw (-1.5,-1.5) node[fill=red!30,rounded corners] {$\mu_0$, $\varepsilon_0$, $\kappa = 0$};

\end{tikzpicture}
\caption{\label{fig:SrfRes}Skin depth and surface resistance}
\end{center}
\end{figure}
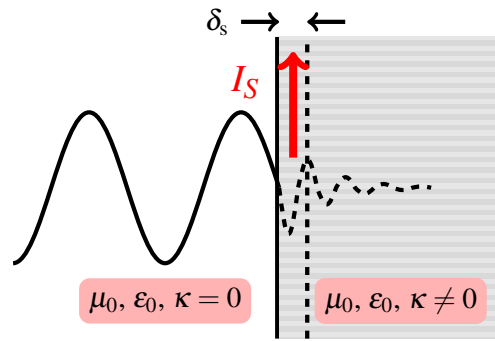

Furthermore, the skin depth allows us to calculate the losses in
the surface easily. For a wave travelling parallel to a conducting
surface, one can define a surface resistance by assuming a
constant current density in a layer of the surface material
equivalent to the skin depth, as shown in Fig.~\ref{fig:SrfRes}:
 \vspace*{-0.5cm} \meqn{
 R_\mathrm{surf} = \frac{1}{\kappa \delta_\mathrm{s}} \left[\Omega\right]
\,.}{surface resistance}{eq:surfResistance}

\medskip
\noindent This value has to be multiplied by $l/w$ to
obtain the full RF resistance, where $l$ is the length of the
conducting wall and $w$ is its width.

\subsection{Energy and transport of energy}
We start this section by presenting Poynting's law, and then
explain its components. Poynting's law states nothing more than
the conservation of electromagnetic energy:

\vspace*{-0.3cm} \meqn{ -\frac{\dd}{\dd t}\int\limits_V w \, \dd V
= \int\limits_A \mathbf{S} \cdot \dd \mathbf{A} + \int\limits_V
\mathbf{E} \cdot \mathbf{J} \, \dd V
\,.}{Poynting's law}{eq:PoyntingLaw}\\

This equation, read from left to right, states that `the rate of
change of stored energy in a volume equals the energy flow out of
the volume (through a surface $\mathbf{A}$ surrounding the volume)
plus the losses within the volume (the work performed on charges
per unit time)'. In the following lines, we shall see that the
components of Poynting's law do indeed correspond to what is
stated in the previous sentence.

\subsubsection*{What is $\mathbf{E} \cdot \mathbf{J} $?}
In order to understand the expression $\mathbf{E} \cdot
\mathbf{J}$, we follow \cite{bib:feynman} and start with the force
acting on a charge moving in an electromagnetic field,

\vspace*{-0.7cm}
\meqn{
    \mathbf{F} = q \left( \mathbf{E} + \mathbf{v} \times \mathbf{B} \right)
\,.}{Lorentz force}{eq:LorentzForce}
 Multiplying this equation by
$\mathbf{v}$ and knowing that $\mathbf{a} \cdot \left( \mathbf{a}
\times \mathbf{b} \right) \equiv 0$, we obtain an expression for
the work done on a charge per unit time,
\begin{align}
    \mathbf{v}\cdot\mathbf{F} &= q \mathbf{v} \cdot \mathbf{E} \,.
\intertext{Assuming $N$ particles per unit volume, we can write}
    N\mathbf{v}\cdot\mathbf{F} &= Nq \mathbf{v} \cdot \mathbf{E}
    =\mathbf{J}\cdot\mathbf{E}\,.
\end{align}
Therefore the expression $\mathbf{J}\cdot\mathbf{E}$ must be equal
to the work done on charges per unit time and unit volume, or, in
other words, the loss of electromagnetic energy per unit volume.

\subsubsection*{The Poynting vector $\mathbf{S}$ and the energy density $w$}
These quantities can be understood by manipulating Maxwell's
equations (compare, e.g., \cite{bib:Henke}). We multiply
Eq.~\eqref{eq:rotH} by $\mathbf{E}$:
\begin{align}
    \mathbf{E}\cdot\mathbf{J} &= \mathbf{E}\cdot\left( \nabla \times \mathbf{H} \right) -
    \mathbf{E} \cdot \frac{\partial \mathbf{D}}{\partial t} \,.
\intertext{Using Eq.~\eqref{eq:va1}, this can be rewritten as}
    \mathbf{E}\cdot\mathbf{J} &= \mathbf{H} \cdot \left(\nabla \times\mathbf{E} \right) -
    \nabla\cdot\left( \mathbf{E} \times \mathbf{H} \right) - \mathbf{E} \cdot \frac{\partial \mathbf{D}}{\partial t} \,.
    \intertext{Using the second of Maxwell's equations \eqref{eq:rotE}
and assuming time-invariant $\mu$ and $\varepsilon$, we can write}
        \mathbf{E}\cdot\mathbf{J} &= -\nabla\cdot \left(\mathbf{E}\times \mathbf{H}\right)
         - \frac{\partial}{\partial t} \left(
        \frac{1}{2} \mathbf{E} \cdot \mathbf{D} + \frac{1}{2} \mathbf{H} \cdot \mathbf{B}\right) \,.
\end{align}

Applying a volume integral together with Gauss's theorem
\eqref{eq:Gauss} and rearranging the elements of the equation, we
end up with

\vspace*{-0.1cm}
\meqn{
    -\frac{\partial}{\partial t} \int_V \left( \frac{1}{2} \mathbf{E} \cdot \mathbf{D} +
    \frac{1}{2} \mathbf{H} \cdot \mathbf{B} \right) \, \dd V \hspace{3cm} \\
    \hspace{3cm}= \int_A \left( \mathbf{E} \times \mathbf{H} \right) \cdot \dd A
    + \int_V \mathbf{E} \cdot \mathbf{J} \, \dd V
\,,}{Poynting's law}{}\\
which can be compared directly with Eq.~\eqref{eq:PoyntingLaw}. On
the left-hand side we have the definition of the energy density,

\vspace*{-0.7cm}
\meqn{
    w = w_\mathrm{el} + w_\mathrm{mag} = \frac{1}{2} \mathbf{E} \cdot \mathbf{D} +
    \frac{1}{2} \mathbf{B} \cdot \mathbf{H}
\,,}{electric and magnetic energy density}{}\\
and from the right-hand side we obtain the definition of the
energy flux density, or the Poynting vector $\mathbf{S}$,

\vspace*{-0.7cm}
\meqn{
    \mathbf{S} = \mathbf{E} \times \mathbf{H}
\,.}{Poynting vector}{}\\

The Poynting vector gives us the direction in which an
electromagnetic wave transports energy, and from the cross product
we understand that this direction is always perpendicular to the
electric and magnetic field components. This is consistent with
Section~\ref{sec:PlaneWaves}, where we found that the field
components ($E_x$, $H_y$) of a plane wave (see
Eq.~\eqref{eq:solharmwave}) are perpendicular to the direction of
propagation ($z$).

In the above derivation, we have used Maxwell's equations in their
general form, meaning with time derivatives. In the case of the
complex notation, the definitions of the energy density and
Poynting vector have to be modified as follows (for a proof, see
\cite{bib:Henke} or \cite{bib:Weiland}):

\vspace*{-0.7cm}
\meqn{
    w = w_\mathrm{el} + w_\mathrm{mag} = \frac{1}{4} \mathbf{E} \cdot \mathbf{D}^* +
    \frac{1}{4} \mathbf{B} \cdot \mathbf{H}^*
\,,}{electric and magnetic energy density in complex notation}{}

\vspace*{-0.7cm}
\meqn{
    \mathbf{S} = \frac{1}{2}\left(\mathbf{E} \times \mathbf{H}^*\right)
\,.}{complex Poynting vector}{eq:PoyCom}

\section{Electromagnetic waves in waveguides}
In this section, we derive the field components of electromagnetic
waves that propagate in waveguides. The same principle can then be
used to calculate the standing-wave pattern in an accelerating
cavity, which is nothing more than a superposition of two waves
travelling in opposite directions.

\subsection{Classification of modes in waveguides and cavities}
Before we start to solve the wave equation, we need to introduce a
classification of the field patterns that can be found in
waveguides and cavities.

\subsubsection*{TM$_{mnp}$ modes, or E$_{mnp}$ modes}
These modes have no magnetic field in the direction of propagation
($z$) and are therefore often called {\it transverse magnetic}, or
TM, modes. On the other hand, they have an electric field
component that is parallel to $z$, hence the equivalent name {\it
E modes}.

\begin{figure}[h!]
\centering
\includegraphics[width=0.7\textwidth]{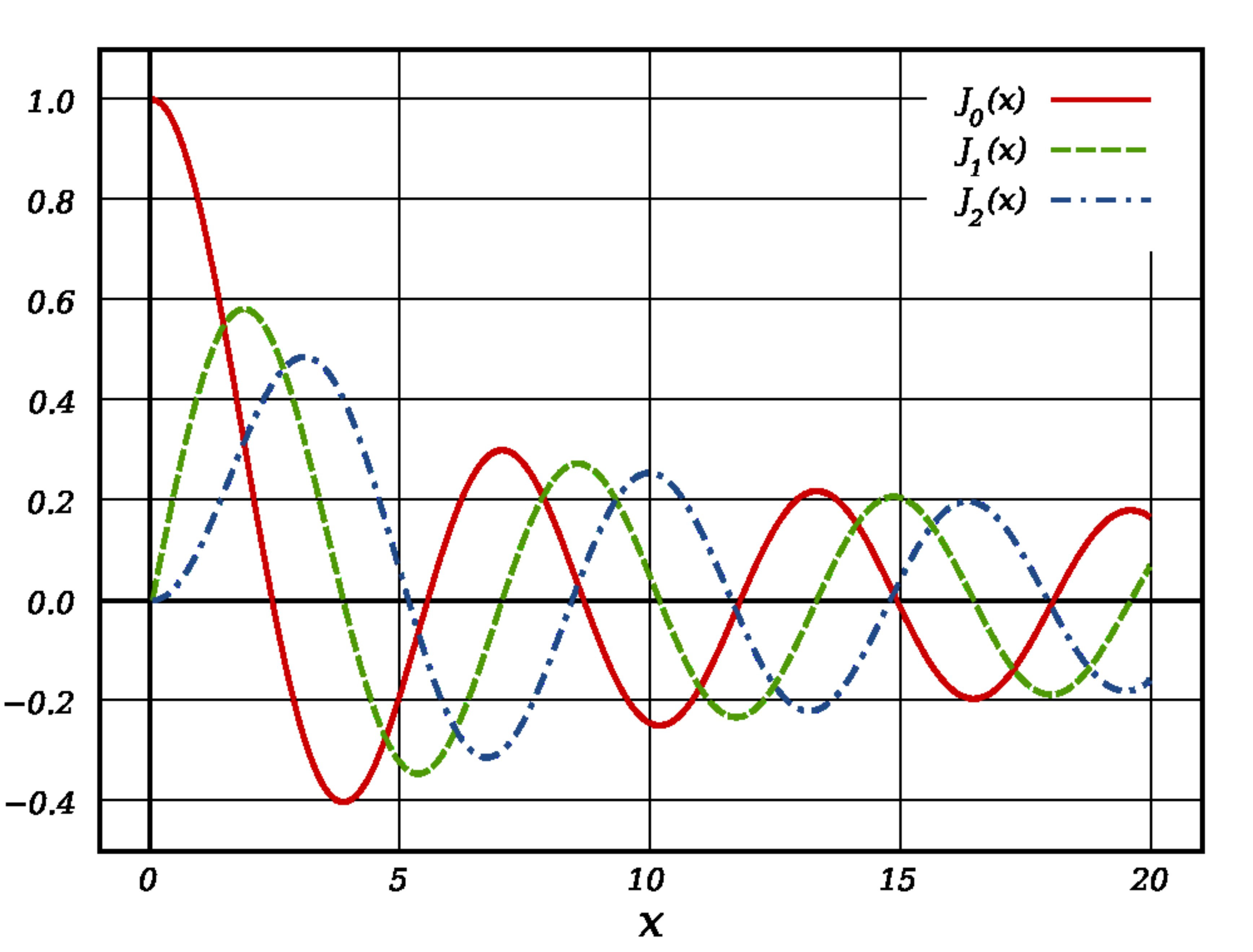}
\caption{Bessel functions of the first kind up to order 2}
\label{fig:Bessel}
\end{figure}

The indices $m$, $n$, $p$ indicate the number of zeros or
variations in the three directions of a coordinate system. In the
case of a waveguide, only the first two indices are used, whereas
in the case of a cavity, owing to the standing-wave pattern along
$z$, all three are needed for a complete description. In the case
of a circular waveguide or cavity, the indices indicate the
following:

\begin{itemize}
    \item[$m$,] number of full-period variations of the field components in the azimuthal direction.
For circularly symmetric geometries, $\mathbf{E}$, $\mathbf{B}
\propto \cos(m\varphi)$, $\sin(m\varphi)$.
    \item[$n$,] number of zeros ($x_{mn}$) of the axial field component in the radial direction.
For circularly symmetric geometries, $E_z$, $B_z \propto
J_m(x_{mn}r/R_\mathrm{c})$.
    \item[$p$,] number of half-period variations of the field components in the longitudinal direction,
 with $\mathbf{E}$, $\mathbf{B} \propto \cos(p\pi z/l)$, $\sin(p\pi z/l)$.
\end{itemize}
The functions $J_m$ introduced above are Bessel functions of the
first kind and of $m$th order, and can be found in mathematical
textbooks. The first three orders are shown in
Fig.~\ref{fig:Bessel}.

\subsubsection*{TE$_{mnp}$ modes, or H$_{mnp}$ modes}
Here, there is no electric field in the direction of propagation
$z$, hence the name \emph{transverse electric}, or TE, modes. In
analogy to the E modes, H modes have a magnetic field component
parallel to $z$. The indices have the same meaning as above.

\subsubsection*{TEM modes}
This class of modes has neither an electric nor a magnetic field
component in the direction of propagation. They can exist between
two isolated conductors, for example in a coaxial line. The
advantage of TEM modes is that waves of any frequency can
propagate, whereas TE and TM modes always have a cut-off
frequency, below which they are damped exponentially (more on this
later). However, the disadvantage of coaxial lines is that the
losses in the two conductors are generally higher than in
rectangular or circular waveguides.

\subsection{Solution of the wave equation in a cylindrical waveguide}
Instead of trying to find solutions for all six vector components
of the electric and magnetic fields, one can simplify the problem
by using a vector potential $\mathbf{A}$ (without any physical
meaning) that has only one component. One can then quickly derive
all six field components from this vector potential.

It can be shown that only two types of modes can exist in
waveguides: TM and TE modes, as introduced above. For
 each mode type, we introduce a vector potential $\mathbf{A}$ as follows. Since $\mathbf{H}$ and $\mathbf{E}$
 are divergence-free, and since $\nabla \cdot \left( \nabla \times \mathbf{a} \right) \equiv 0$, we can write\\

\vspace*{-1cm}
\begin{align}
    \mathbf{H}^\mathrm{TM} &= \nabla \times \mathbf{A}^\mathrm{TM}  \hspace{0.5cm} \mbox{with}& \hspace{0.5cm}  \mathbf{E}^\mathrm{TM}
    &= -\frac{i}{\omega\varepsilon}\nabla \times ( \nabla\times\mathbf{A}^\mathrm{TM}
    )\,,
    && \mbox{\bf vector potential for TM waves} \label{eq:vpTM} \\
    \mathbf{E}^\mathrm{TE} &= \nabla \times \mathbf{A}^\mathrm{TE} \hspace{0.5cm} \mbox{with}& \hspace{0.5cm}  \mathbf{H}^\mathrm{TE}
    &= \frac{i}{\omega\mu}\nabla \times ( \nabla\times\mathbf{A}^\mathrm{TE}
    )\,.
    && \mbox{\bf vector potential for TE waves} \label{eq:vpTE}
\end{align}
\noindent In both cases the vector potential obeys the wave
equation
\begin{align}
    \nabla^2 \mathbf{A} = -k^2\mathbf{A} \hspace{0.5cm} \mbox{with} \hspace{0.5cm}
 k^2 = \omega^2 \mu\varepsilon \,,
\end{align}
\noindent which can then be solved for various coordinate systems
and has only one vector component, in the direction of
propagation:
\begin{align}
    \mathbf{A} = A_z \mathbf{e}_z \,.
\end{align}

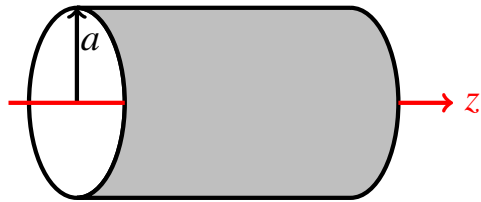
\begin{figure}[h!]
\centering
\begin{tikzpicture}[scale=0.9, ultra thick]

\draw (0,0) ellipse (0.7 and 1.4); \draw[black,fill=gray!50]
(4,1.4) arc (90:-90:0.7 and 1.4) -- (0,-1.4) arc (-90:90:0.7 and
1.4) -- (4,1.4);
\draw[->] (0,0) -- (0,1.4); \draw (0.2,0.9) node {\Large $a$};
\draw[red] (-1,0) -- (0.7,0); \draw[red, ->] (4.7,0) -- (5.5,0)
node [right] {\Large $z$};

\end{tikzpicture}
\caption{\label{fig:circularwaveguide}Geometry of a circular
waveguide}
\end{figure}

\subsection*{Circular waveguides}
In a circular waveguide, as shown in
Fig.~\ref{fig:circularwaveguide}, the vector potentials for the TE
and TM modes are identical:

\begin{align}
    A_z^\mathrm{TM/TE} &= C J_m (k_\mathrm{c} r) \cos(m\varphi) \displaystyle e^{\pm i k_z
    z} \,,
    && \mbox{\bf vector potential for circular waveguide} \label{eq:vpfcw} \\
\intertext{with}
    k_z & = \sqrt{k^2 - k_\mathrm{c}^2} \,. && \mbox{\bf wavenumber in $z$ direction} \label{eq:kzdef}
\end{align}
\noindent Using Eq.~\eqref{eq:vpTM}, we can derive the field components for the TM modes:\\

\vspace{-0.5cm}
\meqn{
\left. \begin{array}{lll}
E_r &= \displaystyle\frac{i}{\omega\varepsilon}\frac{\partial H_{\varphi}}{\partial z} &=
-C \displaystyle\frac{k_z k_\mathrm{c}}{\omega\varepsilon} J'_m(k_\mathrm{c} r) \cos(m\varphi) \vspace{0.2cm} \\
E_{\varphi} &= -\displaystyle\frac{i}{\omega\varepsilon} \frac{\partial H_r}{\partial z} &=
C\displaystyle\frac{m k_z}{\omega\varepsilon r} J_m(k_\mathrm{c} r)\sin(m\varphi)\vspace{0.2cm}\\
E_{z} &= \displaystyle\frac{i k_\mathrm{c}^2}{\omega\varepsilon}A_z &= C \displaystyle\frac{ik_\mathrm{c}^2}
{\omega\varepsilon} J_m (k_\mathrm{c} r) \cos(m\varphi)\vspace{0.2cm}\\
H_r &= \displaystyle\frac{1}{r} \frac{\partial A_z}{\partial \varphi} &= -C\displaystyle\frac{m}{r}
J_m (k_\mathrm{c} r) \sin(m\varphi) \vspace{0.2cm}\\
H_{\varphi} &= -\displaystyle\frac{\partial A_z}{\partial r} &= -C
k_\mathrm{c} J'_m (k_\mathrm{c} r)\cos(m\varphi)
\end{array} \right\}  e^{\pm ik_z z}
\,.}{field components for TM modes in a circular
waveguide}{eq:genfieldscirc}

Now we can use the boundary conditions to specify the cut-off
wavenumber $k_\mathrm{c}$. From Section~\ref{sec:boundary}, we
know that the electric field components parallel to the waveguide
surface have to vanish at the surface, which means
\begin{equation}
    \left. \begin{array}{ll}
    E_{\varphi} (r=a) &= 0 \\
    E_z (r=a) &= 0
    \end{array} \right\} \Rightarrow J_m(k_\mathrm{c} a) = 0 \hspace{0.5cm} \Rightarrow \hspace{0.5cm} k_\mathrm{c}
    = \displaystyle\frac{j_{mn}}{a} \,.
\end{equation}
The $n$th zeros $j_{mn}$ of the Bessel functions of order $m$ are
tabulated in mathematical textbooks (e.g., \cite{bib:Abramowitz}).

Using

\meqn{
    k_\mathrm{c} = \displaystyle\frac{2\pi}{\lambda_\mathrm{c}} = \displaystyle\frac{\omega_\mathrm{c}}{c} \,,
}{}{}\\

\noindent we can define the cut-off frequency of the waveguide,\\

\vspace*{-1cm}
\meqn{
    \omega_\mathrm{c} = c\displaystyle\frac{j_{mn}}{a} \,.
}{cut-off frequency for TM modes in a circular waveguide}{}

\noindent The mode that is most commonly used in circular waveguides is the TM$_{01}$ mode, which has only three field components.
By inserting $m=0$ and $n=1$ into Eq. \eqref{eq:genfieldscirc} and using $J'_0(r) = -J_1(r)$, we obtain\\
\meqn{
\left. \begin{array}{lll}
E_r &= C \displaystyle\frac{k_z k_\mathrm{c}}{\omega\varepsilon} J_1(k_\mathrm{c} r) \vspace{0.2cm} \\
E_{z} &= -C \displaystyle\frac{ik_\mathrm{c}^2}{\omega\varepsilon} J_0 (k_\mathrm{c} r) \vspace{0.2cm}\\
H_{\varphi} &= C k_\mathrm{c} J_1 (k_\mathrm{c} r)
\end{array} \right\} e^{\pm ik_z z}
 \,, }{field components of TM$_{01}$ mode in a circular waveguide}{eq:TM01circ}
with a cut-off frequency $\omega_\mathrm{c} \approx c \times
(2.405/a)$.

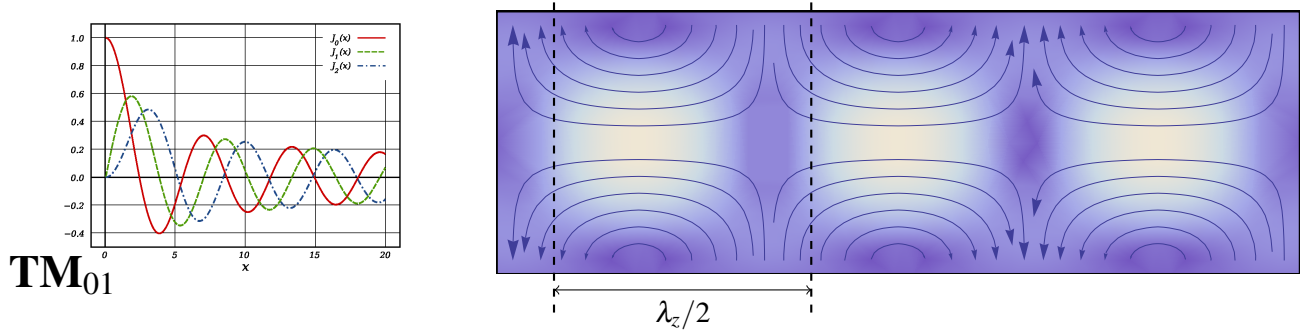
\begin{figure}[h!]
\begin{tikzpicture}
\node[inner sep=0pt,above right]{\includegraphics[height=3.5cm]{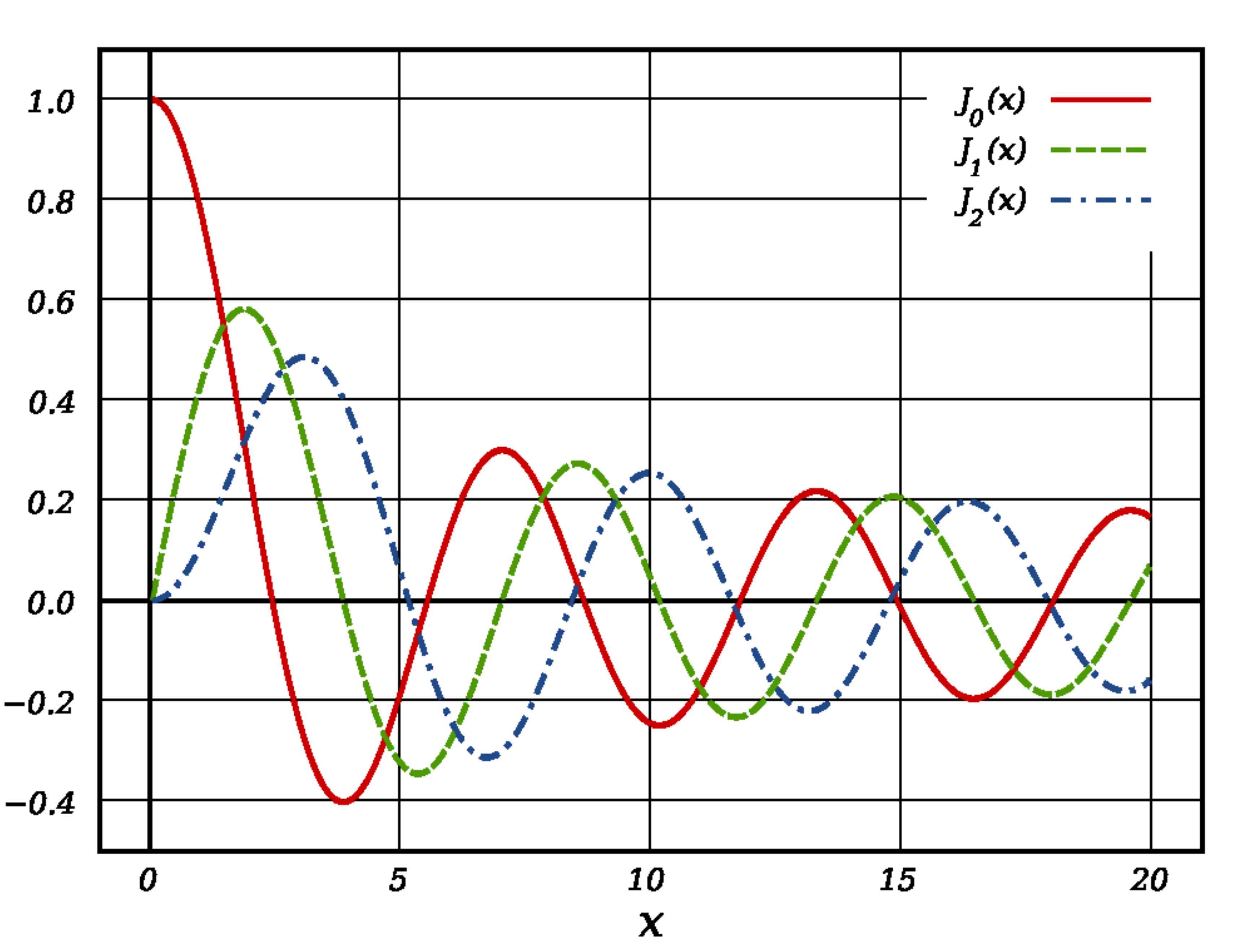}\hspace*{1cm}
\includegraphics[height=3.5cm]{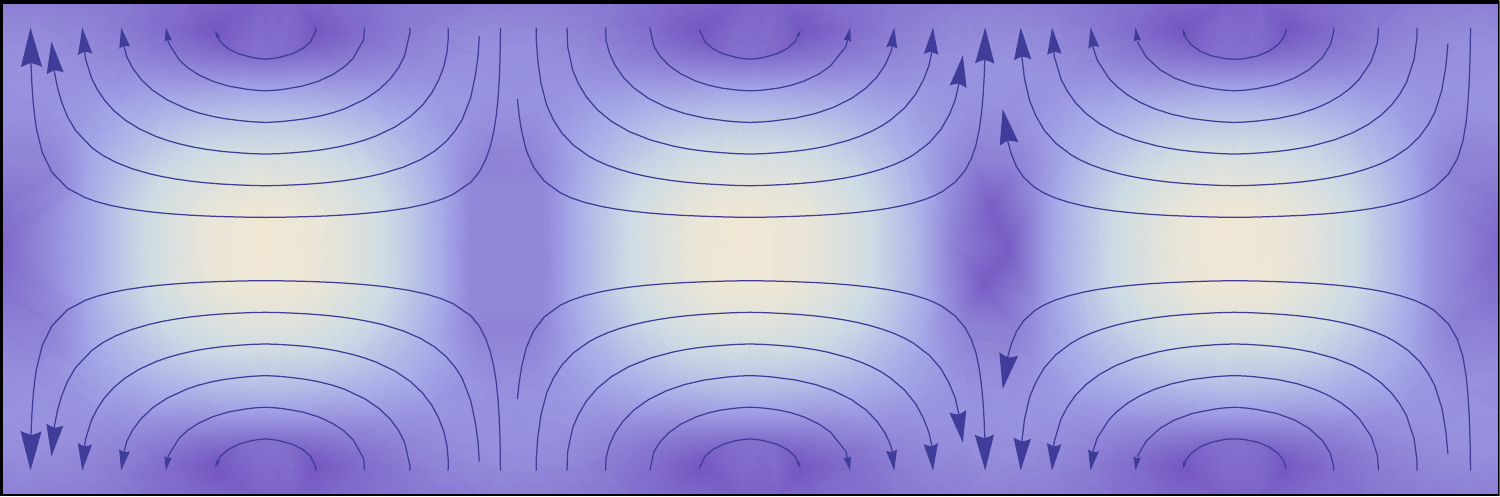}};
\draw[dashed,thick] (6.465,-0.5) -- (6.465,3.7);
\draw[dashed,thick] (9.85,-0.5) -- (9.85,3.7);
\draw[<->] (6.47,-0.2) -- node[below] {$\lambda_z/2$} (9.85,-0.2);
\node (4,3.5) {\LARGE\bf TM$_{01}$};
\end{tikzpicture}
\caption{Field lines of a TM$_{01}$ mode in a circular waveguide
with $\omega = 1.15 \omega_\mathrm{c}$. Solid lines, electric
field lines; dashed lines, magnetic field lines. The brightness of
the background is proportional to the norm of the field vector:
light areas indicate high-field regions of the magnetic field in
the left plot and of the electric field in the right plot.}
\label{fig:circularTM01}
\end{figure}

\begin{figure}[h!]
\begin{tikzpicture}
\node[inner sep=0pt,above right]{\includegraphics[height=3.5cm]{figure13a}\hspace*{1cm}
\includegraphics[height=3.5cm]{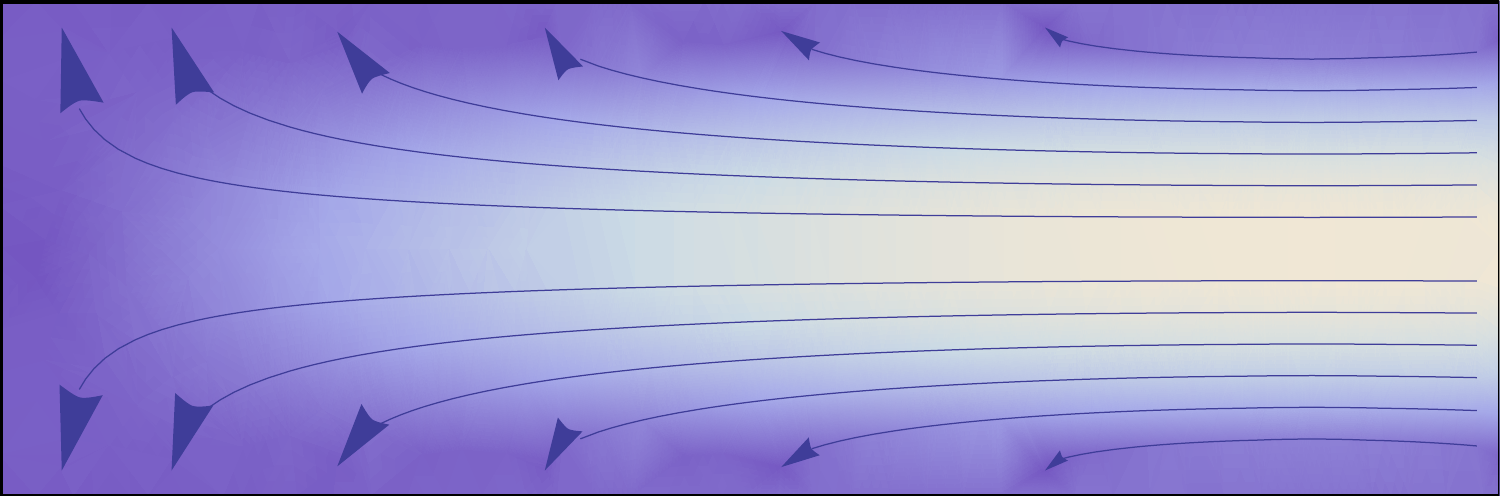}};
\node (4,3.5) {\LARGE\bf TM$_{01}$};
\end{tikzpicture}
\caption{Field lines of a TM$_{01}$ mode in a circular waveguide
with $\omega=1.005 \omega_\mathrm{c}$. Solid lines, electric field
lines; dashed lines, magnetic field lines. The brightness of the
background is proportional to the norm of the field vector: light
areas indicate high-field regions of the magnetic field in the
left plot and of the electric field in the right plot.}
\label{fig:circularabovecutoff}
\end{figure}

The field pattern of the TM$_{01}$ mode is shown in
Fig.~\ref{fig:circularTM01} for a mode frequency 15\% above the
cut-off frequency. The distance between the minima or between the
maxima of the field corresponds to 0.5 times the propagation
wavelength $\lambda_z$. With decreasing mode frequency, $\lambda_z
= 2\pi/k_z$ becomes longer, and finally becomes infinite when the
mode frequency equals the cut-off frequency $\omega_\mathrm{c}$.
This effect is shown in Fig.~\ref{fig:circularabovecutoff}, where
the TM$_{01}$ mode propagates at a frequency just 0.5\% above the
cut-off frequency.
\setlength{\captionmargin}{1cm}

\setlength{\captionmargin}{0cm}
\subsection*{Rectangular waveguides}

The derivation of the fields in a rectangular waveguide follows
the same principle as that used in the previous section for
circular waveguides. In a rectangular waveguide, as shown in
Fig.~\ref{fig:rectangularwaveguide}, two different vector
potentials are needed to describe the TE and TM modes:
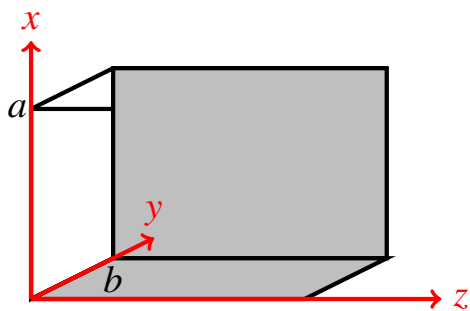
\begin{figure}[h!]
\centering
\begin{tikzpicture}[scale=0.9, ultra thick]

\draw (1.2,-0.8) -- (0,-1.4) -- (0,1.4) -- (1.2,2);
\draw (0,1.4) -- (1.2,1.4);
\draw (1.2,2) -- (5.2,2);
\draw[black,fill=gray!50] (1.2,2) rectangle (5.2,-0.8);
\draw[black,fill=gray!50] (0,-1.4) -- (4,-1.4) -- (5.2,-0.8) -- (1.2,-0.8) -- (0,-1.4);
\draw (-0.2,1.4) node {\Large $a$};
\draw (1.2,-1.1) node {\Large $b$};
\draw[->,red] (0,-1.4) -- (6,-1.4) node [right] {\Large $z$};
\draw[->,red] (0,-1.4) -- (0,2.4) node [above] {\Large $x$};
\draw[->,red] (0,-1.4) -- (1.8, -0.5) node [right,above] {\Large $y$};

\end{tikzpicture}
\caption{\label{fig:rectangularwaveguide}Geometry of a rectangular
waveguide with transverse dimensions $a$ and $b$}
\end{figure}

\meqn{ A_z^\mathrm{TM} = C\sin(k_x x) \sin(k_y y) e^{\pm ik_z z}
\,, } {vector potential for TM waves in a rectangular
waveguide}{ATMr}

\vspace{-0.7cm} \meqn{ A_z^\mathrm{TE} = C\cos(k_x x) \cos(k_y y)
e^{\pm ik_z z} \,, } {vector potential for TE waves in a
rectangular waveguide}{ATEr} \noindent where

\vspace*{-0.7cm} \meqn{ k_z = \sqrt{k^2 - k_\mathrm{c}^2} \,,
\hspace{0.5cm} \mbox{with} \hspace{0.5cm} k_\mathrm{c}^2 = k_x^2 +
k_y^2 \,. } {wavenumber in $\mathbf{z}$ direction}{}

\noindent We note that the position of the origin of the
coordinate system is linked to the sine and cosine terms  in
Eqs.~\eqref{ATMr} and \eqref{ATEr}. The fields derived from the
vector potentials have to fulfil the boundary conditions on the
waveguide walls. So if, for instance, we were to choose the origin
in the centre of the waveguide, then the sine and cosine
expressions would have to be exchanged to account for the changed
symmetries with respect to the coordinate axes. Using
Eq.~\eqref{eq:vpTM} again, we derive the field components for the
TM modes:

\meqnter{
\left. \begin{array}{lll}
E_x &= \displaystyle \frac{i}{\omega\varepsilon}\frac{\partial H_y}{\partial z} &= \pm C \displaystyle\frac{k_z}{\omega\varepsilon}
\cos(k_x x) \sin(k_y y) \vspace{0.2cm} \\
E_y &= -\displaystyle\frac{i}{\omega\varepsilon} \frac{\partial H_x}{\partial z} &= \pm C\displaystyle\frac{k_z}{\omega\varepsilon}
\sin(k_x x) \cos(k_y y) \vspace{0.2cm}\\
E_{z} &= \displaystyle\frac{i (k_z^2-k^2)}{\omega\varepsilon}A_z^\mathrm{TM} &= C \displaystyle\frac{i(k_z^2-k^2)}{\omega\varepsilon}
\sin(k_x x) \sin(k_y y)\vspace{0.2cm}\\
H_x &= \displaystyle \frac{\partial A_z^\mathrm{TM}}{\partial y} &= C\displaystyle k_y \sin(k_x x) \cos(k_y y) \vspace{0.2cm}\\
H_y &= -\displaystyle\frac{\partial A_z^\mathrm{TM}}{\partial x}
&= -C k_x \cos(k_x x) \sin(k_y y)
\end{array} \right\} e^{\pm ik_z z} \,. \hspace{-0.7cm}
 }{field components for TM modes in a rectangular waveguide\hspace{-0.8cm}}{eq:genfieldsrec}

\noindent Using the boundary conditions, we can specify the
wavenumbers $k_x$ and $k_y$:

\begin{equation}
    \left. \begin{array}{l}
    E_y (x=a) = 0   \\
    E_z (x=a) = 0
    \end{array} \right\} \Rightarrow  k_x = \displaystyle\frac{m\pi}{a} \hspace{0.5cm} \mbox{and} \hspace{0.5cm} m=0,1,2,
    \ldots \,,
\end{equation}
\begin{equation}
    \left. \begin{array}{l}
    E_x (y=b) = 0 \\
    E_z (y=b) = 0
    \end{array} \right\} \Rightarrow  k_y = \displaystyle\frac{n\pi}{b} \hspace{0.5cm} \mbox{and} \hspace{0.5cm} n=0,1,2,
    \ldots \,,
\end{equation}

\noindent and the cut-off frequency for a rectangular waveguide is

\vspace*{-0.7cm} \meqn{ \omega_\mathrm{c} = c k_\mathrm{c} = c
\sqrt{k_x^2+k_y^2} = c\pi \sqrt{\left(\frac{m}{a} \right)^2 +
\left(\frac{n}{b} \right)^2}  \,. }{cut-off frequency for TM modes
in a rectangular waveguide}{}

The usual convention is to have $a>b$, and in this case the
TE$_{10}$ mode is the mode with the lowest cut-off frequency. It
is also the only mode that propagates in a relatively large
frequency band, from $f_\mathrm{c}^{\mathrm{TE},10}$ to
$2f_\mathrm{c}^{\mathrm{TE},10}$, which is why it is the mode most
commonly used in rectangular waveguides. The fields of the TE
modes can be derived from the TE vector potential using the same
procedure.

\subsection{Wave propagation and dispersion relation}
In Figs.~\ref{fig:circularTM01} and \ref{fig:circularabovecutoff},
we have seen that the propagation wavelength $\lambda_z$ of a
waveguide mode is determined by its frequency and by how far the
mode frequency is above the cut-off frequency of the waveguide. If
the propagation wavelength depends on the mode frequency, we can
assume that the phase velocity of a particular mode also depends
on the
mode frequency. This relationship is called the dispersion relation and, using the definition of the wavenumber in Eq.~\eqref{eq:kzdef}, we can write\\

\vspace*{-0.7cm} \meqn{ k_z^2 = k^2 - k_\mathrm{c}^2 =
\frac{\omega^2 - \omega_\mathrm{c}^2}{c^2} =
\frac{\omega^2}{v_\mathrm{ph}^2} \,, } {dispersion relation}{}
from which we can immediately see that:
\begin{itemize}
    \item $k_z$ can be real only if the mode frequency $\omega$ is above the cut-off frequency $\omega_\mathrm{c}$;
    \item for $\omega < \omega_\mathrm{c}$, the mode cannot propagate and the fields are exponentially damped.
\end{itemize}
We also have a definition of the phase velocity, which is the
speed at which the maxima and minima of the field
patterns move along the waveguide:\\

\vspace*{-0.7cm} \meqn{ v_\mathrm{ph}= \frac{\omega}{k_z} =
c^2\frac{\omega^2}{\omega^2 - \omega_\mathrm{c}^2}
 \,. }{phase velocity}{}

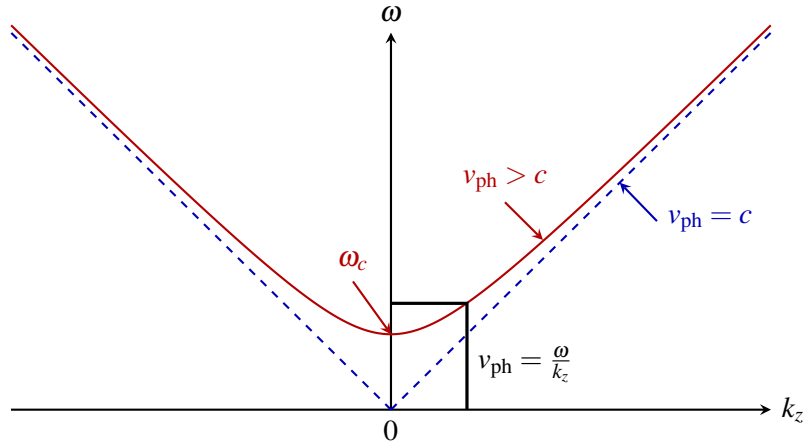
\begin{figure}[h!]
\centering
\begin{tikzpicture}[>=stealth, x=5cm/5,red!70!black,thick]
\draw[->,black](-5,0)  -- (5,0)node[right]{$k_{z}$};
\draw[->,black](0,0)node[below]{$0$}  -- (0,5)node[above]{$\omega$};

\pgfplothandlerlineto
\pgfplotfunction{\x}{0,0.01,...,5}{\pgfpointxy{\x}{sqrt(1+(\x^2))}}
\pgfusepath{stroke}

\pgfplothandlerlineto
\pgfplotfunction{\x}{0,0.01,...,5}{\pgfpointxy{-\x}{sqrt(1+(\x^2))}}
\pgfusepath{stroke}

\draw [->](1.5,2.75)node[above]{$v_{\text{ph}}>c$}--(2,2.25);
\draw [->](-0.5,1.7)node[above]{$\omega_{c}$}--(0,1);

\draw[dashed,blue!70!black,thick] (0,0) -- (5,5);
\draw[dashed,blue!70!black,thick] (0,0) -- (-5,5);
\draw [->,blue!70!black](3.5,2.5)node[right]{$v_{\text{ph}}=c$}--(3.01,3.01);

\draw [black, very thick](1,0)--(1,1.41)--(0,1.41);
\draw [black](1,0.6)node[right]{$v_{\text{ph}}=\frac{\omega}{k_{z}}$};
\end{tikzpicture}
\caption{Dispersion relation in a waveguide. The dotted line shows
the case $v_\mathrm{ph}=c$.} \label{fig:DispRel}
\end{figure}

\noindent This is not to be confused with the speed with which the
wave actually propagates in the waveguide. The dispersion relation
is usually plotted in the form of a `Brillouin diagram', as shown
in Fig.~\ref{fig:DispRel}.

The slope of the dispersion relation is the group velocity\\

\vspace*{-0.7cm} \meqn{ v_\mathrm{gr} = \frac{\dd \omega}{\dd k_z}
 \,, }{group velocity}{}
\noindent which gives the velocity with which a signal or energy
is transported along the waveguide. From Fig.~\ref{fig:DispRel},
we can conclude that:
\begin{itemize}
    \item Each frequency has a certain phase velocity and group velocity, which means that signals with a frequency bandwidth will become deformed
 while travelling along a waveguide. With the help of the dispersion relation, we can easily quantify how much deformation will occur.
    \item The phase velocity $v_\mathrm{ph}$ is always larger than the velocity of light $c$, and at cut-off ($\omega = \omega_\mathrm{c}$) it even
 becomes infinite ($k_z = 0$ and $v_\mathrm{ph} \rightarrow \infty$).
    \item For acceleration, one needs synchronism between the phase velocity (the speed of the field pattern) and
    the velocity of the particles, which implies that acceleration in waveguides is impossible.
    \item Information and therefore energy travel at the group velocity, which is always slower than the speed of light.
\end{itemize}

\subsection{Attenuation of waves (power loss method)}
Up to this point, we have assumed perfect electrical conductors as
the boundaries of our waveguides. Real waveguides and cavities
have a certain resistance, and the fields therefore penetrate into
the conductors, which significantly complicates the solution of
the wave equation. However, we have seen in Section
\ref{sec:skindepth} that the skin depth in metals is very much
smaller than the RF wavelength. This means that we can reasonably
assume that the field patterns in a waveguide with ideal
boundaries and in a waveguide with resistive metal boundaries will
be practically identical (of course, only for good conductors such
as copper or aluminium). In order to calculate the attenuation of
waves, we can therefore use the fields of a waveguide with ideal
electrical boundaries. From the magnetic field, we calculate the
induced current in the waveguide walls, and then apply the
resistance of the real material to calculate the losses and then
the damping of the wave. This principle is called the power loss
method and is a simplified method for calculating RF losses on the
surfaces of good conductors.

We start by defining the power that is lost per unit length along
the longitudinal axis of the wave\-guide,

\begin{align}
P' &= -\frac{\dd P}{\dd z} \,. \\
\intertext{From} E,H &\propto e^{-\alpha z} \hspace{0.5cm}\Rightarrow \hspace{0.5cm} P \propto e^{-2\alpha z} \,,\\
\intertext{we  immediately obtain} P' &= -\frac{\dd P}{\dd z} =
2\alpha P
\end{align}
\noindent and thus the definition of the attenuation constant\\

\vspace*{-0.7cm} \meqn{ \alpha = \frac{P'}{2P} \,. } {attenuation
constant}{eq:att}

In the next steps, we need to derive expressions for the power $P$
transported through the waveguide, and the power loss per unit
length $P'$. Using the field components of the TM$_{01}$ mode
given in Eq.~\eqref{eq:TM01circ} and the definition of the complex
Poynting vector in Eq.~\eqref{eq:PoyCom}, we obtain

\vspace*{-0.5cm}
\begin{align}
P = \frac{1}{2} \int\limits_A \left(\mathbf{E}\times\mathbf{H}^* \right) \cdot \dd \mathbf{A} = \frac{1}{2} \int\limits_0^a \int\limits_0^{2\pi}
E_r H_{\varphi}^* r \, \dd r \, \dd \varphi &= \frac{C^2 k_z k_\mathrm{c}^2\pi a^2 J_1^2(k_\mathrm{c} a)}{\omega\varepsilon} \,, \label{eq:Patt}\\
\intertext{where we have used} \int\limits_0^a J_1^2(k_\mathrm{c}
r) r \, \dd r &= \frac{a^2}{2} J_1^2(k_\mathrm{c} a) \,.
\end{align}

In order to calculate the losses on the waveguide surface, we
first need to know the surface currents that flow within the skin
depth. For this purpose, we make use of Amp\`{e}re's law, as shown
in Fig.~\ref{fig:Amp2}:

\begin{figure}[h!]
\centering
\begin{tikzpicture}[scale=1, very thick]
\draw (0,0) circle (2); \draw[dashed, thick] (0,0) circle (2.2);
\draw[thick, red] (1.32,1.32) -- (1.65,1.65);
 \draw[thick, black,
dashed] (1.65,1.65) -- (2.9,2.9);
 \draw[thick, red,->] (1.65,1.65)
arc (45:0:2.33345);
 \draw (2.33,0) node[right,red]
{$H_{\varphi}=0$};
 \draw[thick, red] (2.33345,0) arc
(0:-45:2.33345);
 \draw[thick, red] (1.65,-1.65) -- (1.32,-1.32);
 \draw[thick, black, dashed] (1.65,-1.65) -- (2.9,-2.9);
 \draw[thick, black, <->, dashed] (2.8,2.8) arc (45:-45:3.960);
 \draw (4,0) node[right] {$a \, \Delta \varphi$};
 \draw[thick, red,
->] (1.32,-1.32) arc (-45:0:1.8668);
 \draw (1.86,0) [left,red]
node {$H_{\varphi}$}; \draw[thick, red] (1.8668,0) arc
(0:45:1.8668);

\draw[->|, thick] (0,2.6) -- (0,2.2);
 \draw[->|, thick] (0,1.6) --
(0,2); \draw (0,2.2) node[above, left] {$\delta_\mathrm{s}$};

\foreach \x/\y in {0.643/0.766,0.643/-0.766,0.766/0.643,0.866/0.5,0.94/0.342,0.985/0.174,1/0,0.985/-0.174,0.94/-0.342,0.866/-0.5,0.766/-0.643}
    \fill[blue] (2.1*\x,2.1*\y) circle (0.06);

\draw (0,-2.7) node[fill=red!30,right,above,rounded corners]
{$\kappa=\kappa_\mathrm{Al}$}; \draw (0,-1)
node[fill=red!30,right,below,rounded corners] {$\kappa=0$};
\end{tikzpicture}
\caption{\label{fig:Amp2}Amp\`{e}re's law applied to calculate the
surface currents in a circular waveguide}
\end{figure}
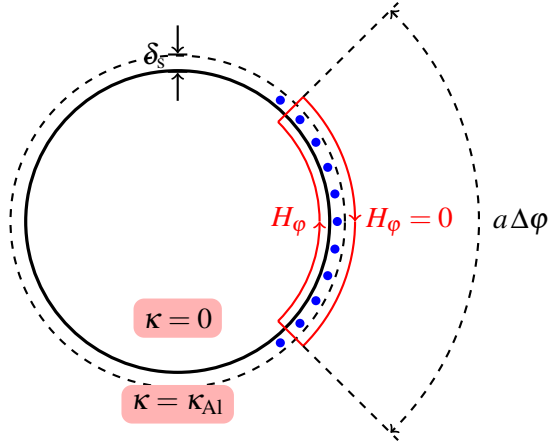

\begin{align}
\oint\limits_c \mathbf{H}\cdot d\mathbf{l} &= I = \oint\limits_c
\mathbf{J} \cdot (\delta_\mathrm{s} \, \dd \mathbf{l}) \,.
&\hspace*{0.75cm}& \mbox{\bf Amp\`{e}re's law}
\end{align}
Since the magnetic field has only an azimuthal
component, we obtain
\begin{align}
H_{\varphi}(r=a,z) &= C k_\mathrm{c} J_1 (k_\mathrm{c} a) e^{-ik_z z} = J_z(z)\delta_\mathrm{s} \,. \label{eq:HpIz}
\end{align}
\begin{align}
\intertext{The power density (in W/m$^3$) in the waveguide wall is
given by} p_v &= \frac{1}{2} \mathbf{E}\cdot \mathbf{J}^* =
\frac{1}{2\kappa} J_z J_z^* = \frac{\partial^3 P}{(\partial r)(r
\, \partial\varphi)(\partial z)} \,,
 && \mbox{\bf power density} \label{eq:powden}\\
\intertext{from which we can write an expression for the power
loss per unit length. Together with Eq.~\eqref{eq:HpIz}, we
obtain} P' &= \frac{\partial P}{\partial z} =  \int\limits_a^{a+
\delta_\mathrm{s}} \int\limits_0^{2\pi} p_v r \, \dd r \, \dd
\varphi =
 \frac{\pi a C^2 k_\mathrm{c}^2 J_1^2(k_\mathrm{c} a)}{\kappa\delta_\mathrm{s}} \,, &&\mbox{\bf power loss per unit length} \label{eq:Ppatt}\\
\intertext{where we have used the fact that $\delta_\mathrm{s} \ll
a$ to simplify the evaluation of the integral. Now we insert
Eqs.~\eqref{eq:Patt} and \eqref{eq:Ppatt} into Eq. \eqref{eq:att}
and obtain an expression for the attenuation of a TM$_{01}$ mode
in a circular waveguide,} \alpha &= \frac{P'}{2P} =
\frac{R_\mathrm{surf}}{Z_0 a \sqrt{1- (f_\mathrm{c}/f )^2}} \,.
&&\parbox{5cm}{\bf attenuation of TM$_{01}$ mode in circular
waveguide}
\end{align}
In the expression above, we have used the definition of the
surface resistance given in Eq.~\eqref{eq:surfResistance} and the
definition of the free-space wave impedance $Z_0$ given in
Eq.~\eqref{eq:ZfreeSpace}.

\begin{figure}[h!]
\centering
\includegraphics[angle=270,width=0.7\textwidth]{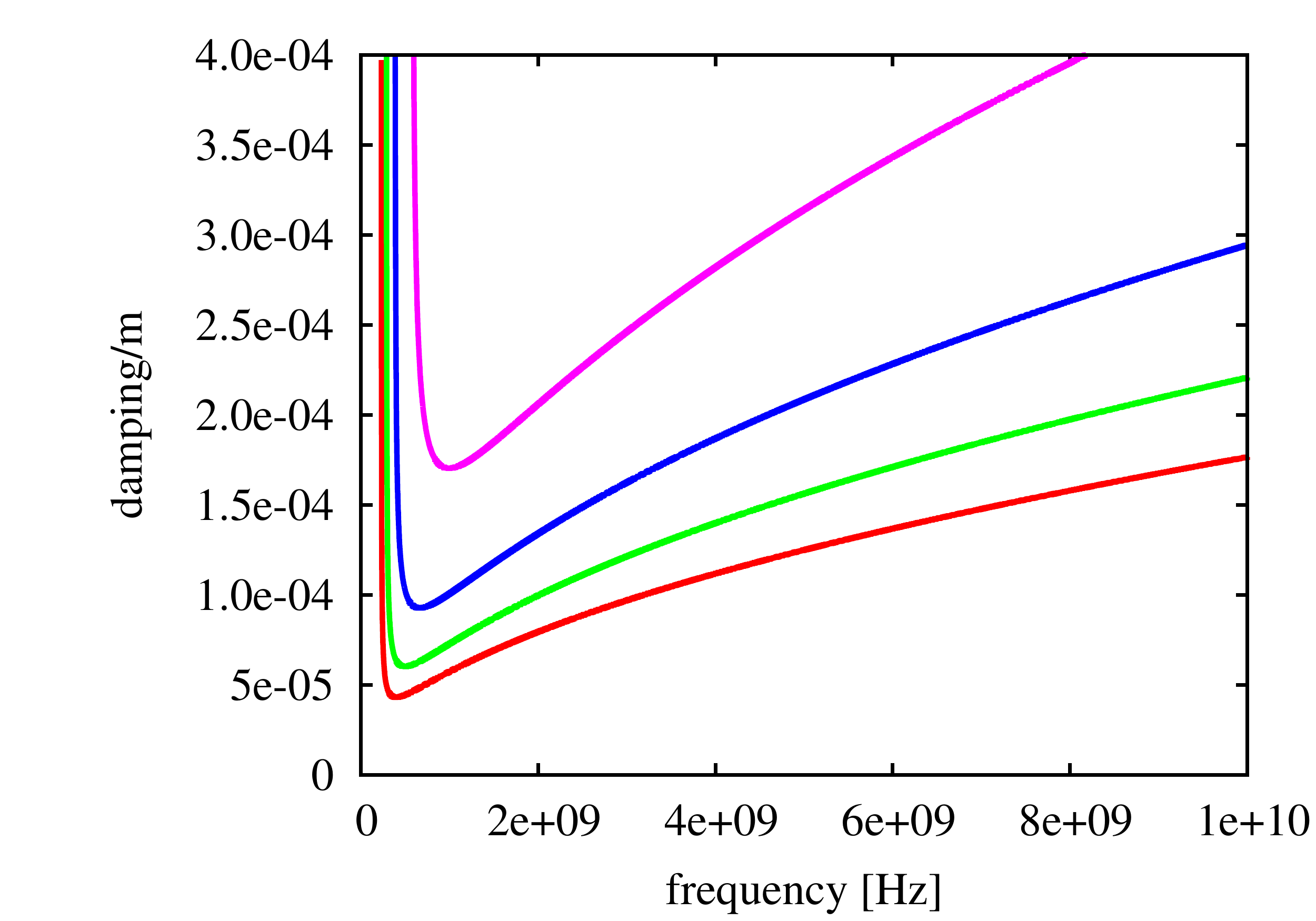}
\caption{Attenuation of the TM$_{01}$ mode in a circular aluminium
waveguide for several different radii: bottom to top, 0.5\,m,
0.4\,m, 0.3\,m, 0.2\,m} \label{fig:AlDamping}
\end{figure}

As an example, we have plotted the attenuation constant for an
aluminium waveguide in Fig.~\ref{fig:AlDamping}, where we can see
that for this type of waveguide:
\begin{itemize}
 \item Large-diameter waveguides result in smaller losses, which
means that a cost optimum has to be found between the cost of the
waveguide, its space requirements, and the losses.
 \item The
minimum losses occur when the operating frequency of the TM$_{01}$
mode is a factor of $\sqrt{3}$ above the cut-off frequency (try to
prove this!).
\end{itemize}

\setlength{\captionmargin}{1cm}

\setlength{\captionmargin}{0cm}
\section{Accelerating cavities}

\subsection{Travelling-wave cavities}
\begin{figure}[h!]
    \begin{center}
        \begin{tikzpicture}[>=stealth, x=5cm/5,red!70!black,thick]
    \draw[->,black,dashed, line width=1pt](-2,0)  -- (2.5,0)node[right]{Beam};
    \foreach \x in {-1.4,-0.7,0,0.7,1.4}
        \draw[black,fill=gray!50] (\x,0) ellipse (0.2cm and 0.8cm);
    \foreach \x in {-1.4,-0.7,0,0.7,1.4}
        \draw[black,fill=white] (\x,0) ellipse (0.12cm and 0.6cm);
    \draw[black] (-1.4,0.8) -- (1.4,0.8) (-1.4,-0.8) -- (1.4,-0.8);
    \draw[<->,black] (-1.4,-1) -- (1.4,-1) node [below,midway] {$L$};
    \draw[<->,black] (-1.8,-0.6) -- (-1.8,0.6) node [left] {$2a$};
    \draw[<->,black] (1.8,-0.8) -- (1.8,0.8) node [right] {$2b$};
    \draw[->,black] (-0.5,1) -- (-0.05,1);
    \draw[->,black] (0.5,1) -- (0.05,1) node [above,midway] {$h$};
    \end{tikzpicture}
    \caption{Geometry of a simple travelling-wave structure}
    \label{fig:discloaded}
    \end{center}
    \end{figure}
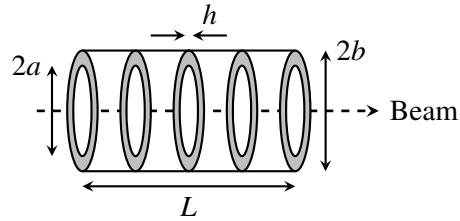

In order to accelerate particles in a `waveguide-like' structure,
the phase velocity in the structure needs to be slowed down, which
can be achieved by putting some `obstacles' into the waveguide. In
Fig.~\ref{fig:discloaded}, we see a simple example of a
disc-loaded waveguide.

\noindent The dispersion relation for such a structure is derived in, for example, \cite{bib:Wangler} as\\
\meqn{
\omega = \frac{2.405 c}{b} \sqrt{1+\kappa (1-\cos(k_z L) e^{-\alpha h})}
 \,, }{dispersion relation of disc-loaded circular waveguide}{eq:dispcirc}
\noindent where
\begin{equation}
\kappa = \frac{4a^3}{3\pi J_1^2(2.405) b^2L} \ll 1\hspace*{0.5cm}
\mbox{and} \hspace*{0.5cm}\alpha \approx \frac{2.405}{a} \,.
\end{equation}

\begin{figure}[h!]
    \begin{center}
    \begin{tikzpicture}[>=stealth]
    \draw[->,black](-5,-0.5)  -- (5,-0.5)node[right]{$k_{z}$};
    \foreach \x in {-4,-3,...,4}
    \draw [x=4*1.57ex](\x,-0.6) -- (\x,-0.4);
    \draw[->,black](0,-0.5)node[below]{$0$}  -- (0,3)node[above]{$\omega$};
    \draw [x=4*1.57ex,dashed](4/3,-0.5)node[below]{$\frac{2\pi}{3L}$}--(4/3,3);
    \draw[x=-4*1.57ex,y=6.5ex,dashed,green!70!black,very thick] (0,0)cos(1,1) sin (2,2);
    \draw[x=-4*1.57ex,y=6.5ex,red!70!black,very thick](2,2)cos (3,1) sin (4,0);
    \draw[x=4*1.57ex,y=6.5ex,red!70!black,very thick] (0,0)cos(1,1) sin (2,2);
    \draw[x=4*1.57ex,y=6.5ex,dashed,green!70!black,very thick](2,2)cos (3,1) sin (4,0);
    \draw [x=4*1.57ex,y=6.5ex,->,green!70!black](3.5,1.5)node[above right]{Reflected wave}--(3,1);
    \draw [dashed,blue!70!black,very thick](0.7,-0.5)--(1.95,3)node[right]{$v_\mathrm{ph}=c$};
    \draw [x=4*1.57ex](-4,-0.5)node[below]{$-\frac{2\pi}{L}$};
    \draw [x=4*1.57ex](4,-0.5)node[below]{$\frac{2\pi}{L}$};
    \draw [x=4*1.57ex](-2,-0.5)node[below]{$-\frac{\pi}{L}$};
    \draw [x=4*1.57ex](2,-0.5)node[below]{$\frac{\pi}{L}$};
    \draw (0.1,0)--(-0.1,0) node[below left]{$\omega_\mathrm{c}$};
    \draw (0.1,2.37)--(-0.1,2.37) node[below left]{$\omega_{\pi}$};
    \end{tikzpicture}
\caption{Dispersion diagram for a disc-loaded travelling-wave
structure. Here, the chosen operating point is $v_\mathrm{ph} = c$
and $k_z = 2\pi/3L$.}
    \label{fig:dispersion-discs}
    \end{center}
    \end{figure}
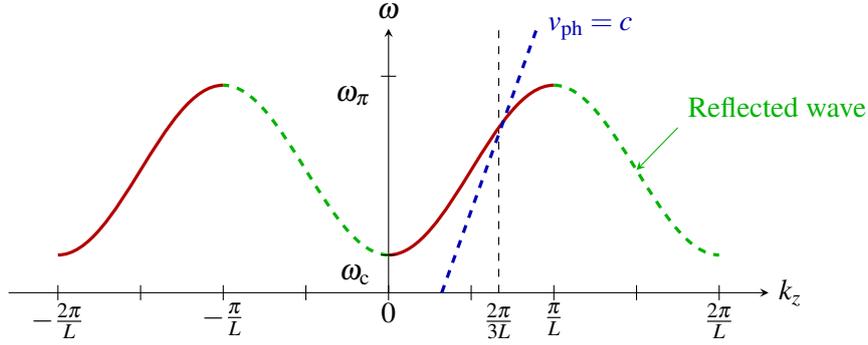

\noindent Plotting Eq.~\eqref{eq:dispcirc} gives us the Brioullin
diagram in Fig.~\ref{fig:dispersion-discs}, where we can see that
we now obtain phase velocities that are equal to or even below the
speed of light. We can also understand why the $2\pi/3$ mode is
often used for acceleration in electron accelerators, because for
this mode (in this example) the phase velocity is just equal to
the speed of light. It should be noted that with different
geometries, it is possible to operate with different modes and
also at velocities $v_\mathrm{ph} < c$. When a structure operates
in the $2\pi/3$ mode, this means that the RF phase shifts by
$2\pi/3$ per cell, or, in other words, one RF period extends over
three cells.

\setlength{\captionmargin}{1cm}

\setlength{\captionmargin}{0cm}

By attaching an input and an output coupler to the outermost cells
of the structure, we obtain a usable accelerating structure. Since
the particles gain energy in every cell, the electromagnetic wave
becomes increasingly damped along the structure. It is then
extracted via the output coupler and dumped in an RF load. If one
is interested in obtaining the maximum possible accelerating
gradient in each cell, then one can counteract the decreasing
fields by changing the bore radius from cell to cell. The idea is
to slow down the group velocity from cell to cell and obtain a
`constant-gradient' structure, rather than a `constant-impedance'
structure where the bore radius is kept constant. Other
optimizations, for example for maximum efficiency, are also
possible.

\subsection{Standing-wave cavities}
One obtains a cylindrical standing-wave structure by simply
closing both ends of a circular waveguide with electric walls.
This yields multiple reflections on the end walls until a
standing-wave pattern is established. Owing to the additional
boundary conditions in the longitudinal direction, we obtain
another `restriction' on the existence of electromagnetic modes in
the structure. Whereas a longitudinally open travelling-wave
structure allows all frequencies and all cell-to-cell phase
variations on the dispersion curve, now only certain `loss-free'
modes (still assuming perfectly conducting walls) with discrete
frequencies and discrete phase changes can exist in a cavity. If
RF power is fed in at a different frequency, then the fields
excited are damped exponentially, similarly to the modes below the
cut-off frequency of a waveguide.

The corresponding dispersion relation for a standing-wave cavity
can again be found in textbooks (see \cite{bib:Wangler} and also
\cite{bib:CASMaurizio}). However, it is necessary to pay attention
to whether the structure under consideration has magnetic or
electric cell-to-cell coupling and what kind of end cell is
assumed in the analysis. The most common form of the dispersion
relation is derived from a coupled-circuit model with $N+1$ cells.
Usually the model has half-cell terminations on both ends of the
chain, representing the behaviour of
an infinite chain of electrically coupled resonators (compare the original paper by Nagle \emph{et al.} \cite{Knapp}):\\

\vspace{-1cm} \meqn{ \omega_n =
\frac{\omega_{0}}{\sqrt{1+k\cos\left(n\pi/N \right)}} \mbox{,
}n=0,1,\ldots ,N\mbox{.}} {dispersion relation for
half-cell-terminated standing-wave structure}{eq:dispstandmag}

\noindent Assuming an odd number of cells, $\omega_{\,0}$ is the
frequency of the $\pi/2$ mode and of an uncoupled single cell; $k$
is the cell-to-cell
coupling constant, and $n\pi/N$ is the phase shift from cell to cell. For $k\ll 1$, which is usually fulfilled, the coupling constant is given by\\

\vspace*{-0.7cm}
\meqn{
 k = \frac{\omega_{\pi \; \mathrm{mode}} - \omega_{0 \; \mathrm{mode}}}{\omega_{0}}\mbox{.}}
 {coupling constant}{eq:couplconstant}

Two characteristics of the dispersion curve are worth noting:

\begin{itemize}
 \item The total width of the frequency band of the mode, $\omega_{\pi \; \mathrm{mode}} - \omega_{0 \; \mathrm{mode}}$,
 is independent of the number of cells,
which means that we can determine the cell-to-cell coupling
constant by measuring the complete structure (but this is only
true if all coupling constants are equal).
 \item For electric coupling, the
0 mode has the lowest frequency and the $\pi$ mode has the
highest. In the case of magnetic coupling, this behaviour is
reversed, and one can find the corresponding dispersion curve by
changing the sign before the coupling constant in
Eq.~\eqref{eq:dispstandmag}.
\end{itemize}

\begin{figure}[h!]
    \begin{center}
    \begin{tikzpicture}[domain=0:7,samples=60]
    \draw[red,very thick, mark=x,mark repeat=10] plot (\x, {80/(sqrt(1-0.05*cos(180*\x /7)))});
    \draw[black] (0,77.5) -- (7,77.5);
    \draw[black] (0,77.5) -- (0,82.5);
    \draw[black] (7,77.5) -- (7,82.5);
    \draw[black] (0,77.5) node [below]{0};
    \draw[black] (7,77.5) node [below]{$\pi$};
    \draw[black] (3.5,77.7) -- (3.5, 77.5) node [below]{$\pi/2$};
    \draw[black] (-0.2,82.078) -- (0,82.078) node [left]{$\frac{\displaystyle\omega_0}{\displaystyle\sqrt{1-k}}$};
    \draw[black] (7.2,82.078) -- (7,82.078);
    \draw[black] (-0.2,78.072) -- (0,78.072) node [left]{$\frac{\displaystyle\omega_0}{\displaystyle\sqrt{1+k}}$};
    \draw[black] (7.2,78.072) -- (7,78.072);
    \end{tikzpicture}
    \caption{Dispersion diagram for a standing-wave structure with seven magnetically coupled cells}
    \label{fig:dispersion-standing}
    \end{center}
    \end{figure}
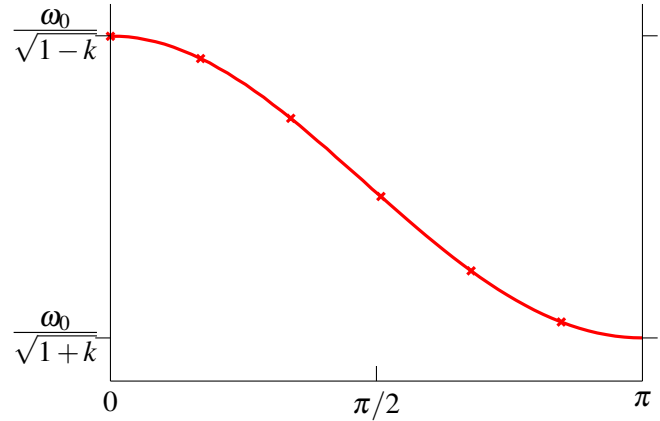

\noindent In Fig.~\ref{fig:dispersion-standing}, we plot the
dispersion curve for a seven-cell (half-cell-terminated)
magnetically coupled structure according to
Eq.~\eqref{eq:dispstandmag}.

In practice, one usually has cavities with full-cell termination,
and in this case one has to detune the frequences of the end cells
to obtain a flat field distribution in the cavity
\cite{Schriber01}. In this case it is possible to have a flat
field distribution for either the $0$ mode or the $\pi$ mode but
not for both at the same time, because the end cells have to be
detuned by different amounts in the two cases \cite{Schriber2}.

\subsection{Standing wave versus travelling wave}
The principal difference between the two types of cavity is in how
and how fast the cavities are filled with RF power.
Travelling-wave structures are filled `in space', which means
that, basically, cell after cell is filled with power. For the
following estimations, we assume a frequency in the range of
hundreds of megahertz. The filling of a travelling-wave structure
typically takes place with a speed of approximately 1--3\% of the
speed of light and results in total filling times in the
submicrosecond range. Standing-wave structures, on the other hand,
are filled `in time': the electromagnetic waves are reflected at
the end walls of the cavity and slowly build up a standing-wave
pattern of the desired amplitude. For normal-conducting cavities,
the time required for this process is typically in the range of
tens of microseconds. For superconducting cavities, the filling
time can easily go into the millisecond range (depending on the
required field level, the accelerated current, and the cavity
parameters). This means that for applications that require very
short beam pulses ($< 1$~\textmu s), travelling-wave structures
are much more power-efficient. For longer pulses ($> n \times
10$~\textmu s), both types of structures can be optimized to
achieve similar efficiencies and costs.

Since one can have extremely short RF pulses in a travelling-wave
structure, one can obtain much higher peak fields than in a
standing-wave structure. This is demonstrated by the accelerating
structures for CLIC \cite{clic}, which have reached values of
approximately 100~MV/m (limited by electrical breakdown), whereas
the design gradient for the superconducting (standing-wave)
cavities for the ILC \cite{ilc} is just slightly above 30~MV/m
(this value is generally limited by field emission and by quenches
caused by the peak magnetic field).

Travelling-wave structures can, theoretically, be designed for
non-relativistic particles. In existing accelerators, however,
they are mostly used for relativistic particles. Low-beta
acceleration is typically performed with standing-wave cavities.

Because of the lack of an obvious criterion (other than the pulse
length or the particle velocity), an optimization and costing
exercise has to be performed for each specific application in
order to decide which structure is more efficient. Two excellent
papers \cite{Miller86, Moiseev00} in which this exercise is
performed can be used as references.

\subsection{The pillbox cavity}
In this chapter, we shall analyse only the simplest TM-mode
cavity, the so-called pillbox cavity. A selection of cavities
using other mode types is described in \cite{bib:RFCASFrank}.

\begin{figure}[h!]
    \begin{center}
    \includegraphics[width=3.5cm]{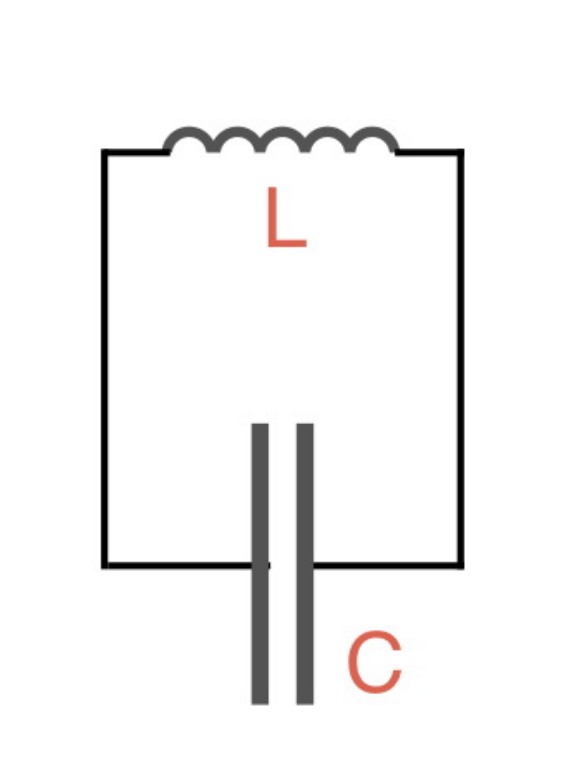}\hspace*{1cm}
    \includegraphics[width=3.5cm]{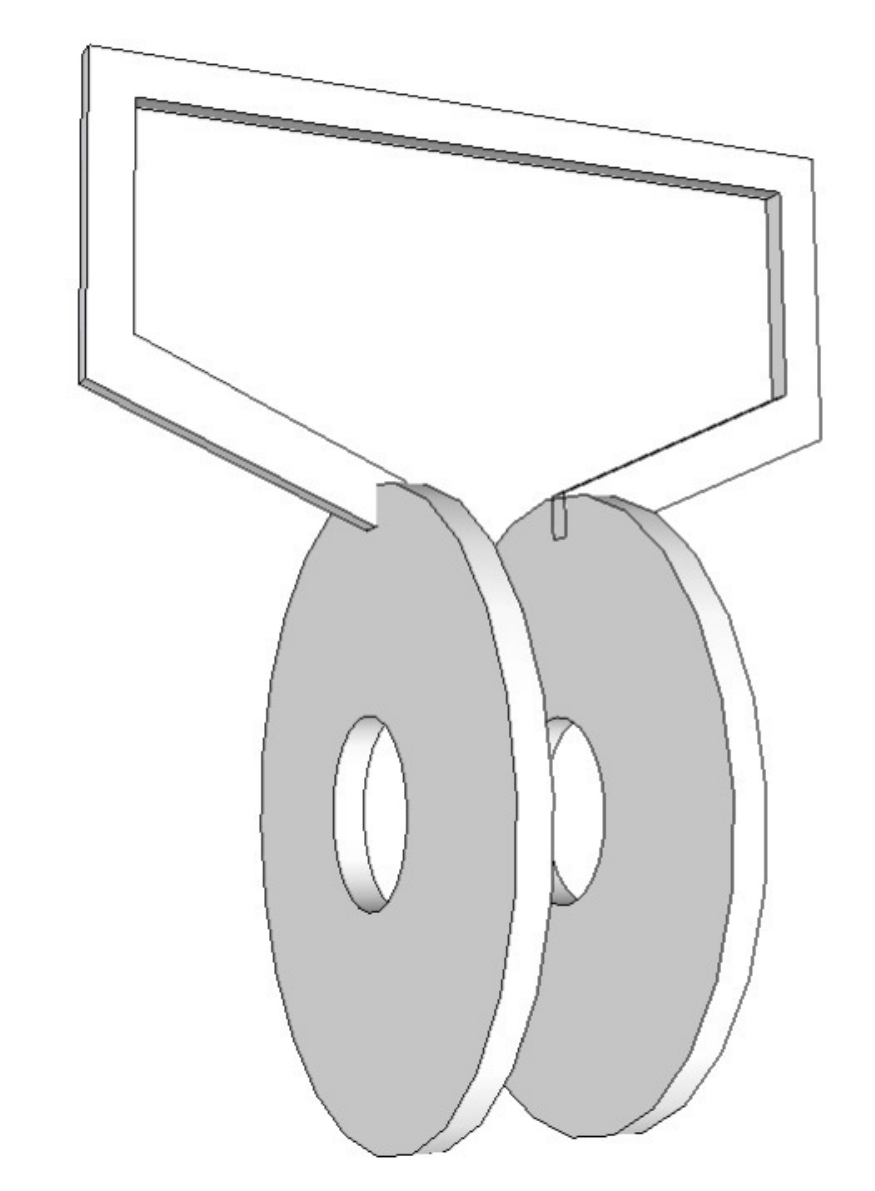}\hspace*{1cm}
    \includegraphics[width=3.5cm]{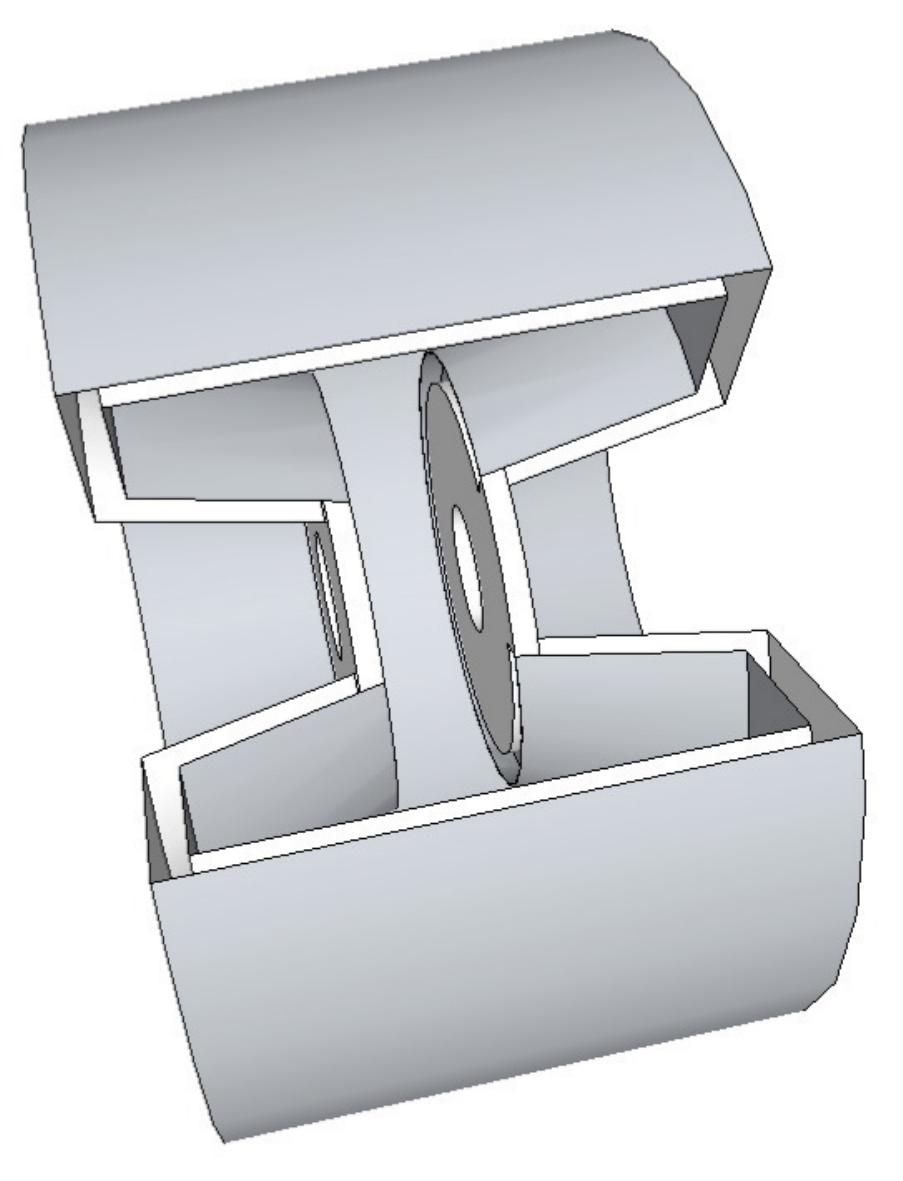}
    \caption{Transition from a lumped resonating circuit to a resonating cavity}
    \label{fig:lumped}
    \end{center}
    \end{figure}

Resonating cavities can be represented conveniently by a
lumped-element circuit consisting of an inductor (for storage of
magnetic energy) and a capacitor (for storage of electric energy).
Looking at Fig.~\ref{fig:lumped}, one can easily imagine how the
lumped circuit can be transformed into a cavity.

The pillbox cavity is nothing more than an empty cylinder with a
conducting inner surface. The simplest mode in this cavity is the
TM$_{\,010}$ mode, which has zero full-period variations in the
azimuthal direction ($\varphi$), one `zero' of the axial field
component in the radial direction ($r$), and zero half-period
variations in the longitudinal ($z$) direction. We can derive the
general field equations by using the vector potential for a
circular waveguide given in Eq.~\eqref{eq:vpfcw} and simply
superimposing two waves, one propagating in the
positive $z$ direction and one in the negative $z$ direction:\\
\meqn{ A_z^\mathrm{TM/TE} = C J_m (k_r r) \cos(m\varphi)
\underbrace{\left(e^{-i k_z z}+ e^{i k_z z} \right)}_{2\cos(k_z
z)} \,. } {vector potential for travelling waves in the positive
and negative $z$ directions}{}

\noindent Using Eq.~\eqref{eq:vpTM}, we derive the TM field components\\
\meqn{
\begin{array}{lcl}
E_r &= {\displaystyle\frac{i}{\omega\varepsilon}} {\displaystyle
\frac{\partial H_{\varphi}}{\partial z}} &= i2C {\displaystyle
\frac{k_z k_r}{\omega\varepsilon}} J'_m(k_r r)
\cos(m\varphi) \sin(k_z z) \,, \vspace{0.2cm}\\
E_{\varphi} &= -{\displaystyle \frac{i}{\omega\varepsilon}}
{\displaystyle \frac{\partial H_r}{\partial z}} &= -i2C
{\displaystyle \frac{m k_z}{\omega\varepsilon r}} J_m(k_r
r)\sin(m\varphi) \sin(k_z z) \,,
 \vspace{0.2cm} \\
E_{z} &= {\displaystyle \frac{i k_r^2}{\omega\varepsilon}} A_z &=
i2C {\displaystyle
\frac{k_r^2}{\omega\varepsilon}} J_m (k_r r) \cos(m\varphi)\cos(k_z z) \,, \vspace{0.2cm} \\
H_r &= {\displaystyle \frac{1}{r}} {\displaystyle \frac{\partial
A_z}{\partial \varphi}} &= -2C {\displaystyle
\frac{m}{r}} J_m (k_r r) \sin(m\varphi) \cos(k_z z) \,, \vspace{0.2cm} \\
H_{\varphi} &= -{\displaystyle \frac{\partial A_z}{\partial r}} &=
-2C k_r J'_m (k_r r)\cos(m\varphi) \cos(k_z z) \,.
\end{array} }
{TM modes in a pillbox cavity}{}

In the case of standing-wave cavities, the term `cut-off'
frequency does not really make sense, so we have replaced the
symbol $k_\mathrm{c}$ by $k_r$, indicating that we have a radial
dependence of the axial field component, which can also be
interpreted as a radial wavenumber.

\begin{figure}[h!]
\centering
\begin{tikzpicture}[scale=0.9, ultra thick]

\draw (0,0) ellipse (0.7 and 1.4);
\draw (2,1.4) arc (90:-90:0.7 and 1.4);
\draw (0,1.4) -- (2,1.4);
\draw (0,-1.4) -- (2,-1.4);
\draw[->] (0,0) -- (0,2.2) node[above] {\Large $r$};
\draw (0.2,1.6) node {\Large $a$};
\draw[dashed] (0,0) -- (2.7,0);
\draw (2.9,0.35) node {\Large $L$};
\draw[->] (2.7,0) -- (3.5,0) node [right] {\Large $z$};

\end{tikzpicture}
\caption{\label{fig:pillbF}Pillbox cavity}
\end{figure}
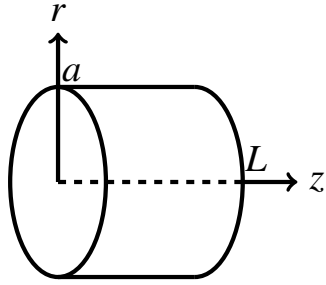

In the next step, we apply the boundary conditions for a pillbox
with radius $a$ and length $L$ as shown in Fig.~\ref{fig:pillbF}.
We obtain

\vspace*{-0.7cm}
\begin{align}
E_r(z=0/L), \; E_{\varphi}(z=0/L) &= 0 \hspace{0.5cm} \Rightarrow k_z = \frac{p\pi}{L} \,, \\
E_{\varphi}(r=a), \; E_z(r=a), \; H_r (r=a) &= 0 \hspace{0.5cm}
\Rightarrow k_r = \frac{j_{mn}}{a} \,.
\end{align}

In the case of the circular waveguide, the transverse boundary
condition made a discrete quantity out of $k_\mathrm{c}$ (which we
now call $k_r$ in the above equations), and thus defined the
cut-off frequency. Now, with the second boundary in the $z$
direction, we obtain a discrete solution for $k_z$ also. The two
boundary conditions together result in a discrete set
of frequencies (the dispersion relation) for our pillbox cavity:\\

\vspace*{-0.7cm} \meqn{ k^2 = \frac{\omega^2}{c^2} = k_z^2 + k_r^2
\hspace{0.5cm} \Rightarrow f_{mnp}^\mathrm{TM}=\frac{c}{2\pi}
\sqrt{\left( \frac{j_{mn}}{a} \right)^2+\left(\frac{p\pi}{L}
\right)^2} \,. } {dispersion relation for TM modes in a pillbox
cavity}{eq:DispPillBox}

We note that the dispersion relation of a single-cell cavity as
given above is different from the dispersion relation that can be
derived for a multicell cavity, as in the case of
Eq.~\eqref{eq:dispstandmag}. The latter is derived from a model of
equivalent lumped circuits, each representing a cell resonating in
the TM$_{010}$ mode and coupled to its neighbours in order to
model the behaviour of a multicell cavity, whereas
Eq.~\eqref{eq:DispPillBox} is directly derived from Maxwell's
equations and describes a field pattern that is created by the
boundary conditions of our pillbox.

The TM mode with the lowest frequency is the TM$_{010}$ mode, with a frequency\\

\begin{figure}[h!]
\centering
\includegraphics[width=2cm]{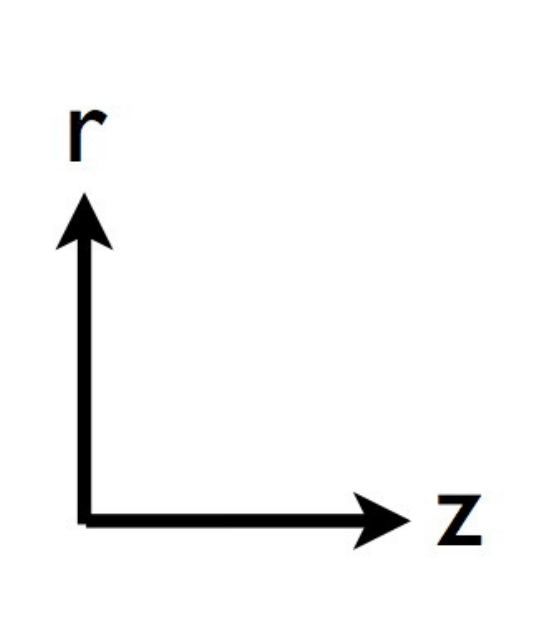}
\includegraphics[width=4cm]{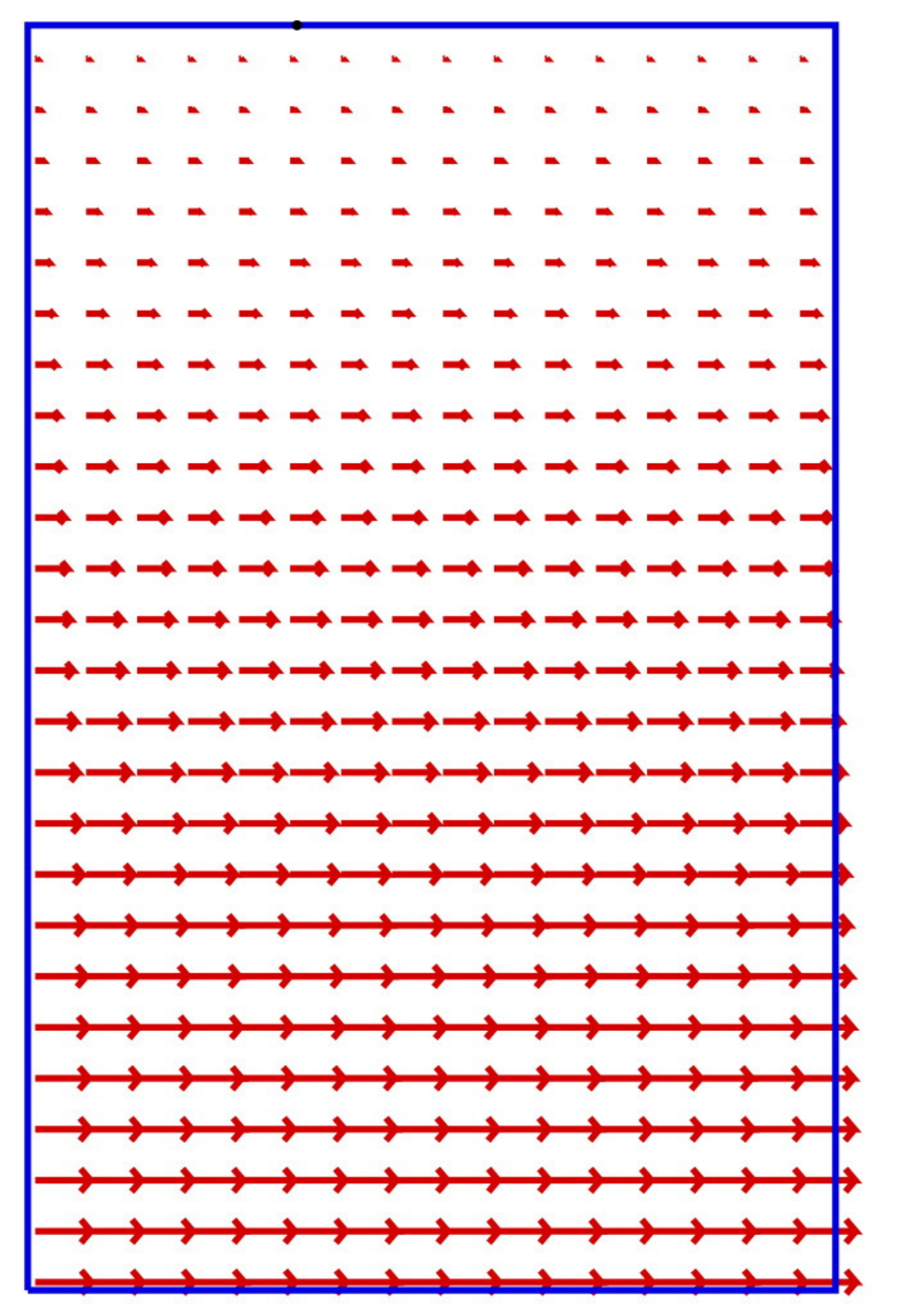}\hspace{2cm}
\includegraphics[width=4cm]{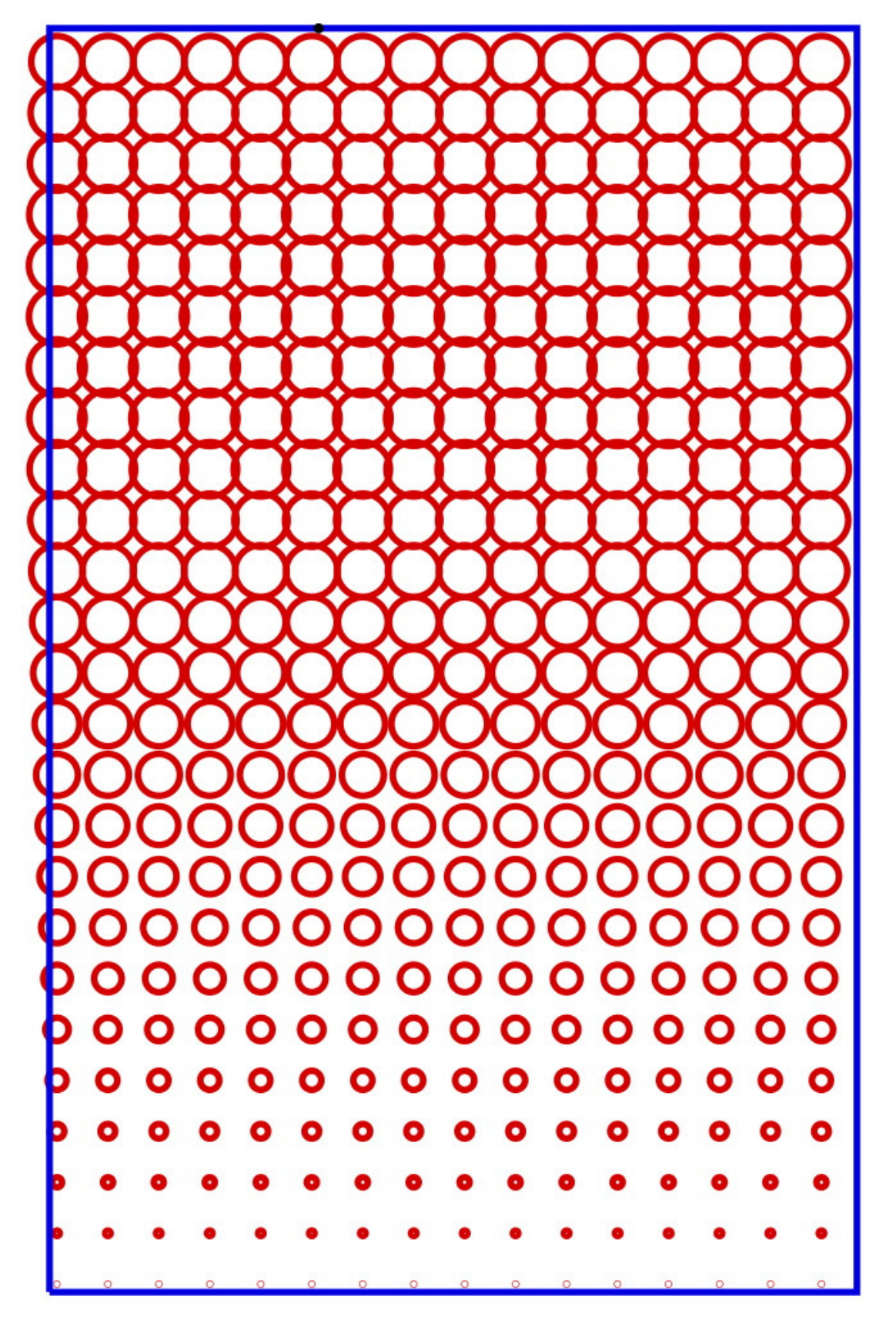}
\caption{\label{fig:PFields}Field pattern of the TM$_{010}$ mode
in a pillbox cavity}
\end{figure}

\vspace*{-0.7cm} \meqn{ f_{010}^\mathrm{TM}= \frac{2.405 c}{2\pi
a} \,, } {frequency of the TM$_{010}$ pillbox mode}{eq:f-pillbox}
and its field components are\\
\meqn{
\begin{array}{lll}
E_z &= -i2C {\displaystyle \frac{j_{01}^2}{a^2\omega\varepsilon}}
J_0 \left({\displaystyle \frac{j_{01}}{a}} r \right) &= E_0 J_0
\left({\displaystyle
\frac{j_{01}}{a}} r \right) \,, \vspace{0.2cm} \\
H_{\varphi} &= \hspace{0.5cm} 2C {\displaystyle \frac{j_{01}}{a}}
J_1 \left({\displaystyle \frac{j_{01}}{a}} r \right) &=
{\displaystyle\frac{E_0}{Z_0}} J_1 \left({\displaystyle
\frac{j_{01}}{a}} r \right) \,.
\end{array} }
{field components of the TM$_{010}$ pillbox mode}{} \noindent
Figure \ref{fig:PFields} shows the field pattern of the TM$_{010}$
mode, simulated by Superfish$^{\copyright}$.

\subsection{Basic cavity parameters}
In order to characterize and optimize cavities, we need some
commonly used figures of merit, which we shall define here in
general terms and then apply to our simple pillbox cavity. In the
following, we assume that we are dealing with an axially symmetric
cavity resonating in the TM$_{010}$ mode.

\subsubsection{Energy gain in a cavity}
For particles traversing a cavity on axis, the electric field generally has the following form:\\
\begin{equation}
E_z (r=0,z,t) = E(0,z) \cos(\omega t + \varphi) \,,
\label{eq:Wgain}
\end{equation}
\noindent which we can use to calculate the energy gain of a particle when it traverses the cavity,\\
\meqn{
\Delta W &= q\int\limits_{-L/2}^{L/2} E(0,z) \cos (\omega t + \varphi)\\
              &= q V_0 T \cos\varphi = qE_0 T L \cos\varphi  \,,
}{energy gain in a cavity (Panofsky equation)}{}

\noindent where the cavity voltage is given by\\
\meqn{ V_0 = \int\limits_{-L/2}^{L/2} E(0,z) \, \dd z = E_0 \,,
}{cavity voltage}{}

\noindent and the `difficult mathematics' has been lumped into the so-called transit time factor\\
\meqnbis{ T = \frac{\int\limits_{-L/2}^{L/2} E(0,z) \cos(\omega t(z))
\, \dd z}{\int\limits_{-L/2}^{L/2} E(0,z) \, \dd z} -
\underbrace{\tan\phi \frac{\int\limits_{-L/2}^{L/2} E(0,z)
\sin(\omega t(z)) \, \dd z}{\int\limits_{-L/2}^{L/2} E(0,z) \, \dd
z}}_{ =0 \; \mbox{\scriptsize  if } E(0,z) \; \mbox{\scriptsize
is symmetric about } z=0} \,. }{transit time factor}{}

\noindent This takes into account the fact that the RF electric
field changes during the passage of the particles. It gives the
ratio between the energy gained in an RF field and in a DC field
and is therefore always less than 1. We note that the Panofsky
equation takes account of the changing velocity of the particles
when they cross the accelerating gap. This makes the integrals in
the above equations difficult to evaluate. Assuming that the
velocity change of the beam particles during their passage is
small, however, one can say that
\begin{equation}
\omega t \approx \omega \frac{z}{v} = \frac{2\pi z}{\beta \lambda}
\,,
\end{equation}
which changes the expression for the transit time factor to (assuming that $E(0,z)$ is symmetric about $z=0$)\\
\meqn{T = \frac{\int\limits_{-L/2}^{L/2} E(0,z) \cos({2\pi z} /
{\beta\lambda}) \dd z}{\int\limits_{-L/2}^{L/2} E(0,z) \, \dd z}
\,. }{transit time factor for small velocity changes}{eq:Tsvc}

The accelerating voltage $V_\mathrm{acc}$ is the voltage that the
particle `sees' when crossing the cavity and should not be
confused with the cavity
voltage $V_0$. We thus define\\
\meqn{ V_\mathrm{acc} = V_0T = E_0 L T
 \,. }{accelerating voltage}{}

\subsubsection{Shunt impedance}
The shunt impedance tells us how much voltage a cavity will
provide when a certain amount of power is dissipated in the cavity
walls. This is one of the parameters to be maximized in cavity design, since a large shunt impedance reduces the power consumption of an RF cavity.
 The general definition is\\

\vspace*{-0.7cm} \meqn{ R_\mathrm{s} = \frac{V_0^2}{P_\mathrm{d}}
 \,. }{shunt impedance (linac definition)}{}

\noindent The benefit of a high shunt impedance can easily be
diminished by having a small transit time factor, because in this
case the cavity voltage cannot be used efficiently  to transfer
energy to the beam. Therefore one usually tries to optimize
both the shunt impedance and the transit time factor, which explains the definition of the effective shunt impedance\\

\vspace*{-0.7cm} \meqn{ R = \frac{(V_0 T)^2}{P_\mathrm{d}}
 \,. }{effective shunt impedance}{}

When comparing multicell structures operating at different
frequencies, one is interested less in the efficiency per cell
(because the cell size depends on, for instance, the frequency
chosen) than in the efficiency per unit length of the accelerating
structure.
For this reason, we define\\

\vspace*{-0.7cm} \meqn{ Z = \frac{R_\mathrm{s}}{L} =
\frac{E_0^2}{P_\mathrm{d}/L} }{shunt impedance per unit length}{}

\noindent and\\

\vspace*{-0.7cm} \meqn{ ZT^2 = \frac{R}{L} = \frac{(E_0
T)^2}{P_\mathrm{d}/L}
 \,. }{effective shunt impedance per unit length}{}

\subsubsection{`Linac' and `circuit' definitions of shunt impedance}
It turns out that different communities of accelerator experts use
different definitions of the shunt impedan\-ce. Linac experts
usually use the definitions presented above, whereas the people
who deal with circular machines generally use a definition that is
derived from the lumped-circuit definition of a resonator (see
Section~\ref{sec:lumped}). In that definition, all
shunt impedances are exactly half as large, following\\

\vspace*{-0.7cm} \meqn{ R_\mathrm{s}^\mathrm{c} = \frac{V_0^2}{2
P_\mathrm{d}}
 \,. }{shunt impedance (circuit definition)}{}
So, before you discuss shunt impedances with anyone, make sure
that you are using the same definition. In order to mark the
difference clearly, we use $R_\mathrm{s}^\mathrm{c}$ in this text
to identify when the circuit definition is being used.

\subsubsection{3~dB bandwidth and quality factor}
The quality factor $Q$ describes the bandwidth of a resonator and
is defined as the ratio of the reactive power (stored energy) to
the real power
that is lost in the cavity walls:\\

\vspace*{-0.7cm} \meqn{ Q=\frac{\omega}{\Delta\omega}=\frac{\omega
W}{P_\mathrm{d}}
 \,. }{quality factor}{eq:qfactor}

\begin{figure}[h!]
\centering
\includegraphics[width=0.6\textwidth]{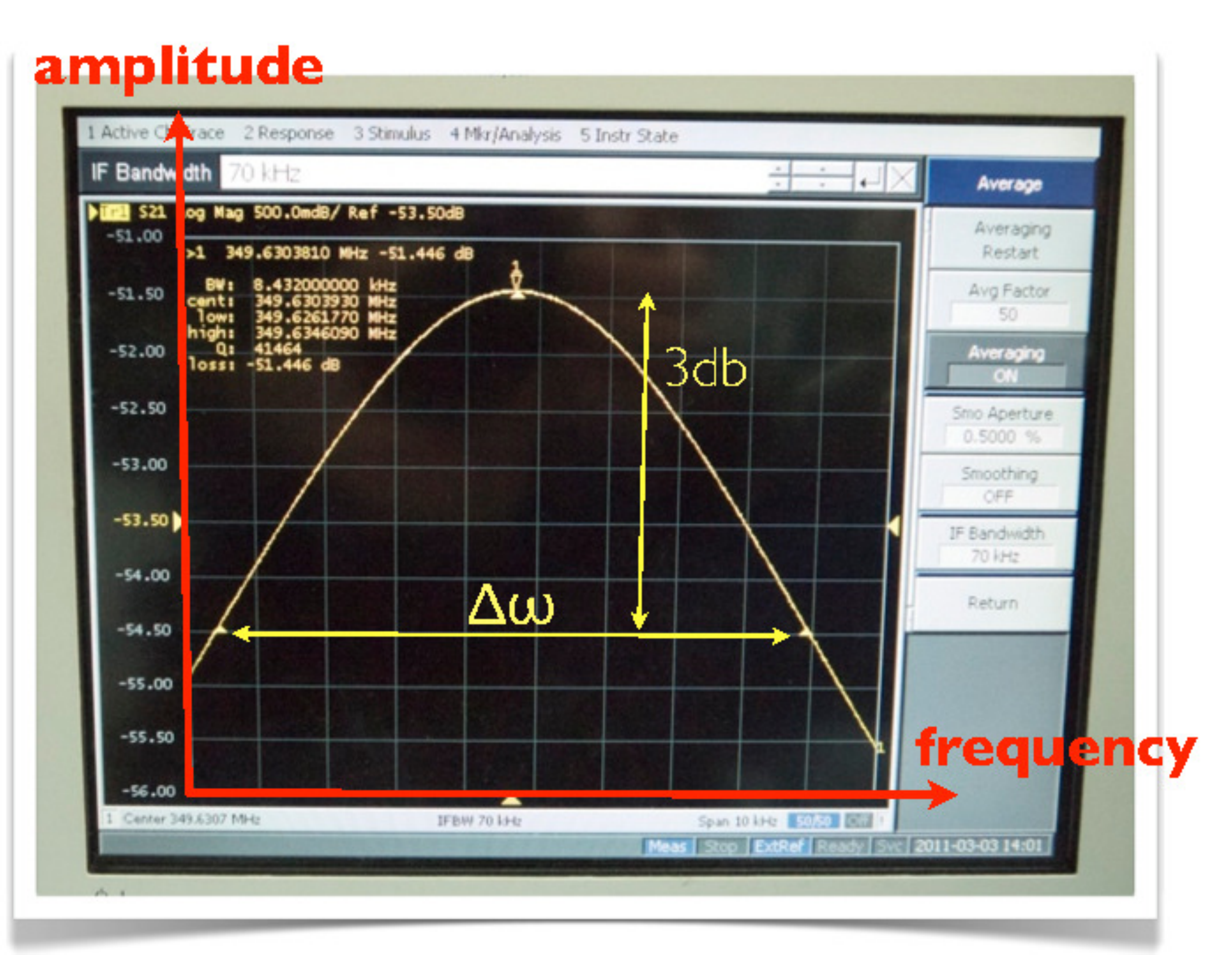}
\caption{Measurement of frequency, 3~dB bandwidth, and Q-factor
with a network analyser} \label{fig:bandwidth}
\end{figure}

If a resonator were built with ideal electrical walls (zero
electrical resistance), the resonance curve would be a delta
function at the resonance frequency. So, the bandwidth
$\Delta\omega$ would be zero and the quality factor would be
infinite. In reality, even superconducting cavities have a certain
surface resistance, which is why all our cavities have a certain
bandwidth and a finite quality factor. Figure~\ref{fig:bandwidth}
shows a typical resonance curve measured with a network analyser.
In a measurement of this kind, two antennas penetrate the cavity.
The first antenna sends an RF signal with a frequency sweep, and
the second picks up the field level in the cavity. As a result, we
obtain a plot of the field level versus frequency. The bandwidth
is defined as the frequency width of the resonance curve, measured
as the distance between the points where the field level has
dropped by 50\% (or $-3$~dB), as shown in
Fig.~\ref{fig:bandwidth}.

Together with the shunt impedance, one can define another figure
of merit, $(R/Q)$, which is used to maximize the energy gain in a
cavity
for a given stored energy:\\

\vspace*{-0.7cm}
 \meqn{
 \left(\frac{R}{Q} \right) = \frac{(V_0 T)^2}{\omega W}
 \,.  }{${(R/Q)}$}{}
$(R/Q)$ is independent of the surface losses of the cavity and can
therefore be used to qualify the geometry of an accelerating
cavity.

\subsubsection{Filling time of a cavity}
This section is a short extract from \cite{bib:RFCASFrank}, which
can be consulted for more details. The dissipated power in a
cavity must be
equal to the rate of change of the stored energy:\\

\vspace*{-0.7cm} \meqn{ P_\mathrm{d} = -\frac{\dd W}{\dd t} =
\frac{\omega_0 W}{Q_0}
 \,. }{}{}

\noindent The solution of the above equation can be written as\\

\vspace*{-0.7cm} \meqn{ W(t) = W_0 e^{-2t/\tau} \,, }{}{}

\noindent which describes an exponential decay of the stored energy with a `filling time constant' $\tau$, where\\

\vspace*{-0.7cm} \meqn{ \tau = \frac{2Q_0}{\omega_0} \,. }{filling
time constant}{}

\noindent If the cavity is equipped with a power coupler, we have
to consider the `loaded $Q$' (which will be derived later),
 and the filling time constant changes to\\

\vspace*{-0.7cm} \meqn{ \tau_\mathrm{l} =
\frac{2Q_\mathrm{l}}{\omega_0} \,. }{filling time constant for a
loaded cavity}{}
 In the above definition, the electric field
decays exponentially with a time constant $1/\tau$, whereas the
stored energy decays with a time constant $2/\tau$. Be aware that
you can often find textbook definitions of the filling time
constant where the stored energy decays with a time constant
$1/\tau$.

\subsection{Basic cavity parameters for a pillbox cavity}

As a small exercise, in this section we calculate the cavity
parameters that were defined in the previous section for a pillbox
cavity of length $L$ and radius $a$. Since the TM$_{010}$ mode has
no $z$ dependence, we can simplify the expression for the transit
time factor \eqref{eq:Tsvc} to\\
\meqn{ T = \frac{\int\limits_{-L/2}^{L/2} E(0,z) \cos({2\pi
z}/{\beta\lambda}) \dd z}{\int\limits_{-L/2}^{L/2} E(0,z) \, \dd
z} = \frac{\sin({\pi L}/{\beta\lambda})} {\pi L/ \beta\lambda} \,.
}{transit time factor of a pillbox for small velocity changes}{}

\noindent In the case of relativistic particles ($\beta\approx 1$)
and a cavity length $L=\lambda/2$, which is often chosen because
the
cavity can then be cascaded into a multicell structure, we obtain\\

\vspace*{-0.7cm}
\meqn{
T = \frac{2}{\pi} = 0.64
 \,. }{transit time factor of a pillbox for relativistic particles}{}
With real cavities, one usually tries to increase the transit time
factor by shortening the accelerating gap. This can be done by
introducing nose cones on the cavity walls, as shown in
Fig.~\ref{fig:lumped}.

We use the power loss method again to calculate the quality factor
of our pillbox cavity. To evaluate Eq.~\eqref{eq:qfactor}, we need
the stored energy and the power lost in the cavity walls. For the
stored energy, we obtain
\begin{equation}
W =W_\mathrm{el}+W_\mathrm{mag} = 2 W_\mathrm{el} = 2 \int\limits_V \frac{1}{4}\mathbf{E}\cdot \mathbf{D}^* \, \dd V \,.
\end{equation}
\noindent With
\begin{equation}
E_z = E_0 J_0 \left(\frac{j_{01}r}{a}\right) \,, \\
\end{equation}
\noindent we obtain
\begin{equation}
 W = \frac{\varepsilon_0}{2} \int\limits_0^a
\int\limits_0^{2\pi} \int\limits_{-L/2}^{L/2}
E_0^2J_0^2\left(\frac{j_{01}r}{a}\right) r \, \dd r \, \dd \varphi
\, \dd z= \frac{1}{2} E_0^2 \varepsilon_0\pi L a^2 J_1^2(j_{01})
\,.
\end{equation}
\noindent To calculate the dissipated power, we integrate
Eq.~\eqref{eq:powden} over a volume that consists of the inner
surface of the pillbox times the skin depth:
\begin{align}
 P_\mathrm{d} &=
\frac{\delta_\mathrm{s}}{2\kappa} \int\limits_{-L/2}^{L/2}
\underbrace{J_z J_z^*}_{(1/ \delta_\mathrm{s})^2
H_{\varphi}^2(r=a,z)} 2\pi a \, \dd z +
\frac{\delta_\mathrm{s}}{\kappa} \int\limits_0^a
\underbrace{J_r J_r^*}_{(1/ \delta_\mathrm{s})^2} H_{\varphi}^2(r,z=0) 2\pi r \, \dd r \\
&=\frac{E_0^2 \pi R_\mathrm{surf} a}{Z_0^2} J_1^2(j_{01}) (a+L) \,, \label{eq:PD}
\end{align}
\noindent where we have made use of
\begin{equation}
H_{\varphi} = \frac{E_0}{Z_0}J_1\left(\frac{j_{01}r}{a} \right) \,. 
\end{equation}
\noindent Putting everything together, we obtain
\begin{equation}
Q_0 =
\frac{\omega W}{P_\mathrm{d}} = \frac{Z_0^2
\omega}{2R_\mathrm{surf}} \frac{La}{L+a} =
\frac{1}{\delta_\mathrm{s}}\frac{La}{L+a} \propto\sqrt{\omega} \,.
\label{eq:Qpillbox}
\end{equation}
As we can see, the quality factor is a function of the material
constants $\kappa$ and $\mu$ (which are contained in
$\rho_\mathrm{s}$), the frequency, and the geometry of the cavity.
We also note that for the same cavity shape, the quality factor
increases with the frequency in proportion to $ \sqrt{\omega}$.

The accelerating voltage in a pillbox cavity is given by\\

\begin{figure}[h!]
\centering
\includegraphics[width=0.7\textwidth]{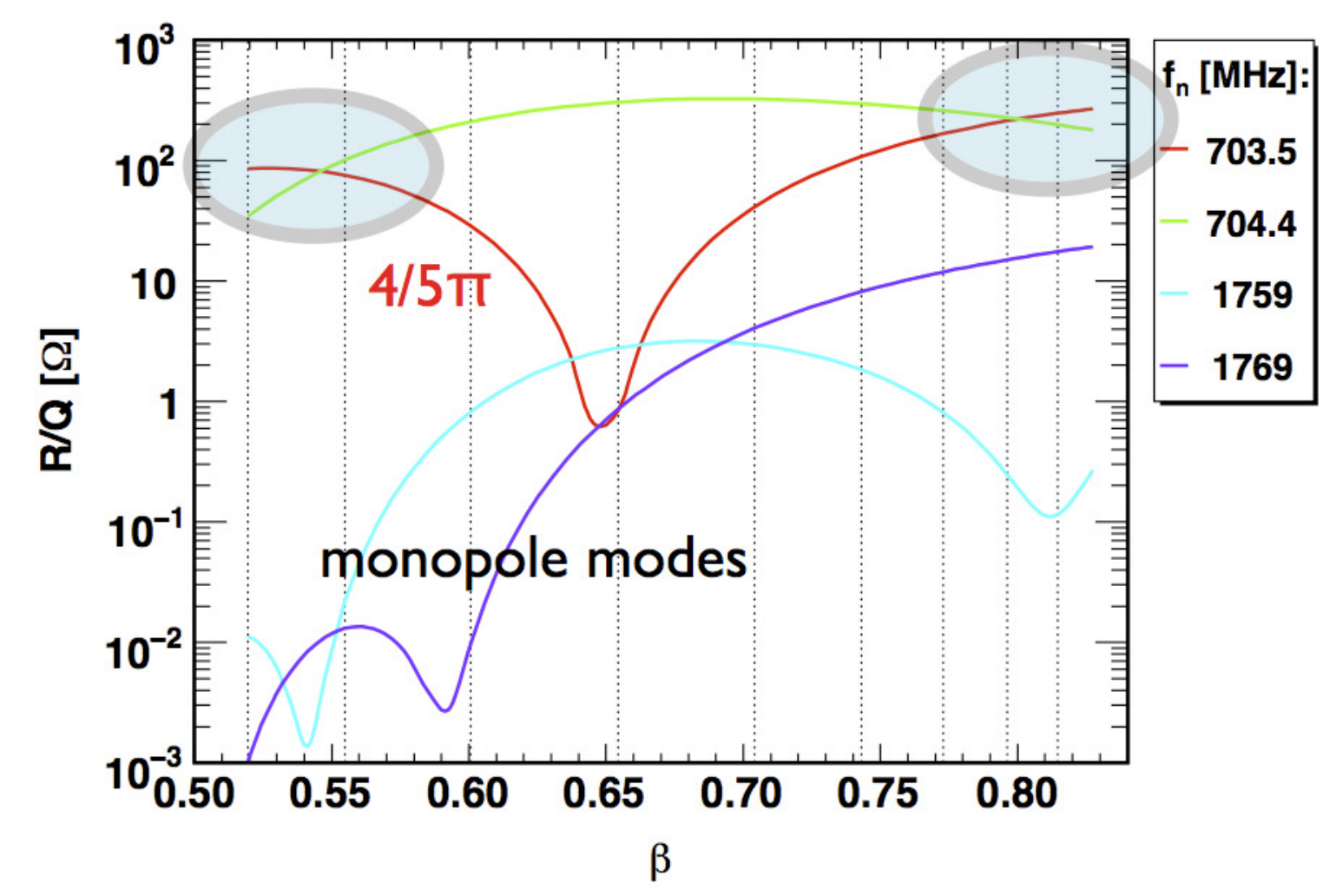}
\caption{Dependence of $(R/Q)$ on particle velocity for a
five-cell superconducting cavity with a geometric $\beta$ of 0.65
and a frequency of 704.4\,MHz. Upper curve, $\pi$ mode (see also
\cite{bib:Schuh}).} \label{fig:rq-beta}
\end{figure}

\vspace*{-0.7cm} \meqn{ V_\mathrm{acc} = V_0 T = E_0 L T = E_0 L
\frac{\sin\left(\pi L / \beta \lambda\right)}{\pi L / \beta
\lambda}
 \,, }{accelerating voltage in pillbox}{eq:V0T}
and is obviously a strong function of the transit time factor. It
therefore depends on the gap length $L$ and the speed of the
particles $\beta$. Owing to their high development costs,
superconducting cavities are often used over large velocity ranges
without changing their cell length, and this results in a
velocity-dependent acceleration efficiency. Figure
\ref{fig:rq-beta} shows $(R/Q) \propto (V_0T)^2$ as a function of
particle velocity for a five-cell superconducting cavity whose
geometric cell length corresponds to a particle speed of
$\beta=0.65$.
\setlength{\captionmargin}{1cm}
\setlength{\captionmargin}{0cm}

If the cavity is used over too large a velocity range, one may
find areas where the passband mode that is closest to the $\pi$
mode (here, the $4/5\pi$ mode) has a higher acceleration
efficiency than the accelerating mode. These areas are highlighted
in Fig.~\ref{fig:rq-beta}, and should be avoided when one is
designing a linac. One should also be aware that the $(R/Q)$ of
the HOMs is highly dependent on the particle velocity.

Using the expressions for the accelerating voltage $V_0T$
(Eq.~\eqref{eq:V0T}) and the dissipated power $P_\mathrm{d}$
(Eq.~\eqref{eq:PD}), we also obtain
 an analytical expression for the effective shunt impedance,\\
\meqn{ R= \frac{(V_0 T)^2}{P_\mathrm{d}} = \frac{Z_0}{\pi
R_\mathrm{surf} J_1^2(j_{01})} \frac{\sin\left( \pi L / \beta
\lambda \right)} {\pi L / \beta \lambda} \frac{L^2}{a(a+L)}
 \,. }{effective shunt impedance of a pillbox}{eq:Rpillbox}

Finally, we calculate the frequency and $(R/Q)$ using Eqs.~\eqref{eq:f-pillbox}, \eqref{eq:Rpillbox}, and \eqref{eq:Qpillbox}:\\

\vspace*{-0.7cm} \meqn{ f_{010}^\mathrm{TM} = \frac{2.405 c}{2\pi
a}
 \,, }{pillbox frequency}{}
\meqn{ \left( \frac{R}{Q} \right) = \frac{2c}{\omega\pi
J_1^2(j_{01})}\frac{\sin\left( \pi L / \beta \lambda \right)}{ \pi
L / \beta \lambda} \frac{L}{a^2}  \,. }{pillbox
$\mathbf{(R/Q)}$}{} As stated before, $(R/Q)$ is indeed
independent of any material parameters. However, it does depend on
the geometry of the cavity and the transit time factor.

\subsection{A cavity as a lumped circuit}
\label{sec:lumped} 
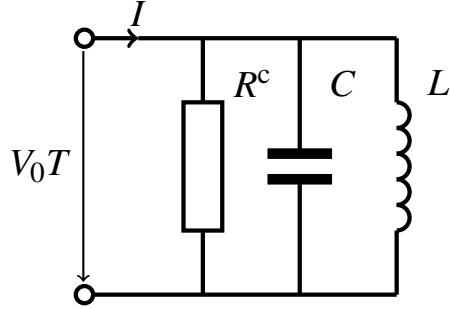
\begin{figure}[h!]
\centering
\begin{tikzpicture}[ultra thick,scale=0.85]

\draw[o-] (-1,0) -- (4,0);
\draw[o->] (-1,4) -- (0,4) node [above] {\Large $I$};
\draw (0,4) -- (4,4);
\draw[->, thick] (-0.85,3.8) -- (-0.85,0.2) node [left, midway] {\Large $V_0T$};

\draw (1,0) -- (1,1);
\draw (0.7,1) rectangle (1.3,3);
\draw (1,3) -- (1,4);

\draw (2.5,0) -- (2.5,1.8);
\draw[line width =4 pt] (2.0,1.8) -- (3,1.8);
\draw[line width =4 pt] (2.0,2.2) -- (3,2.2);
\draw (2.5,2.2) -- (2.5,4);
\draw (4,0) -- (4,1);
\draw (4,3) -- (4,4);
\foreach \x in {1,1.4,1.8,2.2,2.6}
    \draw (4,\x) arc (-90:90:0.2);

\draw (1.3,3.3) node [above, right] {\Large $R^\mathrm{c}$};
 \draw
(2.8,3.3) node [above, right] {\Large $C$};
 \draw (4.3,3.3) node
[above, right] {\Large $L$};

\end{tikzpicture}
\caption{Lumped-circuit equivalent of a resonant cavity}
\label{fig:LCR}
\end{figure}

In the field of RF technology, it is common
practice to describe the behaviour of cavities, RF transmission
lines, and couplers with equivalent lumped circuits. In this
chapter, we shall present only the treatment of a cavity and a
coupler, so that one can understand how to get power into a
cavity. Descriptions of the transmission of RF power and the
associated theory of RF transmission lines can be found in many
textbooks on RF and microwave engineering. We start with the
description of a cavity by a parallel $LCR$ circuit as depicted in
Fig.~\ref{fig:LCR}.

You may remember that the admittance of a parallel circuit is
calculated by adding up the admittances of the individual
elements,
which means that we can write the cavity impedance as\\

\vspace*{-0.7cm} \meqn{ Z^\mathrm{c} = \frac{1}{\displaystyle
i\omega C + 1/ i\omega L + 1/ R^\mathrm{c}}
 \,. }{lumped-circuit cavity impedance}{}

\noindent At resonance ($\omega = \omega_0$), the imaginary parts cancel each other and the cavity impedance becomes real, which means that\\

\vspace*{-0.7cm}
\meqn{
X= \omega_0 L = \frac{1}{\omega_0 C} = \sqrt{\frac{L}{C}}
 \,, }{lumped circuit at resonance}{eq:lcX}
\noindent that the resonance frequency is given by\\

\vspace*{-0.7cm}
\meqn{
\omega_0 = \frac{1}{\sqrt{LC}}
 \,, }{lumped-circuit resonance frequency}{}
\noindent and that the power lost in the resonator is given by\\

\vspace*{-0.7cm} \meqn{ P_\mathrm{d} = \frac{1}{2}
\frac{(V_0T)^2}{R^\mathrm{c}}
 \,. }{lumped-circuit dissipated power}{}

\noindent The stored energy can be written as\\

\vspace*{-0.7cm} \meqn{ W= \frac{1}{2} C (V_0 T)^2 = \frac{1}{2}
\frac{(V_0 T)^2}{\omega_0^2 L}
 \,, }{lumped-circuit stored energy}{eq:lcse}

\noindent and from this we obtain an expression for the quality factor,\\

\vspace*{-0.7cm} \meqn{ Q_0 = \omega_0 \frac{W}{P_\mathrm{d}} =
\omega_0 C R^\mathrm{c} = \frac{R^\mathrm{c}}{\omega_0 L}
 \,. }{lumped-circuit quality factor}{}

Our goal is to relate the lumped elements to the cavity
characteristics, and for this purpose we multiply
Eq.~\eqref{eq:lcse} by $\omega$ and, together with
Eq.~\eqref{eq:lcX}, we obtain
\begin{equation}
\frac{1}{\omega_0 C} = \sqrt{\frac{L}{C}} = \frac{(V_0
T)^2}{2\omega_0 W} = \left( \frac{R^\mathrm{c}}{Q} \right) =
\frac{1}{2} \left( \frac{R}{Q} \right) \,.
\end{equation}
From this, we can understand the difference between the `circuit
ohm' and the `linac ohm', and it also provides a lumped-circuit
description of a cavity, as summarized in
Table~\ref{tab:lumpedRCL}. As we can see, three quantities are
sufficient to describe a resonator. Instead of using $R$, $L$, and
$C$, one can also use the parameters $\omega_0$, $Q_0$, and
$(R/Q)$ to completely characterize an RF cavity, as in
Table~\ref{tab:3quan}.

\begin{table}[h!]
\caption{Lumped-circuit elements of a cavity}
\centering
\begin{tabular}{lc}
\hline\hline
{\bf Lumped circuit} & {\bf Field description} \\
\hline
\parbox[0pt][1.2cm][c]{1cm}{$R^\mathrm{c}$} & \parbox[0pt][1.2cm][c]{3cm}{$\displaystyle\frac{1}{2}R$} \\
\parbox[0pt][1.2cm][c]{1cm}{$C$} & \parbox[0pt][1.2cm][c]{3cm}{$\displaystyle\frac{2}{\omega_0 (R/Q)}$}  \\
\parbox[0pt][1.2cm][c]{1cm}{$L$} & \parbox[0pt][1.2cm][c]{3cm}{$\displaystyle \frac{1}{2\omega_0} \left(\frac{R}{Q} \right)$} \\
\hline\hline
\label{tab:lumpedRCL}
\end{tabular}
\end{table}

\begin{table}[h!]
\caption{Three characteristic quantities of a cavity}
\centering
\begin{tabular}{lc}
\hline\hline
{\bf Lumped circuit} & {\bf Field description} \\
\hline
\parbox[0pt][1.4cm][c]{4cm}{$\displaystyle\omega_0 = \frac{1}{\sqrt{LC}}$} & \parbox[0pt][1.4cm][c]{3cm}{$\displaystyle\frac{2.405 c}{a}$ (pillbox)} \\
\parbox[0pt][1.4cm][c]{4cm}{$\displaystyle Q_0 =
\omega_0 C R^\mathrm{c} = \frac{R^\mathrm{c}}{\omega_0 L}$} & \parbox[0pt][1.4cm][c]{3cm}{$\displaystyle Q_0 =
 \frac{\omega_0 W}{P_\mathrm{d}}$}  \\
\parbox[0pt][1.4cm][c]{4cm}{$\displaystyle\left( \frac{R^\mathrm{c}}{Q}\right) = \sqrt{\frac{L}{C}} =
\frac{1}{2} \left(\frac{R}{Q}\right)$} & \parbox[0pt][1.4cm][c]{3cm}{$\displaystyle \left(\frac{R}{Q} \right) = \frac{(V_0T)^2}{\omega_0 W}$} \\
\hline\hline \label{tab:3quan}
\end{tabular}
\end{table}

\subsection{Getting power into a cavity: couplers}
\begin{figure}[h!]
\centering
    \begin{tikzpicture}[scale=1, ultra thick]
    \draw (-0.5,0) arc (90:100:9);
    \draw (0.5,0) arc (90:80:9);
    \draw (-0.5,-0.03) -- (-0.5,0.7);
    \draw (0.5,-0.03) -- (0.5,0.7);
    \draw[line width = 3pt] (0.47,0.7) -- (0.85,0.7);
    \draw[line width = 3pt] (-0.47,0.7) -- (-0.85,0.7);
    \draw[line width = 3pt, blue!50] (-0.47,0.78) -- (-0.85,0.78);
    \draw[line width = 3pt, blue!50] (0.47,0.78) -- (0.85,0.78);
    \draw[blue!50] (0.5, 0.78) -- (0.5, 1.5);
    \draw[blue!50] (-0.5, 0.78) -- (-0.5, 1.5);
    \draw[blue!50,fill=blue!50] (-0.1,-0.5) rectangle (0.1,1.5);
    \draw (4.5,0) arc (90:100:9);
    \draw (5.5,0) arc (90:80:9);
    \draw (4.5,-0.03) -- (4.5,0.7);
    \draw (5.5,-0.03) -- (5.5,0.7);
    \draw[line width = 3pt] (5.47,0.7) -- (5.85,0.7);
    \draw[line width = 3pt] (4.53,0.7) -- (4.15,0.7);
    \draw[line width = 3pt, blue!50] (4.53,0.78) -- (4.15,0.78);
    \draw[line width = 3pt, blue!50] (5.47,0.78) -- (5.85,0.78);
    \draw[blue!50] (5.5, 0.78) -- (5.5, 1.5);
    \draw[blue!50] (4.5, 0.78) -- (4.5, 1.5);
    \draw[line width = 7pt,blue!50] (5,-0.01) -- (5,1.5);
    \draw[line width = 7pt,blue!50] (5,0) arc (0:-178:0.5);
    \end{tikzpicture}
\caption{Example of an antenna-type coupler (left) and a loop-type
coupler (right)} \label{fig:antennaloop}
\end{figure}
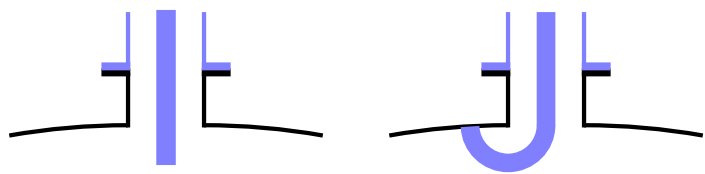

In this section, we shall extend the circuit model to include the
power coupler and also extend our basic equations to describe the
process of coupling power into a cavity. There are two basic types
of couplers that are used in standing-wave cavities:

\begin{itemize}
\item \emph{Antenna/loop couplers:} here, the coupler is usually
some kind of coaxial line, with the outer conductor connected to
the cavity wall and the inner conductor either penetrating into
the cavity volume or connected in a loop to the inner surface of
the cavity (Fig.~\ref{fig:antennaloop}).
 \item \emph{Iris couplers:} here,
the fields in a waveguide are coupled to the cavity fields via an
opening that connects the waveguide to the cavity.
\end{itemize}

When designing a coupler, one has to keep in mind the principle of
reciprocity: the coupler has to produce a field pattern in the
area of the coupling port that is very similar to the field
pattern of the mode that will be excited in the cavity. Looking at
Fig.~\ref{fig:antennaloop}, one can imagine that an antenna-type
coupler would be very effective on the end walls of our pillbox,
where it would couple electrically to the axial electric field
lines. On the cylindrical surface of the pillbox, a loop coupler
would be a better choice, with the loop oriented such that the
azimuthal magnetic field penetrates the loop.

\begin{figure}[h!]
\centering
\includegraphics[width=0.5\textwidth]{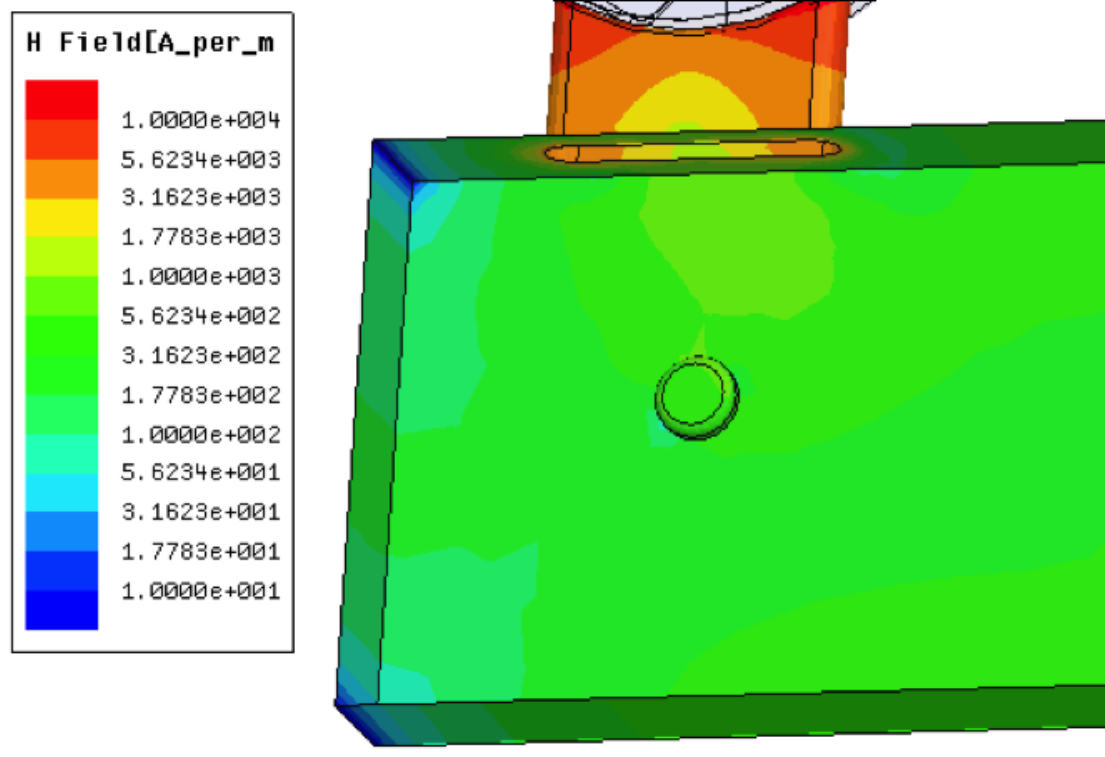}
\caption{Waveguide coupler connected to a Linac4 cavity}
\label{fig:TaCo}
\end{figure}

Figure~\ref{fig:TaCo} shows an example of a `tuner-adjustable
(waveguide) coupler' (TaCo) \cite{bib:TaCo}, as used for the
Linac4 \cite{bib:Linac4} cavities at CERN. In this case a
short-circuited rectangular waveguide is coupled to a
standing-wave cavity via a racetrack-shaped coupling iris. The
coupling factor (more on this later) here is a function of the
position of the short circuit (left side), the height of the
racetrack-shaped coupling channel between the cavity and the
waveguide (on the top), the size of the coupling slot, and the
position of a stub tuner, which is used to fine-tune the coupling.

\begin{figure}[h!]
\centering
\begin{tikzpicture}[ultra thick,scale=0.85]

\draw[-] (-1,0) -- (4,0);
\draw[->] (-1,4) -- (0,4) node [above] {\Large $I$};
\draw (0,4) -- (4,4);
\draw[->, thick] (0.5,3.8) -- (0.5,0.2) node [left, midway] {\Large $V_0T$};

\draw (-1,0) -- (-1,1);
\draw (-1,3) -- (-1,4);
\foreach \x in {1,1.4,1.8,2.2,2.6}
    \draw (-1,\x) arc (270:90:0.2);
\draw (-1.7,0) -- (-1.7,1);
\draw (-1.7,3) -- (-1.7,4);
\foreach \x in {1,1.4,1.8,2.2,2.6}
        \draw (-1.7,\x) arc (-90:90:0.2);
\draw[-o] (-1.7,0) -- (-2.7,0); \draw[-o] (-1.7,4) -- (-2.7,4);
\draw (-1.35,4) node [above] {\LARGE $1 : n$};
 \draw[->,
thick] (-2.8,3.8) -- (-2.8,0.2) node [left, midway] {\Large
$V_\mathrm{gen}$};

\draw[dashed] (-2.7,0) -- (-5.5,0);
\draw[dashed] (-2.7,4) -- (-5.5,4);

\draw[o-] (-5.5,0) -- (-8.5,0); \draw[o-] (-5.5,4) -- (-7.75,4)
node [above] {\Large $I_\mathrm{gen}$}; \draw[->] (-8.5,4) --
(-7.75,4); \draw (-7,0) -- (-7,1); \draw (-7,4) -- (-7,3); \draw
(-7.3,1) rectangle (-6.7,3); \draw (-8.5,0) -- (-8.5,1.5); \draw
(-8.5,2.5) -- (-8.5,4); \draw (-8.5,2) circle (0.5) node {RF};

\draw (1,0) -- (1,1);
\draw (0.7,1) rectangle (1.3,3);
\draw (1,3) -- (1,4);

\draw (2.5,0) -- (2.5,1.8);
\draw[line width =4 pt] (2.0,1.8) -- (3,1.8);
\draw[line width =4 pt] (2.0,2.2) -- (3,2.2);
\draw (2.5,2.2) -- (2.5,4);

\draw (4,0) -- (4,1);
\draw (4,3) -- (4,4);
\foreach \x in {1,1.4,1.8,2.2,2.6}
    \draw (4,\x) arc (-90:90:0.2);

\draw (1.3,3.3) node [above, right] {\Large $R^\mathrm{c}$};
 \draw
(2.8,3.3) node [above, right] {\Large $C$};
 \draw (4.3,3.3) node
[above, right] {\Large $L$};
 \draw (-6.7,3.3) node [above, right]
{\Large $Z_0$};
 \draw (-4.5,3.3) node [above, right] {\Large
$Z_0$};

\draw[dashed, red] (-9.5,-1) -- (-9.5,5); \draw[dashed, red]
(-5.65,-1) -- (-5.65,5); \draw[thick, red,<->] (-9.5,-0.8) --
(-5.65,-0.8) node [below, midway] {Matched generator};
\draw[dashed, red] (-2.55,-1) -- (-2.55,5); \draw[thick, red,<->]
(-5.65,-0.8) -- (-2.55,-0.8) node [below, midway] {Waveguide};
\draw[dashed, red] (-0.7,-1) -- (-0.7,5); \draw[thick, red,<->]
(-2.55,-0.8) -- (-0.7,-0.8) node [below, midway] {Coupler};
\draw[dashed, red] (5.2,-1) -- (5.2,5); \draw[thick, red,<->]
(-0.7,-0.8) -- (5.2,-0.8) node [below, midway] {Cavity};
\draw[line width = 3pt, red,->] (-3.55,5.5) -- (-2.55,5.5) node
[above, midway] {\Large $Z'_\mathrm{c}$}; \draw[line width = 3pt,
red,->] (-1.7,5.5) -- (-0.7,5.5) node [above, midway] {\Large
$Z_\mathrm{c}$};
\end{tikzpicture}
\caption{Lumped-element circuit for RF power source, waveguide,
power coupler, and cavity} \label{fig:pclumped}
\end{figure}
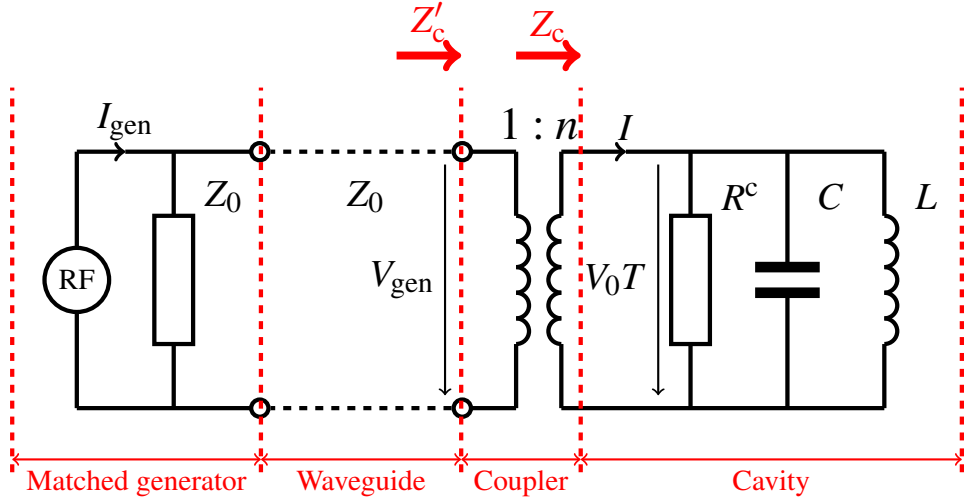

In the ideal case, the power coupler is matched to the
(beam-)loaded cavity, which means that there is no reflected power
returning from the cavity towards the RF power source. Here,
`matched' means that the coupler acts like an ideal transformer
that transforms the impedance $Z_\mathrm{c}$ of the cavity into
the impedance $Z_0$ of the attached waveguide. To keep things
simple, let us assume that the RF generator is also matched to
$Z_0$ so that we can establish a lumped-element circuit as shown
in Fig.~\ref{fig:pclumped}.

Considering the coupler as a transformer, we can write that
\begin{equation}
\left.
\begin{array}{ll}
V_0T &= nV_\mathrm{gen} \\
I &= \displaystyle\frac{I_\mathrm{gen}}{n}
\end{array}
\right\} \Rightarrow Z_\mathrm{c} = \frac{V_0T}{I} = n^2
Z'_\mathrm{c} \,,
\end{equation}
\noindent which means that the cavity impedance\\

\vspace*{-0.7cm} \meqn{ Z_\mathrm{c} = \frac{1}{\displaystyle
i\omega C + 1/ i \omega L + 1/ R^\mathrm{c}}
  }{cavity impedance}{}
\noindent is transformed into\\

\vspace*{-0.7cm} \meqn{ Z'_\mathrm{c} = \frac{1}{\displaystyle
i\omega n^2 C + n^2 / i \omega L + n^2 / R^\mathrm{c}} \,,
}{cavity + coupler impedance}{}
which is the impedance `seen' from the waveguide. The stored energy in the resonator, expressed in lumped-circuit values, becomes\\

\vspace*{-0.7cm} \meqn{ W= \frac{C}{2 (V_0T)^2} = n^2 \frac{C}{2
V_\mathrm{gen}^2} \,, }{stored energy}{}
\noindent and the dissipated power can be written as\\

\vspace*{-0.7cm} \meqn{ P_\mathrm{d} =
\frac{(V_0T)^2}{2R^\mathrm{c}} = n^2
\frac{V_\mathrm{gen}^2}{2R^\mathrm{c}}
 \,. }{dissipated power}{}

Now we can define the quality factor of the unloaded cavity with lumped-circuit elements: \\

\vspace*{-0.7cm} \meqn{ Q_0 = \frac{\omega_0 W}{P_\mathrm{d}} =
\omega_0 R^\mathrm{c} C
 \,. }{unloaded $Q$}{}
\noindent When the generator is switched off, not only will the
stored energy in the cavity be dissipated in the cavity walls, but
a power $P_\mathrm{ex}$ will also leak out through the power coupler, where\\

\vspace*{-0.7cm} \meqn{ P_\mathrm{ex} =
\frac{V_\mathrm{gen}^2}{2Z_0}
 \,. }{}{eq:pex}

\noindent Using $P_\mathrm{ex}$, one can define the quality factor of the external load. The external $Q$ is thus defined as\\

\vspace*{-0.7cm} \meqn{ Q_\mathrm{ex} = \frac{\omega_0
W}{P_\mathrm{ex}} = n^2 \omega_0 Z_0 C
 \,. }{external $Q$}{}

\subsubsection{Undriven cavity}
In order to understand the power balance and matching for a driven
cavity with beam, we start with a simple case, assuming that
the RF is switched off and that there is no beam in the cavity. The power balance is then\\

\vspace*{-1.1cm}
\meqn{
 P_\mathrm{tot} = P_\mathrm{d} + P_\mathrm{ex}
 \,,  }{power balance of undriven cavity}{}
 \noindent with which we can define the so-called `loaded $Q$' of the ensemble of cavity and coupler by\\

 \vspace*{-0.9cm}
 \meqn{
 \frac{1}{Q_\mathrm{l}} = \frac{1}{Q_\mathrm{ex}} + \frac{1}{Q_0}
 \,.  }{loaded $Q$}{}
\noindent The coupling between the cavity and the waveguide is described by the coupling factor $\beta$, where\\

\vspace*{-0.9cm} \meqn{ \beta = \frac{P_\mathrm{ex}}{P_\mathrm{d}}
= \frac{Q_0}{Q_\mathrm{ex}} = \frac{R^\mathrm{c}}{n^2 Z_0}
 \,. }{coupling factor}{}
\noindent Optimum power transfer between the cavity (+ coupler)
and the waveguide takes place when the impedance at the coupler
input equals the waveguide impedance at the resonance frequency of
the cavity. We know that the cavity impedance becomes real at
resonance, which means that
\begin{equation}
Z_\mathrm{c} = R^\mathrm{c} =n^2Z'_\mathrm{c} \stackrel{!}{=}
n^2Z_0 \hspace{0.5cm} \Rightarrow \hspace{0.5cm} \beta=1 \,.
\end{equation}
It is important to keep in mind that the `matching condition'
$\beta=1$ is only valid for a cavity without beam.

\subsubsection{RF on, beam on}
Once we take the beam loading into account, the power needed in
the cavity increases and will yield a different value for the
coupling factor $\beta$ at the point of optimum power transfer. A
simple way to introduce the beam is to treat it as an
additional loss in the cavity, which can be added to the power dissipated in the cavity walls:\\

\vspace*{-0.7cm} \meqn{ P_\mathrm{db} = P_\mathrm{d} +
P_\mathrm{b}
 \,. }{dissipated power + beam power}{}
As in the case without beam, maximum power transfer to the cavity
is achieved when the input impedance of the coupler equals the
impedance of the waveguide. This condition yields zero reflection
and also implies that the power needed in the cavity,
$P_\mathrm{bd}$ (for losses and beam), has to be equal to
$P_\mathrm{ex}$ as defined in Eq.~\eqref{eq:pex}. This means that
\begin{equation}
\frac{P_\mathrm{ex}}{P_\mathrm{db}} = 1 =
\frac{Q_\mathrm{0b}}{Q_\mathrm{ex}} \hspace{0.5cm} \Rightarrow
\frac{P_\mathrm{ex}}{P_\mathrm{d}} =
1+\frac{P_\mathrm{b}}{P_\mathrm{d}} \,,
\end{equation}
where we have introduced a quality factor $Q_\mathrm{0b}$ for the cavity plus beam. For the matched condition, we therefore obtain a coupling factor of\\

\vspace*{-0.9cm} \meqn{  \beta = 1+
\frac{P_\mathrm{b}}{P_\mathrm{d}}
 \,, }{matched coupling factor with beam}{}
and the following quality factors:

\vspace*{-0.5cm} \meqn{ Q_\mathrm{ex} = Q_\mathrm{0b} =
\frac{\omega_0 W}{P_\mathrm{b} + P_\mathrm{d}} = \frac{Q_0}{1 +
P_\mathrm{b} / P_\mathrm{d}}= \frac{Q_0}{\beta}
 \,, }{external $Q$ with beam}{}

\vspace*{-0.7cm} \meqn{ Q_\mathrm{l} = \frac{Q_0}{1 + \beta} =
\frac{Q_0}{2 + P_\mathrm{b} / P_\mathrm{d}}
 \,. }{loaded $Q$ with beam}{}
In the case of a superconducting cavity, one can generally assume
that $P_\mathrm{b} \gg P_\mathrm{d}$, which means that the
coupling factor for the matched condition can be written as

\vspace*{-0.5cm} \meqn{ \beta = 1 +
\frac{P_\mathrm{b}}{P_\mathrm{d}} \approx
\frac{P_\mathrm{b}}{P_\mathrm{d}}
 \,. }{matched coupling factor for SC cavity + beam}{}
Using

\vspace*{-0.7cm} \meqn{ P_\mathrm{b} = I_\mathrm{beam} V_0 T
\cos\phi_\mathrm{s}
 \,, }{}{}\\
we can write a simple expression for calculating the loaded and
external $Q$ values for a superconducting cavity as follows:

\vspace*{-0.5cm} \meqn{ Q_\mathrm{l} \approx Q_\mathrm{ex} \approx
\frac{Q_0}{P_\mathrm{beam}/P_\mathrm{d}} =
\frac{V_0T}{(R/Q)I_\mathrm{beam} \cos\phi_\mathrm{s}}
 \,. }{$Q_\mathbf{l/ex}$ for SC cavity}{}
The results in this paragraph are summarized in
Table~\ref{tab:QPlex}.

\begin{table}[h!]
\centering
\caption{Definitions of $Q$ values and coupling factors for driven and undriven cavities}
\begin{tabular}{l|c|c}
\hline\hline
 & {\bf Undriven cavity} & {\bf Driven cavity} \\
 \hline  & \multicolumn{2}{c}{\parbox[0pt][1.2cm][c]{4cm}{\centerline{$\displaystyle \frac{1}{Q_\mathrm{l}} = \frac{1}{Q_\mathrm{ex}} + \frac{1}{Q_0}$}}} \\
 {\bf General} & \multicolumn{2}{c}{\parbox[0pt][1.2cm][c]{4cm}{\centerline{$\displaystyle \beta = \frac{P_\mathrm{ex}}{P_\mathrm{d}}
   = \frac{Q_0}{Q_\mathrm{ex}}$}}} \\
  & \multicolumn{2}{c}{\parbox[0pt][1.2cm][c]{4cm}{\centerline{$\displaystyle Q_\mathrm{l} = \frac{Q_0}{1 + \beta}$}}} \\
  \hline
  & \parbox[0pt][1.2cm][c]{4.5cm}{\centerline{$\displaystyle \frac{P_\mathrm{ex}}{P_\mathrm{d}} = \frac{Q_0}{Q_\mathrm{ex}} = 1 \Rightarrow \beta = 1$}}
   & \parbox[0pt][1.2cm][c]{4.5cm}
  {\centerline{$\displaystyle \frac{P_\mathrm{ex}}{P_\mathrm{db}} = \frac{Q_\mathrm{0b}}{Q_\mathrm{ex}} = 1 \Rightarrow \beta =
  1+ \frac{P_\mathrm{b}}{P_\mathrm{d}}$}} \\
 {\bf Matched case} & \parbox[0pt][1.2cm][c]{4.5cm}{\centerline{$\displaystyle Q_\mathrm{ex} = Q_0 = \frac{\omega_0 W}{P_\mathrm{d}}$}}
   & $\displaystyle Q_\mathrm{ex} = Q_\mathrm{0b} = \frac{\omega_0 W}{P_\mathrm{d} + P_\mathrm{b}}$ \\
  & \parbox[0pt][1.2cm][c]{4.5cm}{\centerline{$\displaystyle Q_\mathrm{l} = \frac{Q_0}{2}$}} & $\displaystyle Q_\mathrm{l} = \frac{Q_0}{2
  + P_\mathrm{b} / P_\mathrm{d}}$ \\
  \hline\hline
\end{tabular}
\label{tab:QPlex}
\end{table}

\subsection{`Matching' a cavity}
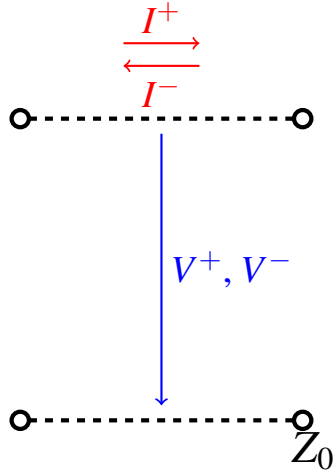
\begin{figure}[h]
\centering
\begin{tikzpicture}[ultra thick]

\draw[dashed, o-o] (0,0) -- (4,0) node [below] {\LARGE $Z_0$};
\draw[dashed, o-o] (0,4) -- (4,4);
\draw[thick,->,blue] (2,3.8) -- (2,0.2) node [right, midway] {\Large $V^{+}$, $V^{-}$};
\draw[thick,->,red] (1.5,5) -- (2.5,5) node [above, midway] {\Large $I^{+}$};
\draw[thick,<-,red] (1.5,4.7) -- (2.5,4.7) node [below, midway] {\Large $I^{-}$};

\end{tikzpicture}
\caption{\label{fig:tline}Voltages and currents along a transmission line}
\end{figure}

In the last section, it was claimed that part of an
electromagnetic wave is reflected when it `sees' a change in
impedance during its propagation. In fact, the whole purpose of
the power coupler was to transform the impedance of the waveguide
into the impedance of the cavity. In this last section, we shall
see why this is so. For this purpose, we look at a transmission
line as shown in Fig.~\ref{fig:tline}.

This transmission line is representative of a waveguide, a coaxial
line, or any other kind of transport geometry used to guide
electromagnetic waves. Since waves can travel in the positive and
negative $z$ directions, a sign convention is introduced for the
associated voltages and currents, as shown in
Fig.~\ref{fig:tline}, where the voltage vectors of the forward and
reflected waves have the same direction and the current vectors
have opposite directions.

Using the same time and location dependence as for the electric
and magnetic fields in a wave\-guide, we can write
\begin{align}
V &= V_0 e^{i(kz - \omega t)} + \Gamma V_0 e^{i(-kz - \omega t)} \,,\\
I &= \frac{V_0}{Z_0} e^{i(kz - \omega t)} - \Gamma \frac{V_0}{Z_0}
e^{i(-kz - \omega t)} \,,
 \intertext{where we have introduced a
reflection coefficient $\Gamma$. If we connect a cavity to an
impedance $Z'_\mathrm{c}$ at $z=0$, the expressions above simplify
to}
V &= V_0 e^{-i\omega t} (1+ \Gamma) \,, \\
I &= \frac{V_0}{Z_0} e^{-i\omega t} (1- \Gamma) \,,
 \intertext{and
the cavity impedance can be expressed in terms of the transmission
line impedance $Z_0$ and the reflection coefficient $\Gamma$:}
Z'_\mathrm{c} &= \frac{V}{I} = Z_0 \frac{1+\Gamma}{1-\Gamma} \,.
 \intertext{We can
then rearrange the equation for the reflection coefficient and
obtain} \Gamma &= \frac{Z'_\mathrm{c} - Z_0}{Z'_\mathrm{c} + Z_0}
= \frac{1 - \beta}{1 + \beta} \,.
\end{align}
From this equation, we can see that the reflection disappears only
for $Z'_\mathrm{c} = Z_0$, the `matched condition', where the
waveguide impedance equals the cavity impedance. In the case
without beam, this corresponds to a coupling factor of $\beta=1$.

In the context of matching, we therefore have to consider the
following points:
\begin{itemize}
\item At the resonance frequency, the power coupler transforms the
cavity impedance into the impedance of the waveguide.
 \item If the
cavity is resonating off-resonance or if the coupler is
mismatched, power is reflected and travels back to the RF source.
 \item
Since the cavity impedance depends on the $Q$ of the cavity, and
since in reality most cavities have different $Q$ values, every
cavity needs a different matching.
 \item Beam loading increases the
power needed in the cavity and changes the loaded $Q$ and the
cavity impedance. Power couplers are usually matched for the case
with beam loading.
 \item During the start of an RF pulse (before the
arrival of the beam), when the cavity is being `filled' with RF
power, the cavity is always mismatched, which means we need to
make sure that the reflected power does not damage the RF source
(e.g., by using a circulator between the cavity and the RF
source).
\end{itemize}

\begin{figure}
\centering
\includegraphics[width=0.7\textwidth]{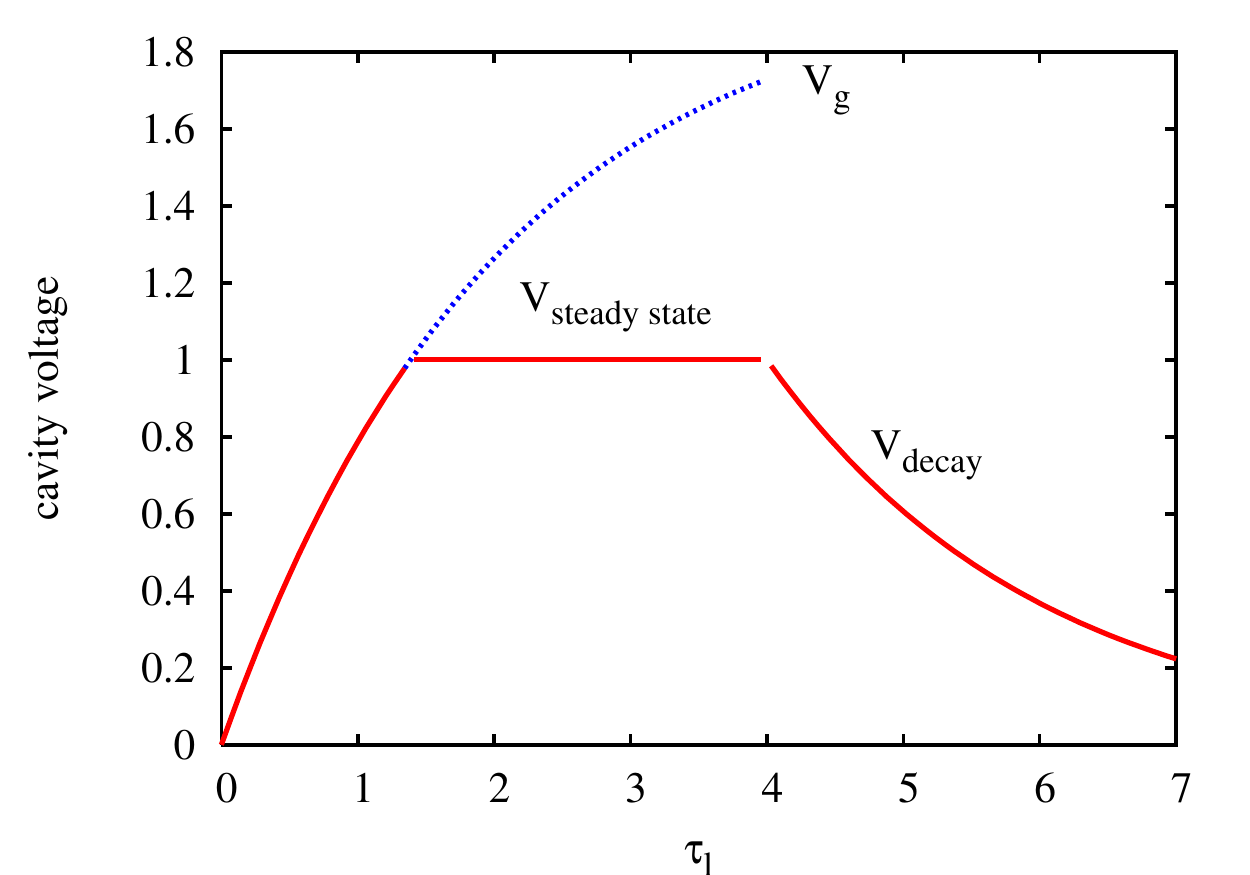}
\caption{\label{fig:SCpulsed}Voltage profile in a pulsed
superconducting cavity}
\end{figure}

The last point is especially important in the case of
superconducting cavities, where the dissipated power is negligible
with respect to the power taken by the beam. In this case one has,
basically, full reflection of the RF wave at the beginning of the
RF pulse before the cavity field increases to its nominal level.
At that point the beam should enter the cavity, and from then
onwards the RF generator is matched to the power needs of the RF
cavity. After the RF signal is switched off, the cavity voltage
decays exponentially, as shown in Fig.~\ref{fig:SCpulsed} (more
details can be found in \cite{bib:FGpowcon}).

\section*{Acknowledgements}

In the preparation of this chapter, I have made extensive use of
the material listed in the Bibliography below.

\section*{Bibliography}
\begin{itemize}
 \item T. Weiland, M. Krasilnikov, R. Schuhmann, A. Skarlatos, and
M. Wilke, Review of theory (I, II, III), CAS RF Engineering,
Seeheim, Germany (2005).
 \item T. Wangler, \emph{Principles of RF Linear Accelerators} (Wiley-VCH, Weinheim,
2004).
 \item A. Wolski, Theory of
electromagnetic fields, CAS RF Engineering, Ebeltoft, Denmark
(2010).
 \item H. Henke, Theoretische Elektrotechnik, German
script of lectures on electrodynamics at the Technical University
of Berlin (1992)
 \item H. Henke, Basic concepts I and II, CAS RF
Engineering, Seeheim, Germany (2005).
 \item K. Simonyi, \emph{Foundations
of Electrical Engineering}, Vol. 3 (Pergamon Press, New York,
1963). [Hungarian edition, \emph{Elméleti
 villamossagtan Tankonyvkiado} (Budapest, 1973); German edition, \emph{Theoretische Elektrotechnik} (VEB
Deutscher Verlag der Wissenschaften, 1973).]
 \item H. Padamsee,
J. Knobloch, and T. Hays, \emph{RF Superconductivity for
Accelerators} (Wiley, New York, 2008).
\end{itemize}

\appendix

\section{Cartesian coordinates ($x$, $y$, $z$)}

\subsection{Differential elements}

\eqn{\hfill
\dd\mathbf{l} = \threevector
{\dd x} {\dd y}{\dd z} \,, \hfill}{path element}{}\\
\eqn{\hfill \dd V = \dd x \, \dd y \, \dd z \,. \hfill}{volume
element}{}

\subsection{Differential operators}
\eqn{\hfill\nabla\phi=\threevector{\frac{\partial\phi}{\partial x}}{\frac{\partial\phi}{\partial y}}{\frac{\partial\phi}{\partial z}} \,,
\hfill}{gradient}{}\\
\eqn{\hfill\nabla\cdot\mathbf{a} = \displaystyle\frac{\partial
a_x}{\partial x} +\frac{\partial a_y}{\partial y} + \frac{\partial
a_z}{\partial z} \,, \hfill}{divergence}{}

$\left. \right.$\\
\eqn{\hfill\nabla\times\mathbf{a} = \threevector{\frac{\partial a_z}
{\partial y} - \frac{\partial a_y}{\partial z}}{\frac{\partial a_x}{\partial z}-\frac{\partial a_z}{\partial x}}
{\frac{\partial a_y}{\partial x}-\frac{\partial a_x}{\partial y}} \,, \hfill}{curl}{}\\
\eqn{\hfill\Delta\phi =
\displaystyle\frac{\partial^2\phi}{\partial x^2} +
\frac{\partial^2\phi}{\partial y^2} +
\frac{\partial^2\phi}{\partial z^2} \,. \hfill}{Laplace}{}

\section{Cylindrical coordinates ($r$, $\phi$, $z$)}
\label{sec:cylindrical}

\subsection{Transformations}

\meqn{
x &= r \cos\varphi  \,, \\
y &= r \sin\varphi \,, \\
z &= z  \,,}{}{}

with $ 0 \le r \le \infty$, $0 \le \varphi \le 2\pi $.

\subsection{Differential elements}
\eqn{\hfill \dd\mathbf{l} = \threevector{\dd r}{r \, \dd
\varphi}{\dd z} \,, \hfill}
{path element}{}\\
\eqn{\hfill \dd V = r \, \dd r \, \dd \varphi \, \dd z
 \,. \hfill}{volume element}{}

\subsection{Differential operators}

$\left. \right.$\\
\eqn{\hfill
\nabla\phi = \threevector{
    \frac{\partial \phi}{\partial r}}
{
    \frac{1}{r}\frac{\partial \phi}{\partial \varphi}}
{
    \frac{\partial \phi}{\partial z}}
 \,, \hfill}{gradient}{}\\
\eqn{\hfill \nabla \cdot \mathbf{a} = \displaystyle
\frac{1}{r}\frac{\partial (r a_r)}{\partial r} +
\frac{1}{r}\frac{\partial a_{\varphi}}{\partial\varphi} +
\frac{\partial a_z}{\partial z}  \,, \hfill}{divergence}{}

$\left. \right.$\\
\eqn{\hfill
\nabla\times \mathbf{a} = \threevector{
    \frac{1}{r}\frac{\partial a_z}{\partial\varphi} - \frac{\partial a_{\varphi}}{\partial z}}
{
    \frac{\partial a_r}{\partial z} - \frac{\partial a_z}{\partial r}}
{
    \frac{1}{r} \left(
        \frac{\partial (ra_{\varphi})}{\partial r} -
        \frac{\partial a_r}{\partial\varphi}
\right)}
 \,, \hfill}{curl}{}\\
\eqn{\displaystyle\hfill \Delta\phi = \frac{\partial^2
\phi}{\partial r^2} + \frac{1}{r}\frac{\partial\phi}{\partial r} +
\frac{1}{r^2}\frac{\partial^2\phi}{\partial \varphi^2}
 + \frac{\partial^2\phi}{\partial z^2} \,. \hfill}{Laplace}{}

\section{Spherical coordinates ($r$, $\vartheta$, $\varphi$)}
\label{sec:spherical}

\subsection{Transformations}
\meqn{
x &= r\sin\vartheta\cos\varphi  \,, \\
y &= r\sin\vartheta\sin\varphi  \,, \\
z &= r\cos\vartheta  \,,}{}{}

with $0 \le r \le \infty$, $0\le \vartheta \le \pi$, $0\le\varphi
\le 2\pi$.

\subsection{Differential elements}
\eqn{\hfill\dd\mathbf{l} = \threevector{\dd r}{r \, \dd \vartheta}{r\sin\vartheta \, \dd \varphi} \,, \hfill}{path element}{}\\
\eqn{\hfill\dd V = r^2\sin\vartheta \, \dd r \, \dd \vartheta \,
\dd\varphi \,. \hfill}{volume element}{}

\subsection{Differential operators}
\eqn{\hfill
\nabla\phi = \threevector{\frac{\partial\phi}{\partial r}}
{\frac{1}{r}\frac{\partial\phi}{\partial\vartheta}}
{\frac{1}{r\sin\vartheta}\frac{\partial\phi}{\partial\varphi}}
 \,, \hfill}{gradient}{}\\
\eqn{\hfill \nabla\cdot\mathbf{a} =
\displaystyle\frac{1}{r^2}\frac{\partial(r^2a_r)}{\partial r} +
\frac{1}{r\sin\vartheta}
\frac{\partial(a_{\vartheta}\sin\vartheta)}{\partial\vartheta} +
\frac{1}{r\sin\vartheta}\frac{\partial
a_{\varphi}}{\partial\varphi}
 \,, \hfill}{divergence}{}\\
\eqn{\hfill \nabla\times\mathbf{a} =
\threevector{\frac{1}{r\sin\vartheta}\left(
\frac{\partial(a_{\varphi}\sin\vartheta)}{\partial\vartheta} -
\frac{\partial
a_{\vartheta}}{\partial\varphi}\right)}{\frac{1}{r}\left(
\frac{1}{\sin\vartheta}\frac{\partial a_r}{\partial\varphi} -
\frac{\partial(r a_{\varphi})}{\partial r}
\right)}{\frac{1}{r}\left( \frac{\partial (r
a_{\vartheta})}{\partial r} - \frac{\partial
a_r}{\partial\vartheta}   \right)}
 \,, \hfill}{curl}{}\\
\meqn{ \Delta\phi &= \displaystyle\frac{\partial^2\phi}{\partial
r^2} + \frac{2}{r}\frac{\partial\phi}{\partial r} +
\frac{1}{r^2\sin\vartheta}\frac{\partial}{\partial\vartheta}\left( \sin\vartheta\frac{\partial\phi}{\partial\vartheta} \right)\\
&+
\frac{1}{r^2\sin^2\vartheta}\frac{\partial^2\phi}{\partial\varphi^2}
 \,.}{Laplace}{}

\section{Useful relationships}

\begin{align}
\nabla\cdot\left(\mathbf{a}\times\mathbf{b}\right) &= \mathbf{b} \cdot\left(\nabla\times\mathbf{a} \right) - \mathbf{a}
\cdot\left(\nabla\times\mathbf{b}\right) \,, \label{eq:va1} \\
\nabla \cdot \left( \nabla \times \mathbf{a} \right) &= 0 \,.
\end{align}



\end{document}